\documentclass{article}
\usepackage{graphicx} 
\usepackage[margin=1.0in]{geometry}
\usepackage{amsmath}
\usepackage{amsbsy}
\usepackage{amsthm}
\usepackage{amssymb}
\usepackage{mathtools}
\usepackage[dvipsnames]{xcolor}
\usepackage{hyperref}
\usepackage{caption}

\newcommand{\ie}[0]{\textit{i.e. }}
\newcommand{\eg}[0]{\textit{e.g. }}

\usepackage[
    backend=biber,
    style=numeric,
]{biblatex}
\title{Characteristic Bending in Incompressible Flows}
\author{Matthew Blomquist\thanks{Corresponding author: mblomquist@ucmerced.edu}, Stéphane Gaudreault, Maxime Theillard}

\begin{document}

\maketitle

\begin{abstract}
%
We present the Characteristic Bending (CB) method, a general framework for advecting quantities under incompressible velocity fields. The method builds on standard semi-Lagrangian advection by interpreting the backward-in-time characteristic reconstruction as the construction of a reference map, a diffeomorphism between the current and initial geometries of the advected space. From this viewpoint, the CB method applies a volume-preserving projection to the map, systematically removing spurious compressible errors arising from time integration, interpolation, or from velocity fields that are only approximately divergence-free. This projection “bends” the characteristics toward the divergence-free space, preserving mass and geometric features of the advected fields, even in the presence of significant error. We demonstrate the method in both two and three dimensions using benchmark problems and for multiphase flows governed by the incompressible Navier–Stokes equations. The results show that the CB method serves as a drop-in replacement for traditional semi-Lagrangian schemes and as an augmentation of reference map formulations, offering improved robustness and accuracy in incompressible flow simulations.
\end{abstract}

\section{Introduction}
%
Advection under incompressible velocity fields arises in a wide range of applications, from oceanic transport to multiphase and interfacial flows. A defining property of incompressibility is that volumes are preserved exactly as they are transported. Enforcing this property numerically, however, is challenging. Errors from time integration, interpolation, or from velocity fields that are only approximately divergence-free introduce spurious compressibility, which accumulates over the course of a simulation. This can result in mass loss, distortion of interfaces, and reduced fidelity in simulations intended to resolve fine-scale structure. Although a variety of numerical strategies have been developed to mitigate these issues, these methods typically assume that the advecting velocity is incompressible and thus may not eliminate all sources of error.

In this work, we address these challenges by developing the Characteristic Bending (CB) method, a volume-preserving advection scheme designed specifically for incompressible flows. Our approach builds on standard numerical advection schemes, introducing a correction mechanism that systematically removes spurious compressibility arising from discretization error or how incompressibility is enforced in a larger numerical scheme. The resulting method is broadly applicable for incompressible transport problems and can even be used to augment existing advection methods. To place our approach in context, we first review several common numerical strategies for advection, beginning with the semi-Lagrangian (SL) method. 

The SL method is a standard numerical approach for advection problems that circumvents traditional timestep restrictions imposed by the Courant–Friedrichs–Lewy (CFL) condition \cite{robert1981stable, staniforth1991semi}. The method advances the solution by tracing characteristics of the velocity field backward in time and interpolating the advected quantity at the departure point from the previous timestep. For scalar advection with linear interpolation, the SL method is unconditionally stable \cite{falcone1998slconvergence}, which makes it particularly effective for simulations that require long-time advection \cite{staniforth1991semi} or for adaptive mesh refinement (AMR) settings, where small cut-cells can otherwise impose severe timestep constraints \cite{rosatti2005semi}. The primary drawback is that SL methods are not inherently conservative, and their accuracy is dominated by interpolation and integration errors, which accumulate systematically over repeated steps.

A wide range of conservative extensions of the semi-Lagrangian method have been developed to address the lack of inherent conservation. Flux-form conservative formulations, such as SLICE and CSLAM, recast the method in a finite-volume framework to enforce mass conservation through flux reconstruction \cite{zerroukat2002slice, lauritzen2010cslam, priestley1993quasi}. Related approaches include the cubic interpolated propagation method with conservative semi-Lagrangian corrections (CIP–CSL) \cite{yabe2001exactly, xiao2001completely}, trajectory correction schemes based on the Monge-Ampère equation \cite{cossette2014monge}, and finite-volume variations adapted to cut-cell grids \cite{rosatti2005semi}. Lagrange–Galerkin methods are an alternate approach that solves the conservative form of the advection equation using a variational finite-element framework, with conservation enforced through a mapping of upstream volumes \cite{morton1985generalised}. These schemes achieve both conservation and high accuracy, and when paired with spectral elements (see, e.g. \cite{giraldo2000lagrange, giraldo2000lagrange2}), can resolve fine-scale features beyond the reach of flux-form formulations. Their primary drawback is the high computational expense associated with conservative mapping, which has limited their adoption in large-scale applications.

While conservative semi-Lagrangian methods for incompressible advection improve accuracy and mass conservation, they typically rely on the assumption that the velocity field is exactly divergence-free (\textit{e.g.}, in \cite{lauritzen2010cslam}). When the velocity field is only approximately divergence-free, such as in approximate projection methods \cite{almgren1996numerical, almgren1997cartesian, almgren2000approximate}, semi-Lagrangian schemes will still follow characteristics of the inexact field. In the context of Navier–Stokes solvers, this rarely causes significant issues when the deviations from the divergence-free space are small, for example, in \cite{blomquist2024stable}. In other settings, a projection based on the Helmholtz–Hodge decomposition can be used to clean the velocity field, as in the classical MAC projection \cite{almgren1998conservative}.

An additional limitation of semi-Lagrangian schemes is that, after every timestep, the advected quantity is interpolated and the reference to the original configuration is lost. Even if this process is performed in a conservative manner, repeated interpolation gradually degrades fine-scale features unless very high-order methods are employed. This compounding error motivates an alternative perspective. Rather than advecting and interpolating the field directly, we can advect a map of the characteristics and compute the solution using a reference configuration.

The reference map technique (RMT) introduced in \cite{kamrin2012reference} makes this idea more precise. Originally developed to bridge Eulerian and Lagrangian formulations in fluid–structure interaction (FSI), RMT advances a reference configuration map that encodes material deformation. The construction parallels classical solid mechanics, where deformation gradients are the fundamental kinematic variable, while simultaneously providing a consistent link to Eulerian transport. While the formulation in \cite{kamrin2012reference} focused on constitutive modeling for solids, the underlying mechanism naturally extends to more general advection problems.

The Characteristic Mapping Method (CMM) \cite{mercier2020characteristic, yin2021characteristic} and the Coupled Level Set Reference Map (CLSRM) \cite{bellotti2019rm} are two general approaches that extend the reference map perspective specifically to advection problems. In CMM, the characteristic map (\textit{i.e.} the reference map) is evolved under the advecting velocity, and the advected quantity is reconstructed by composition with this map. Accuracy is maintained by decomposing the global deformation into near-identity submaps, evolved on a coarse grid and periodically remapped by interpolation onto a finer grid. In contrast, CLSRM advances a single reference map and avoids interpolation-based remapping, instead introducing a selective restarting procedure that reinitializes the map to prevent a loss of bijectivity. Despite this difference in strategy, both methods can achieve very high accuracy when characteristics are well resolved numerically.

An important feature of the reference map formulation is that the map gradient encodes deformation metrics of the flow. This parallels the role of deformation gradients in solid mechanics, but here it provides a measure of how numerical advection distorts volume. Building on this observation, Theillard \cite{theillard2021vprm} introduced a volume-preserving projection for reference maps. The central idea is to measure the deviation of the map from the volume-preserving space and then apply a projection that enforces this constraint. For analytic velocity fields, where characteristics are well resolved and the velocity is exactly divergence-free, the projection provides little benefit and can even slightly degrade interface accuracy. However, for more practical applications, such as rising droplets in density-stratified media, the projection substantially improves performance, reducing mass loss by two orders of magnitude with negligible impact on interface location. These results highlight that the projection is most effective when the advecting velocity is inexact and contains small compressible components, and this insight motivates the framework developed in the present work.

An alternative strategy for enforcing volume preservation within the Characteristic Mapping framework was proposed in \cite{holman2024volume}, where a symplectic time integration scheme was employed to guarantee exact volume conservation during advection. While this approach is robust and structure-preserving, it assumes that the advecting velocity field is already divergence-free and the symplectic integrator itself does not correct residual compressibility that may arise, for example, from approximate projection methods. Similarly, \cite{lee2019robust} introduced a volume-conserving extension of the level set method for free-surface flows, achieving strict mass conservation through a geometric reconstruction procedure. Although effective, both approaches do not remove spurious compressible modes from the underlying velocity field.

In this work, we present the CB method. This method builds on foundational ideas of the CMM, CLSRM, and the VPRM to create a generalized method for advecting flows with incompressible velocity fields. We begin by recasting the classic SL scheme in the reference map setting and using this formulation to produce near-identity maps of the characteristics at each time step. Next, we apply the volume-preserving projection of \cite{theillard2021vprm} to correct numerical errors from time integration and interpolation and filter any compressible modes present in the advecting velocity. The central idea is to view the projection as bending the characteristics toward the divergence-free space. Applying this to near-identity maps ensures that corrections remain minimal, well posed, and effective over long integrations. When used as the primary advecting scheme, CB can further augment the reference map framework by creating long-time maps through the composition of CB-corrected steps. This yields a natural extension of the CLSRM method in which a single reference map is advected using CB to filter spurious compressibility through a volume-preserving transformation. The result is a robust methodology that can be viewed either as a drop-in replacement for SL advection or as a volume-preserving extension of reference map approaches.

The remainder of this paper is organized as follows. Section~\ref{sec:preliminaries} introduces the governing equations, the classic semi-Lagrangian scheme, general reference map advection, and the construction of the volume preserving projection. In Section~\ref{sec:cb}, we provide the full details of our novel CB algorithm and discuss implementation details for non-graded, adaptive quadtree and octree grids. Section~\ref{sec:results} presents verification and validation studies, highlighting the key performance differences compared to other schemes. Additionally, in Section~\ref{sec:multiphase} we show how the CB can be incorporated into a two-phase, incompressible Navier-Stokes solver and highlight the associated improvements. Finally, Section~\ref{sec:conclusions} summarizes our findings and outlines future directions.

\section{Preliminaries}\label{sec:preliminaries}
%
%
In this section, we present the scalar advection equation and discuss its solutions when using an incompressible advecting velocity. Next, we review the traditional semi-Lagrangian method as our primary tool for solving the advection equation and discuss how the SL scheme is a natural way of constructing the reference map. The key takeaway from this section is that the SL method reconstructs individual characteristics, while the reference map represents the collection of characteristics. With this framework, we can leverage the reference map to design a volume-preserving correction for advecting quantities under incompressible fields. 

\subsection{Governing Equations}
%
The Characteristic Bending method is fundamentally a numerical scheme for solving the advection equation that builds on ideas from the semi-Lagrangian (SL) method \cite{staniforth1991semi} and the reference map technique (RMT) \cite{kamrin2012reference, rycroft2020reference}. The scalar advection equation, using an Eulerian frame of reference, describes the transport of a scalar quantity (\eg temperature, pressure, level set functions, etc.) by a velocity field and has the form,
\begin{align}
    \frac{\partial \phi}{\partial t} + \nabla \cdot \left ( \mathbf{u} \phi \right ) & = 0 \text{,} 
\end{align}
subject to the initial conditions;
\begin{align}
    \phi \left ( \mathbf{x}, 0 \right ) = \phi_0 (\mathbf{x}) \text{,} \quad \forall \mathbf{x} \in \Omega \text{.}
\end{align}
Here $\phi = \phi (\mathbf{x}, t)$ is the scalar quantity, $\mathbf{u}=\mathbf{u}(\mathbf{x},t)$ is the advecting velocity, $\mathbf{x}$ is the spatial coordinate in the domain $\Omega$, and time $t \ge 0$. We are primarily concerned with incompressible velocity fields (\ie $\nabla \cdot \mathbf{u} = 0$) for the work presented herein and can simplify the advection equation as
\begin{align}
    \frac{\partial \phi}{\partial t} + \mathbf{u} \cdot \nabla \phi = 0 \text{.} \label{eq:advection_incompressible}
\end{align}

If we consider a Lagrangian frame of reference, we can write Equation \eqref{eq:advection_incompressible} in Lagrangian form as,
\begin{align}
    \frac{D \phi}{D t} = 0 \text{,}
\end{align}
where $D/Dt$ is the material derivative with the advection velocity,
\begin{align}
    \frac{D \mathbf{x}}{D t} = \mathbf{u} \text{.} \label{eq:characteristics}
\end{align}
In this form, we can clearly see that our advected quantity, $\phi$, is constant along the characteristics of the velocity field. This invariance property of the advection equation naturally enforces conservation of the transported scalar. Furthermore, the solution at any point in space and time can be obtained by tracing the characteristic trajectory backwards to the initial state, $\phi_0$.

\subsection{Semi-Lagrangian Method}
\begin{figure}
    \centering
    \includegraphics[width=0.5\linewidth]{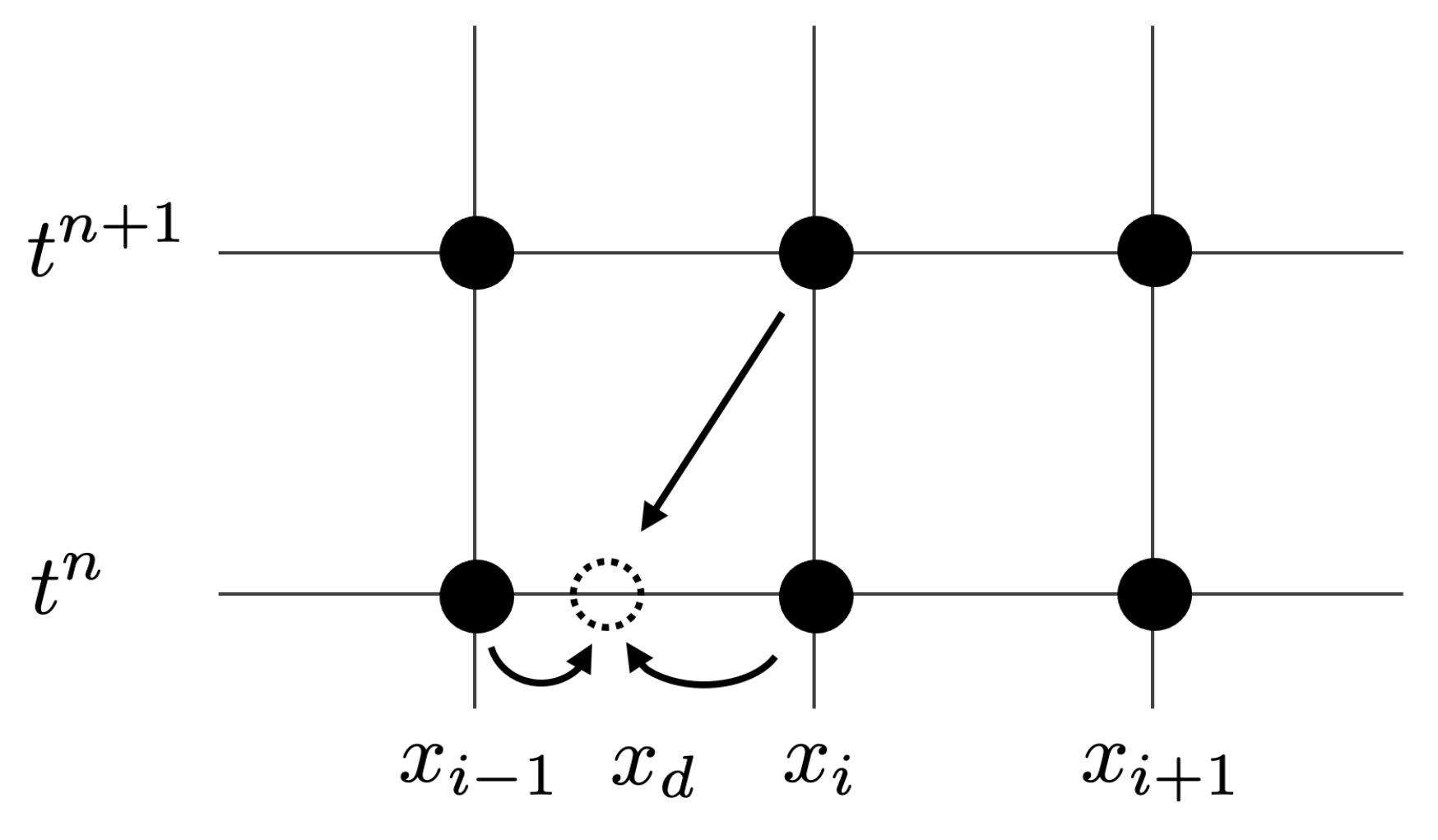}
    \caption{Schematic of the semi-Lagrangian method where the characteristic passing through grid point $x_i$ at time $t^{n+1}$ is traced backward to the departure point $x_d$. Additionally, the value of the advected quantity at the departure point is computed through interpolation using the neighboring points, $x_{i-1}$ and $x_i$ at time $t^n$.}
    \label{fig:sl_schematic}
\end{figure}

%
The semi-Lagrangian method, coined by Sawyer \cite{sawyer1963semi}, is a numerical scheme for solving the advection equation by tracing characteristics backward in time using an Eulerian grid. The basic algorithm can be viewed as a two-step procedure. First, the backward trajectory of a grid point is computed to locate its departure point at the previous time. Since this departure point rarely coincides with a grid node, the second step involves evaluating the advected quantity at that location using an interpolation procedure. A schematic of this process is shown in Figure \ref{fig:sl_schematic}. The interested reader is referred to  \cite{fletcher2019semi} (or the classic text \cite{boyd2001}) for details of the historical development of semi-Lagrangian methods.

The semi-Lagrangian methodology offers two notable advantages. First, unlike forward-in-time integration schemes, it is not restricted by a traditional Courant–Friedrichs–Lewy (CFL) condition \cite{robert1981stable}. Second, when linear interpolation is used at departure points, the scheme is unconditionally stable for linear advection \cite{falcone1998slconvergence}. These methods are essential in numerical weather prediction, which often uses very large time steps \cite{staniforth1991semi, boyd2001}. They are also valuable in situations where time-step restrictions are severe. For example, semi-Lagrangian approaches have been applied to turbulence simulations, where they relax stability limits while retaining high-order spatial accuracy \cite{xiu2001semi}, and in cut-cell finite volume methods, where they alleviate the restrictive constraints imposed by small cells \cite{rosatti2005semi}.

These advantages, however, come at a cost. By construction, the semi-Lagrangian method does not enforce conservation, and low-order time integration and interpolation schemes can introduce excessive dissipation. In fact, with linear interpolation and a CFL number less than one, the scheme is equivalent to the first-order upwind method, which is well known to be diffusive \cite{robert1981stable}. To address these shortcomings, a number of conservative formulations have been proposed \cite{yabe2001exactly, priestley1993quasi, xiao2001completely}, including Lagrange–Galerkin variations \cite{morton1985generalised}, though these generally increase computational cost as noted in \cite{giraldo1998lagrange}. Similarly, higher-order time integration and interpolation schemes have been developed to improve accuracy, but these are no longer unconditionally stable \cite{falcone1998slconvergence}.

While considerable developments have been made to improve semi-Lagrangian schemes, their accuracy ultimately remains limited by the choice of time integration and interpolation. At its core, the semi-Lagrangian method is a single time-level scheme, where each step traces characteristics over one interval and updates the advected field through interpolation. This repeated tracing and interpolation compound errors at every step. An alternative approach is to consider the advection of the characteristics themselves, building a smooth mapping that evolves from the initial configuration to the current time.

\subsection{Reference Map Advection}
%
The Reference Map Technique (RMT), originally proposed by Kamrin \cite{kamrin2008stochastic} and independently by Cottet \textit{et al.} \cite{cottet2008eulerian, maitre2009applications}, solves advection problems by advecting the characteristics themselves. It does so by introducing a static Eulerian grid together with a reference map that encodes the mapping from the current configuration back to the original one. The notion of advecting characteristics through an auxiliary mapping is not new, however. Early formulations appeared in the Arbitrary Lagrangian–Eulerian (ALE) framework \cite{hirt1974arbitrary, hughes1981lagrangian, donea1982arbitrary}, where the computational mesh is deformed to follow the flow or large displacements, but requires periodic remeshing to maintain mesh quality. Although modern ALE formulations have significantly advanced remeshing and mesh optimization strategies (e.g., see \cite{dobrev2012high, vargas2025multi}), the RMT avoids these issues altogether by maintaining a fixed Eulerian mesh and reconstructing fields through interpolation at the roots of characteristic curves.

The Characteristic Mapping Method (CMM) \cite{mercier2020characteristic} is a related technique developed specifically for advection problems. It builds on the mapping approach of \cite{kohno2013new} and employs the Gradient Augmented Level Set (GALS) method of \cite{nave2010gradient} to advect the reference map on a fixed Eulerian grid. A similar idea, introduced for level set advection, is the Coupled Level Set–Reference Map (CLSRM) of Bellotti and Theillard \cite{bellotti2019rm}, in which the standard level set method is augmented by advecting the reference map and reconstructing the level set function at each time step via pullback. This formulation reduces the frequency of level set reinitialization, thereby reducing mass loss and interface error. The CLSRM framework provides a natural and flexible formulation for general advection problems, and we adopt this viewpoint for the presentation herein.

The central idea of the reference map approach is to replace the direct advection of a scalar field $\phi$ with the advection of an auxiliary vector field, the reference map $\mathbf{\xi}$. At any location $\mathbf{x}$ and time $t$, the value $\mathbf{\xi} \left (\mathbf{x}, t \right )$ gives the initial position $\mathbf{x}_0$ along the characteristic curve that passes through $\mathbf{x}$ at time $t$. The reference map evolves according to
\begin{align}
    \frac{\partial \mathbf{\xi}}{\partial t} + \mathbf{u} \cdot \nabla \mathbf{\xi} = 0,
\end{align}
with the initial condition,
\begin{align}
    \mathbf{\xi} \left ( \mathbf{x} , 0 \right ) = \mathbf{x} \text{,} \qquad \forall \mathbf{x} \in \Omega.
\end{align}
Once $\mathbf{\xi}$ is known, we can then construct the solution of Equation~\eqref{eq:advection_incompressible} through the composition of the reference map and $\phi_0$ as,
\begin{align}
    \phi \left ( \mathbf{x}, t \right ) = \phi_0 \left ( \mathbf{\xi} \left (\mathbf{x},t \right ) \right ) \text{,} \quad \forall t \ge 0 \text{,} \quad \forall \mathbf{x} \in \Omega.
\end{align}

\begin{figure}
    \centering
    \includegraphics[width=0.95\linewidth]{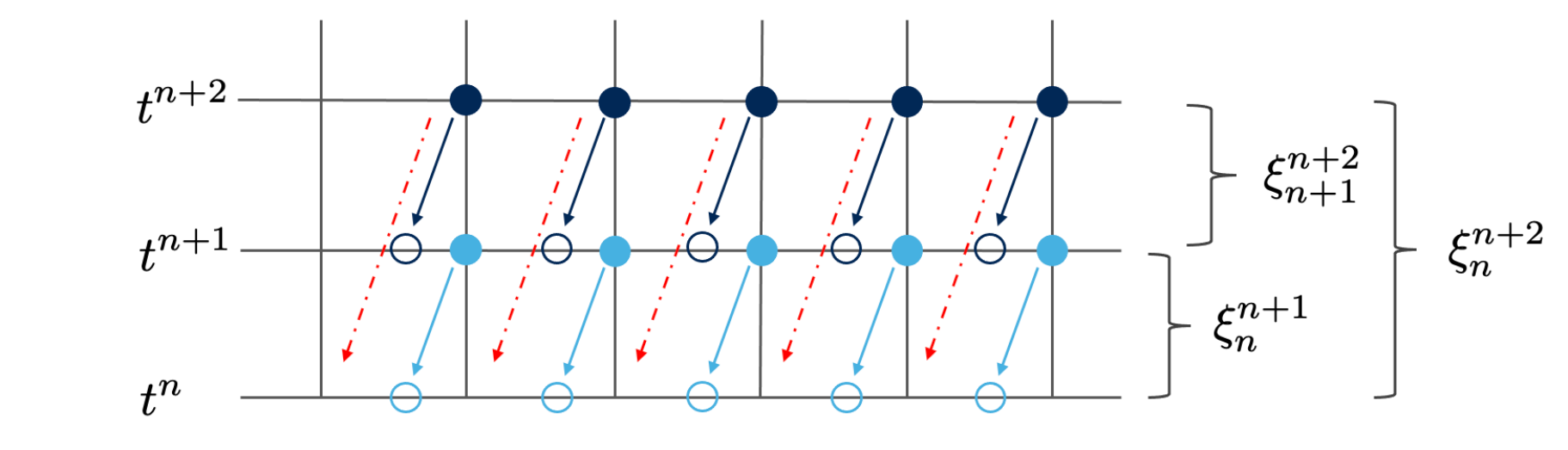}
    \caption{A schematic comparison of the semi-Lagrangian and Reference Map Advection methods using the reference map, $\mathbf{\xi}$. The semi-Lagrangian method can be seen as a single time step variation of the reference map method (\textit{e.g.} using $\mathbf{\xi}_n^{n+1}, \; \mathbf{\xi}_{n+1}^{n+2}$), whereas the reference map method traces the characteristics over multiple time steps (\textit{e.g.} using $\mathbf{\xi}_n^{n+2}$).}
    \label{fig:SLvRM}
\end{figure}

This formulation makes it clear that the reference map is essentially a collection of the characteristics of the advection problem. In this sense, the semi-Lagrangian method can be interpreted as a special case of reference map advection over a single time step, where the “initial condition” is redefined at each update. The advantage of evolving $\mathbf{\xi}$ directly is that the initial configuration of the scalar field is preserved, avoiding error accumulation from repeated interpolation. Moreover, the reference map is typically a smoother object than the scalar field, making it easier to interpolate accurately. A schematic comparison of the two approaches is shown in Figure~\ref{fig:SLvRM}.

\subsection{Volume Preserving Correction}
%
Thus far, we have introduced the reference map as a collection of characteristics used to reconstruct advected fields. A more formal viewpoint, however, reveals additional structure that can be used to quantify errors and guide corrections to the reference map formulation. In what follows, we adopt the treatment of \cite{theillard2021vprm} to develop this perspective.

We consider the deformed domain, $\mathcal{B} \left ( t \right )$, and scalar field, $\phi(\mathbf{x},t)$, defined for $t \ge 0$. We denote the initial domain as $\mathcal{B}_0$ and the initial scalar field as $\phi_0(\mathbf{x},0)$. We assume that this domain is deformed under the action of the velocity field, $\mathbf{u}$, and we define the motion map $\chi \left (\cdot, t \right )$ as the morphism that transforms the initial domain $\mathcal{B}_0$ into the deformed one $\mathcal{B} \left ( t \right )$. Specifically, this maps every material point $\mathbf{x}_0 \in \mathcal{B}_0$ to its corresponding image on $\mathcal{B} \left ( t \right )$,
\begin{align}
    \mathbf{x} \left ( t \right ) = \chi \left ( \mathbf{x}_0, t \right ) \quad \forall t \ge 0 \text{,} \quad \forall \mathbf{x}_0 \in \mathcal{B}_0.
\end{align}

\begin{figure}
    \centering
    \includegraphics[width=0.85\linewidth]{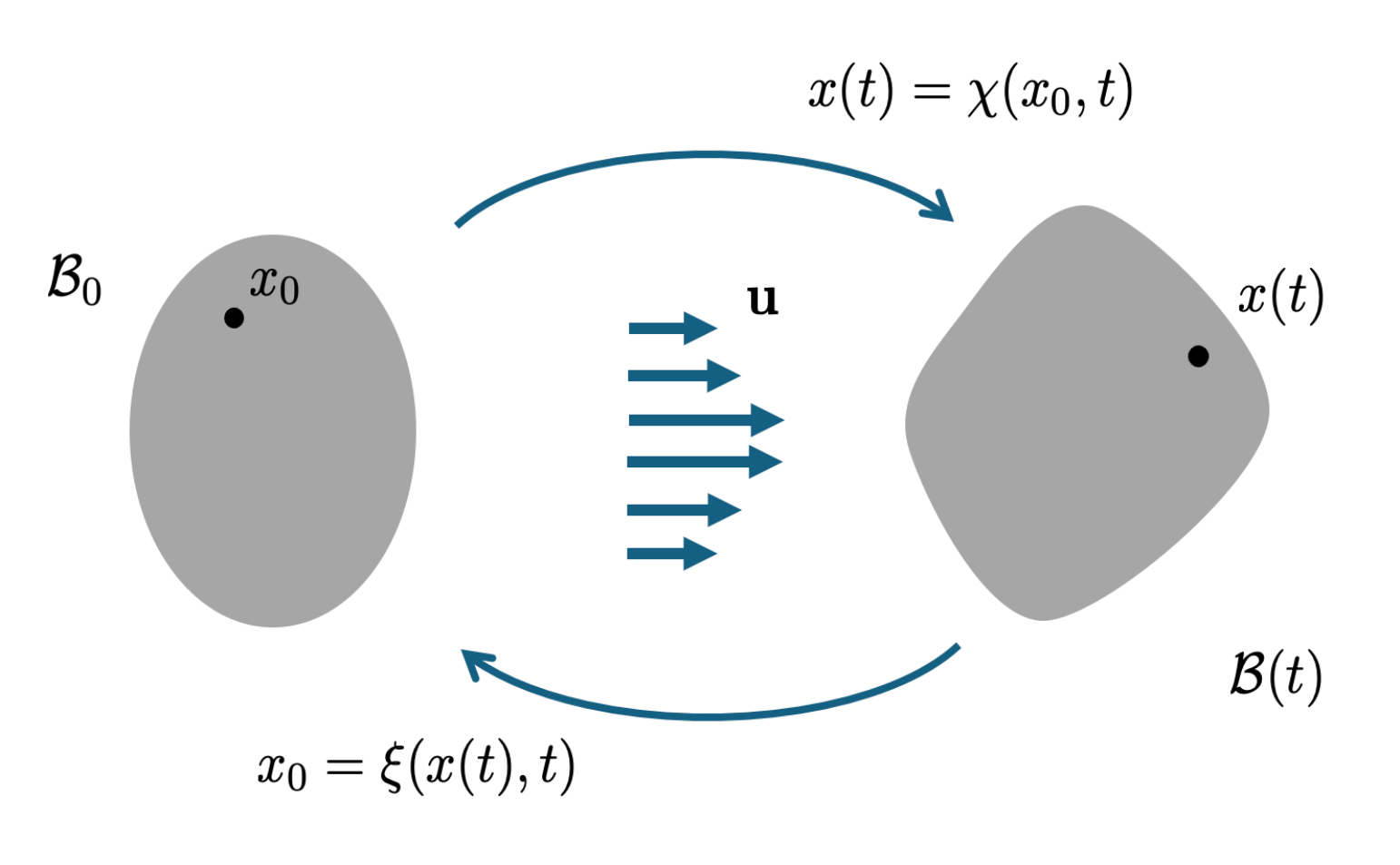}
    \caption{The basic schematic of the reference map method. We consider the initial domain, $\mathcal{B}_0$ deformed by the velocity field $\mathbf{u}$. The motion map, $\chi(\cdot,t)$, transforms any point $\mathbf{x}_0$ in the domain, $\mathcal{B}_0$ to its associated image, $\mathbf{x}(t)$, in the deformed domain, $\mathcal{B}(t)$. The reference map, $\xi(\cdot,t)$, is the inverse of the motion map and transforms any point $\mathbf{x}(t)$ to its initial position, $\mathbf{x}_0$, in the domain $\mathcal{B}_0$. }
    \label{fig:rm_geo}
\end{figure}

Assuming that $\chi \left (\cdot , t \right )$ is a smooth invertible mapping (which holds provided the velocity field is well-behaved, \textit{i.e.} it does not generate shocks, rarefactions, or intersecting trajectories), the inverse mapping $\mathbf{\xi} $ is defined as
\begin{align}
    \mathbf{x}_0 = \mathbf{\xi} \left (\mathbf{x}, t \right ) \quad \forall t \ge 0 \text{,} \quad \forall \mathbf{x} \in \mathcal{B} \left ( t \right ).
\end{align}
Here, $\mathbf{\xi}$ is the reference map, which maps each point $\mathbf{x}$ to the corresponding point $\mathbf{x}_0$ in the initial domain, $\mathcal{B}_0$. With this view, the deformation of the domain, $\mathcal{B} \left ( t \right )$, is captured by the advection of the reference map
\begin{align}
    \frac{\partial \mathbf{\xi}}{\partial t} + \mathbf{u} \cdot \nabla \mathbf{\xi} & = 0 \quad \forall t \ge 0 \text{,} \quad \forall \mathbf{x} \in \mathcal{B} \left ( t \right ) \label{eq:ref_map_adv} \\
    \mathbf{\xi} \left (\mathbf{x}, 0 \right ) & = \mathbf{x} \qquad \qquad \quad \forall \mathbf{x} \in \mathcal{B}_0.
\end{align}
The solution to the original advection problem can then be reconstructed from the composition of the reference map and the initial scalar field as
\begin{align}
    \phi \left (\mathbf{x}, t \right ) = \phi_0 \left ( \mathbf{\xi} \left (\mathbf{x},t \right ) \right ) \quad \forall \mathbf{x} \in \mathcal{B} \left ( t \right ).
\end{align}
A schematic of the relation between the motion map, $\chi$, and the reference map, $\mathbf{\xi}$, is shown in Figure~\ref{fig:rm_geo}.

\begin{figure}
    \centering
    \includegraphics[width=0.85\linewidth]{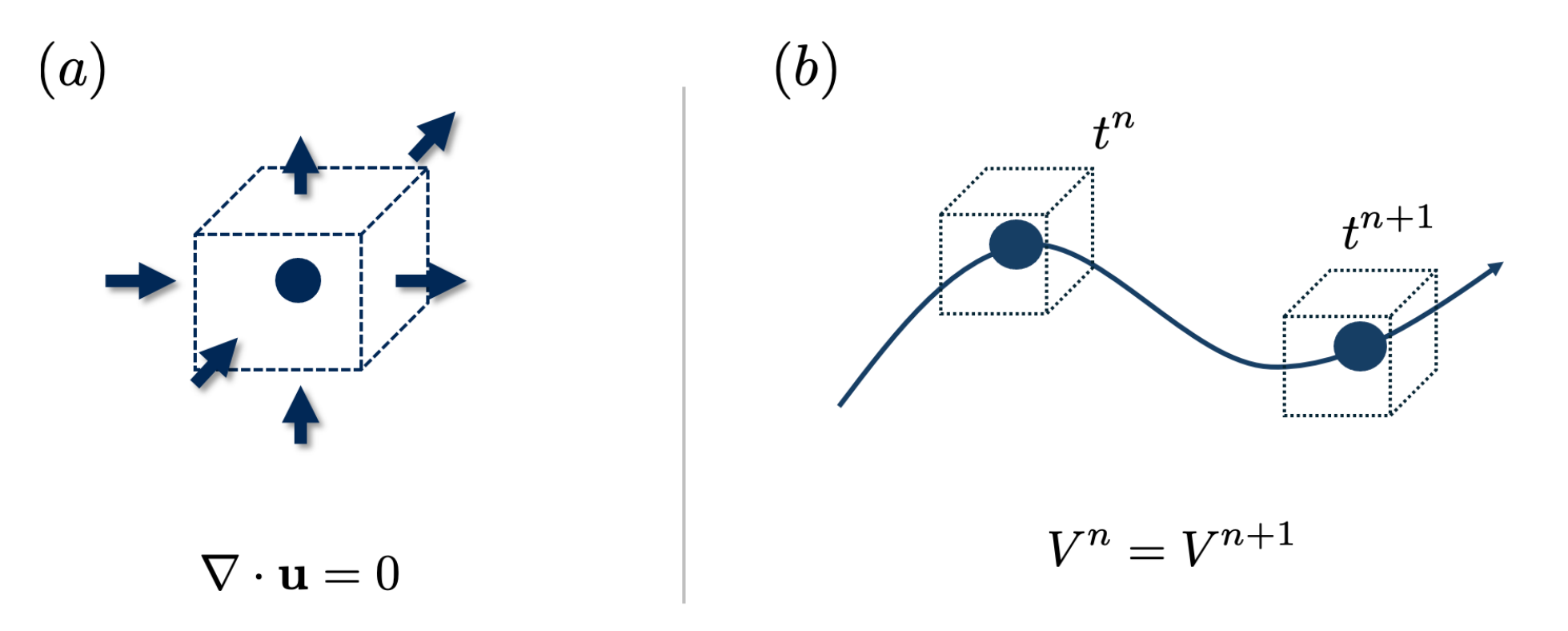}
    \caption{The Eulerian vs. Lagrangian perspective of incompressibility. In the Eulerian frame (a), incompressibility corresponds to a divergence-free velocity field, $\nabla \cdot \mathbf{u}=0$. In the Lagrangian frame (b), the equivalent statement is that a material control volume preserves its volume during motion, $V^n = V^{n+1}$.}
    \label{fig:Euler_v_Lagrange}
\end{figure}

With this formalism, we can now use the reference map, $\xi$, to probe properties of the transformation itself. In the Eulerian frame, incompressibility corresponds to a divergence-free velocity field (assuming constant material properties). In the Lagrangian frame, the equivalent statement is that the transformation preserves volume (see Figure~\ref{fig:Euler_v_Lagrange}). Using the reference map, this condition is expressed as
\begin{align}
\det \left ( \nabla \mathbf{\xi}(\mathbf{x},t) \right ) = 1 \quad \forall \mathbf{x} \in \mathcal{B}(t). \label{eq:vp_map}
\end{align}
While this condition holds in theory, the numerical construction of the reference map (time integration and interpolation) introduces errors that cause the map to drift away from the volume-preserving space. Furthermore, for velocity fields that are only approximately divergence-free, such as those obtained with approximate projection methods, the characteristics themselves can produce deviations from the volume preserving space.

In \cite{theillard2021vprm}, these errors are addressed by introducing a volume-preserving projection for the reference map. The key idea is that the numerical evolution of the map may drift from the volume-preserving space, even when advected with an incompressible velocity field. By quantifying these deviations, we can design a projection that corrects the map by projecting it back into the admissible space. To formalize this, we denote the intermediate map as $\xi^* (\mathbf{x},t)$, which is obtained directly from Equation~\eqref{eq:ref_map_adv}. Since $\xi^*$ is not guaranteed to satisfy \eqref{eq:vp_map}, we seek a correction operator $\gamma^{-1}$ such that the composition satisfies
\begin{align}
    \text{det} \left ( \nabla \gamma^{-1} \left (\mathbf{\xi}^* (\mathbf{x},t) \right ) \right ) = 1.
\end{align}
Here, the correction $\gamma^{-1}$ is assumed to be a near identity (see \cite{theillard2021vprm}) transformation as the deviations from the volume-preserving space are typically on the order of the numerical accuracy of the scheme used to solve \eqref{eq:vp_map} (\textit{e.g.} $\mathcal{O} \left (\Delta t^\alpha + \Delta x^\beta \right )$ for the semi-Lagrangian method, where $\alpha$ and $\beta$ represent the order of the backward time integration and spatial interpolation, respectively). Informally, $\gamma^{-1}$ can be interpreted as re-mapping the reference map to restore volume preservation.

The correction $\gamma^{-1}$ is not arbitrary, but arises naturally as the solution to a constrained optimization problem. Specifically, \cite{theillard2021vprm} shows that the correction is the unique minimizer of a saddle-point Lagrangian that seeks the minimal correction to $\mathbf{\xi}^*$ in the $L^2$ sense, subject to the volume-preserving constraint. This optimality condition reduces to a Poisson equation for the adjoint $\lambda$,
\begin{align}
    - \Delta \lambda & = 1 - \text{det} \left ( \nabla \mathbf{\xi}^* \right ) \quad \forall \mathbf{x} \in \Omega \text{,} \label{eq:lambda_poisson}\\
    \lambda & = 0 \qquad \qquad \qquad \; \; \forall \mathbf{x} \in \partial \Omega \text{,} \label{eq:lambda_poisson_bc}
\end{align}
for which the correction is, to leading order,
\begin{align}
    \gamma^{-1} \left ( \mathbf{x} \right ) = \mathbf{x} - \nabla \lambda. \label{eq:correction}
\end{align}
Conceptually, this projection is a direct analog of the projection step used in incompressible Navier-Stokes solvers, carried out in the Lagrangian reference frame. 

The Volume-Preserving Reference Map (VPRM) of \cite{theillard2021vprm} incorporates this projection directly into the numerical evolution of the reference map. Their method, designed for the evolution of level set-defined interfaces, projects the reference map onto the volume-preserving space after every time step. This approach significantly reduces mass loss, a long-standing challenge in level set methods, but introduces additional errors in the predicted interface position. For simple advection problems with analytically divergence-free velocities, the benefits are limited and the method deteriorates the interface accuracy. Additionally, the projection step becomes the most time-consuming part of the overall algorithm. In contrast, for multiphase flows where the interface velocity is inexact, the VPRM formulation yields appreciable improvements.

It is important to highlight several key assumptions in the derivation of the volume-preserving correction in \cite{theillard2021vprm}. The first is that the correction remains close to the identity, under the expectation that deviations from the volume-preserving space are of the same order as the discretization error of the underlying advection scheme. While this assumption is valid for a single step, it is less clear that it remains valid when the reference map is evolved over many steps, as in the VPRM algorithm. This concern is closely tied to the linearization of the determinant constraint (see \cite{theillard2021vprm}), since small deviations are repeatedly encoded into the reference map at every step. In practice, these subtleties are partly obscured by the combination of projection and restarting the map (see \cite{theillard2021vprm}), but they highlight important considerations for the long-time performance of the volume-preserving scheme.

\section{Characteristic Bending} \label{sec:cb}
%
In this section, we present the characteristic bending method, a generalized framework for reference map advection. We begin with a high-level description of the algorithmic structure and the motivation for reformulating existing approaches. Next, we outline the specific numerical discretizations used to construct the reference map with the semi-Lagrangian method and to perform the volume-preserving projection. Finally, we briefly describe where to find the additional implementation details required to incorporate the method on adaptive quadtree and octree grids.

The key takeaway from this section is to interpret the volume-preserving projection as a means of bending characteristics toward the divergence-free space. This operation simultaneously corrects numerical errors from characteristic tracing and filters compressible modes in the advecting velocity. By applying the projection only to maps that remain close to the identity, the bending step operates in the region where the volume-preserving constraint can be treated linearly, yielding a method that is both robust and accurate.

\subsection{Overall Method}
%
The characteristic bending method is a general framework for advection with incompressible velocity fields. This method can be applied as a standalone advection scheme or it can be used to augment other numerical methods, such as in reference map advection. The basic CB algorithm is presented in Figure~\ref{fig:basic_cb_algo}, and its extension to reference map advection is shown in Figure~\ref{fig:rm_cb_algo}.

\begin{figure}[!ht]
    \scriptsize
    \begin{tabular}{c }
    \hline \\ {\begin{minipage}[c]{0.9\textwidth}
    \begin{description}
    \item[1. Advection Step] \hfill \\
    Advect an intermediate reference map, $\xi^*$, from time $t^n$ to $t^{n+1}$ by solving the advection equation
    \begin{align*}
        \begin{rcases}
            \begin{aligned}
                \frac{\partial \xi^*}{\partial t} + \mathbf{u} \cdot \nabla \xi^* & = 0  \;\\
                \xi^* \left ( \mathbf{x}, t^n \right ) & = \mathbf{x} \;
            \end{aligned}
        \end{rcases}
        \forall \mathbf{x} \in \Omega \text{.}
    \end{align*}

    \item[2. Correction Step] \hfill \\
    Compute the correction to the map by solving the Poisson system for $\lambda$
    \begin{align*}
        - \Delta \lambda & = 1 - \text{det} \nabla \xi^*  \quad \forall \mathbf{x} \in \Omega \text{,} \\
        \lambda & = 0 \qquad \qquad \quad \forall \mathbf{x} \in \partial \Omega \text{,}
    \end{align*}
    and the correction as
    \begin{align*}
        \gamma^{-1}(\mathbf{x}) = \mathbf{x} - \nabla \lambda \text{.} 
    \end{align*}
    The volume-preserving map, $\xi$, is then constructed as
    \begin{align*}
        \xi(\mathbf{x},t^{n+1}) = \xi^*(\gamma^{-1}(\mathbf{x}),t^{n+1}).
    \end{align*}
    
    \item[3. Reconstruction] \hfill \\
    Reconstruct the solution of the scalar field, $\phi$, at time $t^{n+1}$ as
    \begin{align*}
        \phi(\mathbf{x},t^{n+1}) = \phi \left ( \xi \left ( \mathbf{x}, t^{n+1} \right ), t^n \right ) \text{.}
    \end{align*}
\end{description}
\end{minipage}}
\\
\\
\hline
\end{tabular}
\caption{The basic characteristic bending algorithm. At each time step, an intermediate reference map is advected, corrected through a volume-preserving projection, and used to reconstruct the scalar field.}
\label{fig:basic_cb_algo}
\end{figure}

The basic characteristic bending algorithm proceeds in three stages. In the advection step, an intermediate reference map $\xi^*$ is advanced from $t^n$ to $t^{n+1}$ by solving an advection equation with velocity $\mathbf{u}$. The map is initialized as the identity, $\xi^*(\mathbf{x},t)=\mathbf{x}$, so that after advection it represents the characteristic curves traced backward from $t^{n+1}$ to $t^n$. In the correction step, deviations of $\xi^*$ from the volume-preserving space form the right hand side of a Poisson equation for the adjoint $\lambda$. The gradient of $\lambda$ is used to determine the correction $\gamma^{-1}$ and the corrected, volume-preserving map is obtained by composing this correction with $\xi^*$. Importantly, because the intermediate map is initialized as the identity at each step, numerical errors are confined to the characteristic tracing and interpolation within a single time step. Finally, in the reconstruction step, the scalar field $\phi$ at time $t^{n+1}$ is obtained by composing $\phi(\cdot,t^n)$ with the corrected, volume-preserving map. 
 
\begin{figure}[!ht]
    \scriptsize
    \begin{tabular}{c }
    \hline \\ {\begin{minipage}[c]{0.9\textwidth}
    \begin{description}
    \item[1. Advection Step] \hfill \\
    Advect an intermediate reference map, $\psi^*$, from time $t^n$ to $t^{n+1}$ by solving the advection equation
    \begin{align*}
        \begin{rcases}
            \begin{aligned}
                \frac{\partial \psi^*}{\partial t} + \mathbf{u} \cdot \nabla \psi^* & = 0  \;\\
                \psi^* \left ( \mathbf{x}, t^n \right ) & = \mathbf{x} \;
            \end{aligned}
        \end{rcases}
        \forall \mathbf{x} \in \Omega \text{.}
    \end{align*}

    \item[2. Correction Step] \hfill \\
    Compute the correction to the map by solving the Poisson system for $\lambda$
    \begin{align*}
        - \Delta \lambda & = 1 - \text{det} \nabla \psi^*  \quad \forall \mathbf{x} \in \Omega \text{,} \\
        \lambda & = 0 \qquad \qquad \quad \forall \mathbf{x} \in \partial \Omega \text{,}
    \end{align*}
    and the correction as
    \begin{align*}
        \gamma^{-1}(\mathbf{x}) = \mathbf{x} - \nabla \lambda \text{.} 
    \end{align*}
    The volume-preserving map, $\psi$, is then constructed as
    \begin{align*}
        \psi(\mathbf{x},t^{n+1}) = \psi^*(\gamma^{-1}(\mathbf{x}),t^{n+1}).
    \end{align*}
    
    \item[3. Reference Map Update] \hfill \\
    Update the reference map, $\xi$, by composition as
    \begin{align*}
        \xi ( \mathbf{x}, t^{n+1} ) = \xi ( \psi ( \mathbf{x}, t^{n+1} ), t^n ) \text{.}
    \end{align*}

    \item[4. Reconstruction] \hfill \\
    Reconstruct the solution of the scalar field, $\phi$, at time $t^{n+1}$ as
    \begin{align*}
        \phi(\mathbf{x},t^{n+1}) = \phi \left ( \xi \left ( \mathbf{x}, t^{n+1} \right ), t^n \right ) \text{.}
    \end{align*}
\end{description}
\end{minipage}}
\\
\\
\hline
\end{tabular}
\caption{The characteristic bending method applied to reference map advection. Here,     the basic CB algorithm is embedded within the reference map framework. Instead of directly reconstructing the scalar field from the corrected single-step map, the long-time reference map is updated by composition with the corrected near-identity map.}
\label{fig:rm_cb_algo}
\end{figure}

\begin{figure}[!htb]
    \centering
    \includegraphics[width=0.85\linewidth]{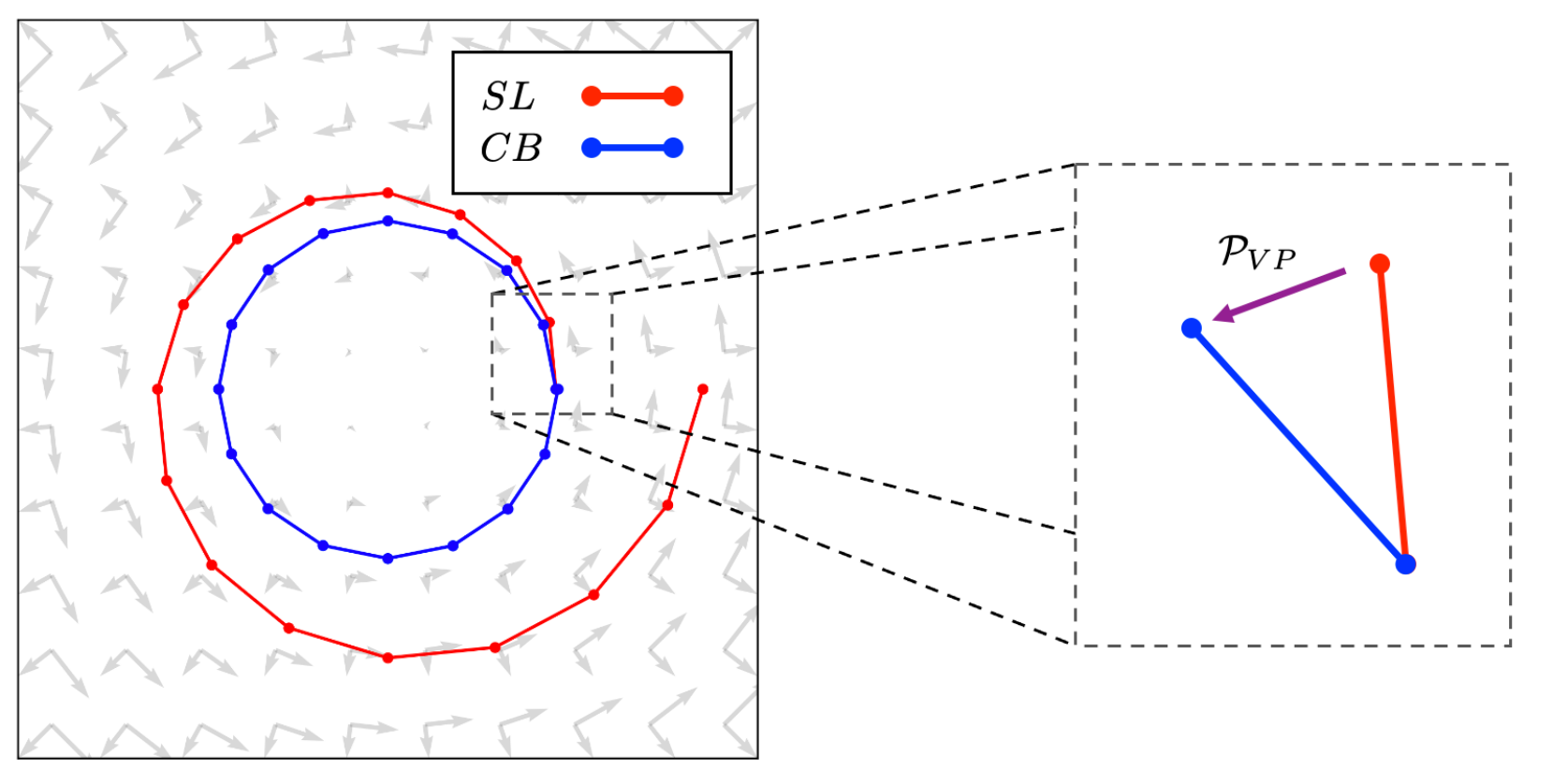}
    \caption{Schematic illustration of the SL (red) and the CB (blue) methods tracing the characteristics for solid-body rotation with artificial expansion. The SL method spirals outward following the modified velocity field, while the CB method remains concentric. The projection triangle, shows the correction and the difference between the two methods.}
    \label{fig:cb_characteristics}
\end{figure}

The characteristic bending method differs from existing approaches in how it applies the volume-preserving projection. The semi-Lagrangian scheme, which CB uses to construct the intermediate reference map, is efficient and stable, but it does not inherently conserve volume or eliminate compressible modes introduced by numerical error. Similarly, the CLSRM framework \cite{bellotti2019rm} advects the characteristics of the velocity field but lacks any internal mechanism to enforce incompressibility. For analytically divergence-free velocities, CLSRM is highly accurate, but in practice it is prone to errors in the velocity field unless frequent restarts are employed. The characteristic bending method is closely related to the VPRM and, in fact, is equivalent when the map is restarted every iteration. The distinction between the two approaches becomes clear when considering multi–time-step reference maps. The VPRM applies its projection directly to the evolving long-time map, which can correct small numerical errors but may accumulate distortion and fail to fully filter compressible modes introduced by the velocity field. By contrast, characteristic bending applies the projection only to near-identity single-step maps, ensuring the correction is minimal and well-posed, and then composes these corrected maps into the long-time reference map.

The characteristic bending method provides a general framework for advection with incompressible velocity fields that builds on the semi-Lagrangian method, reference map advection, and a volume-preserving projection. The central novelty is to apply the projection only to near-identity maps, ensuring that corrections remain minimal, consistent, and free from the long-time drift associated with projecting the full reference map. This viewpoint clarifies CB as a principled recombination of existing building blocks that substantially improves robustness and accuracy. In the following subsections, we provide the discretization details for the advection step, the projection step, and reconstruction of the solution for implementation on adaptive quadtree and octree grids.

\begin{figure}
    \centering
    \includegraphics[width=0.85\linewidth]{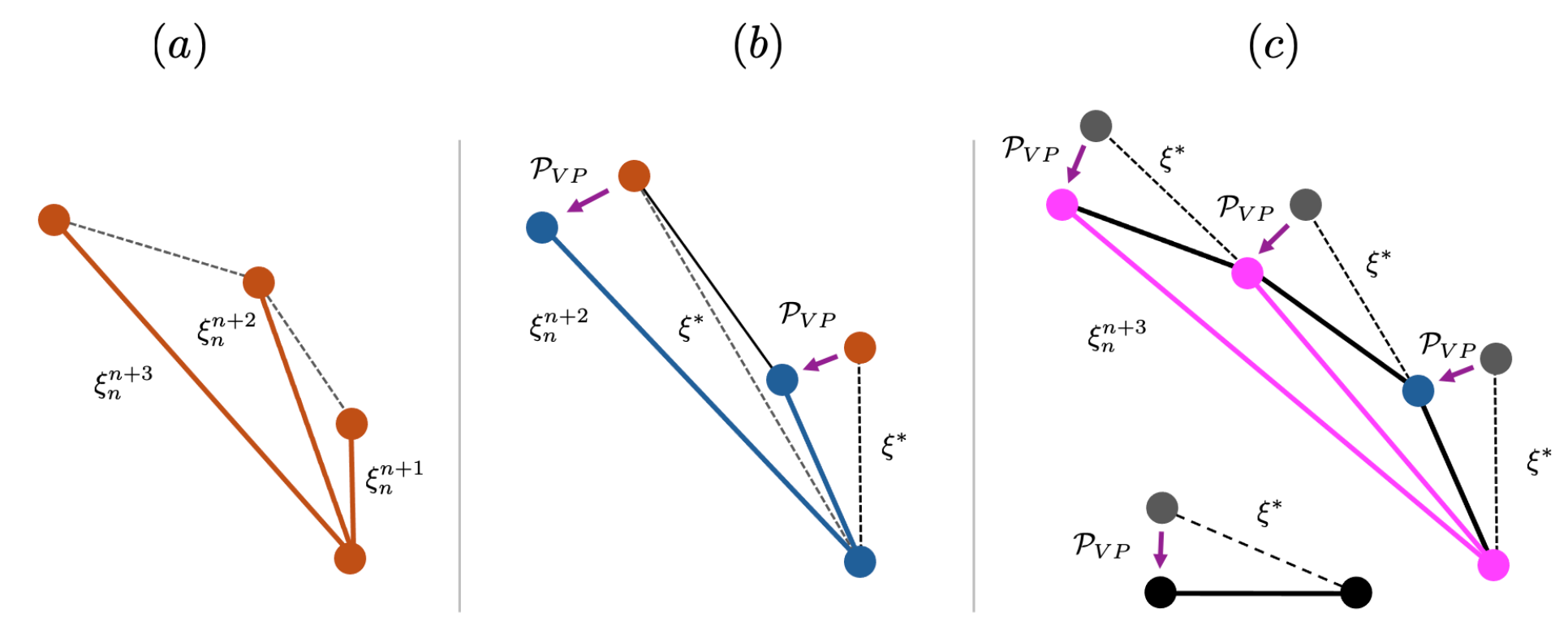}
    \caption{(a) Construction of the reference map (CLSRM) for solid-body rotation with artificial expansion. The CLSRM maps trajectories back to the starting point. (b) The VPRM projects the entire reference map. (c) RMCB projects only a temporary, single-timestep reference map and then composes this with the full reference map.}
    \label{fig:rm_comp}
\end{figure}

\subsection{Advection Step}
%
The first step of the characteristic bending method advances the intermediate reference map $\xi^*$ from $t^n$ to $t^{n+1}$ using the semi-Lagrangian method. Since $\xi^*$ is initialized as the identity at each time step, this advection step amounts to tracing characteristics backward from grid points at $t^{n+1}$ to their departure locations at $t^n$.

The departure point $\mathbf{x}^n_d$ is computed using a second-order backward Runge–Kutta (RK2) scheme,
\begin{align}
\hat{\mathbf{x}} &= \mathbf{x}^{n+1} - \frac{\Delta t^{n}}{2} \mathbf{u}^{n+1}(\mathbf{x}^{n+1}), \label{eq:xnd_intermediate} \\
\hat{\mathbf{u}}^{n+1/2} \left (\hat{\mathbf{x}} \right ) &= \left(1 + \frac{\Delta t^{n}}{2\Delta t^{n-1}} \right)\mathbf{u}^n(\hat{\mathbf{x}}) - \frac{\Delta t^{n}}{2\Delta t^{n-1}}\mathbf{u}^{n-1}(\hat{\mathbf{x}}), \\
\mathbf{x}^n_d &= \mathbf{x}^{n+1} - \Delta t^{n} \hat{\mathbf{u}}, \label{eq:xnd_departure}
\end{align}
where $\mathbf{u}^n$ is the velocity field at time $t^n$. To evaluate velocities at off-grid points, we use quadratic interpolation with WENO-limited second derivatives. Specifically, if $\mathbf{x}=(x,y)$ lies in a cell $\mathcal{C}$, here taken for illustration as $[0,1]^2$, then
\begin{align}
\phi(x,y) = \phi^{00}(1-x)(1-y) + \phi^{01}(1-x)y + \phi^{10}x(1-y) + \phi^{11}xy - \phi_{xx}\frac{x(1-x)}{2} - \phi_{yy}\frac{y(1-y)}{2}, \label{eq:quadENO}
\end{align}
where $\phi^{ij}$ denote the nodal values at the corners of $\mathcal{C}$, and the second derivatives are computed in a WENO fashion \cite{liu1994weighted}, e.g.
\begin{align}
\phi_{xx} = \frac{1}{W_x}\sum_{n \in \text{nodes}(\mathcal{C})} \phi_{xx}^n w^n_x, \qquad w^n_x = \frac{1}{|\phi_{xx}^n|^2}, \qquad W_x = \sum_{n \in \text{nodes}(\mathcal{C})} w_x^n.
\end{align}

With the departure point defined, we next address the approximation of the velocity $\mathbf{u}^{n+1}$. Although it is common in SL methods to approximate $\mathbf{u}^{n+1}$ by $\mathbf{u}^n$ (\textit{e.g.}, \cite{boyd2001, karniadakis2005spectral}), we instead adopt the trajectory reconstruction proposed in \cite{blomquist2024stable},
\begin{equation}
\mathbf{u}^{n+1}(\mathbf{x}^{n+1}) = \mathbf{u}^n(\mathbf{x}^{n+1}) + \frac{\Delta t^n}{\Delta t^{n-1}} \Big( \mathbf{u}^{n}(\mathbf{x}^{n+1}) - \mathbf{u}^{n-1}(\mathbf{x}^{n+1}) \Big) + \mathcal{O}(\Delta t^2),
\end{equation}
which has been shown to improve stability and accuracy, particularly at large CFL numbers. 

We note that using a forward-time extrapolated velocity field for $\mathbf{u}^{n+1}$ may lead to instabilities in the standard two–time-level semi-Lagrangian method. In \cite{hortal2002development}, the authors identify this extrapolation as the source of these instabilities and propose the SETTLS scheme to correct the issue while maintaining second-order accuracy in time. In the examples presented here, $\mathbf{u}^{n+1}$ is either known a priori or, in the case of incompressible multiphase flows, the simulation is diffusion-dominated, which appears to suppress these instabilities. In pure advection or strongly advection-dominated flows, a predictor–corrector formulation is often used, with $\mathbf{u}^{n+1} = \mathbf{u}^{n}$ in the predictor step. For the incompressible Euler and incompressible multiphase Navier–Stokes examples considered here, we did not observe any difference between the predictor–corrector and time-extrapolation approaches and therefore used the method presented in \cite{blomquist2024stable}.

Once the departure points $\mathbf{x}^n_d$ are computed, the intermediate reference map $\xi^*$ is updated by interpolating its values from $t^n$ at those locations. The combination of the second-order RK scheme with WENO interpolation yields a second-order accurate method in both space and time. For further analysis of the accuracy of SL methods, we refer the reader to \cite{falcone1998slconvergence}.

\subsection{Correction Step}
%
In the projection step, we solve the Poisson problem \eqref{eq:lambda_poisson}–\eqref{eq:lambda_poisson_bc} and compute the correction \eqref{eq:correction}. The system is solved using the supra-convergent solver of Min \textit{et al.} \cite{min2006supra}, designed for quadtree and octree grids. This solver achieves second-order accuracy for both the solution and its gradient, ensuring that the correction is computed with second-order accuracy.

The projection step is the most computationally expensive component of the characteristic bending algorithm, as it requires a global elliptic solve. As noted in \cite{theillard2021vprm}, this expense can exceed that of alternative correction strategies such as the Coupled Level Set/Volume-of-Fluid method \cite{sussman2000coupled, sussman2003second} or the Conservative Level Set method \cite{olsson2005conservative, olsson2007conservative, desjardins2008accurate, owkes2013discontinuous}. Those approaches are designed primarily to conserve volume or mass and often impose stricter stability constraints. Furthermore, they do not eliminate compressible modes that may arise from numerical errors in the advecting velocity. The projection step in characteristic bending, by contrast, directly filters such modes by enforcing that the reference map remains volume-preserving under the assumption of incompressible flow. This filtering capability is central to the robustness of the characteristic bending framework and sets it apart from prior semi-Lagrangian correction schemes.

\subsection{Numerical Implementation}
Details of the numerical implementation, including the data structures used for the quadtree/octree grids and the adaptive mesh refinement (AMR) methodology, are presented in Appendix A for conciseness. These aspects follow standard practices established in previous studies (\textit{e.g.}, \cite{guittet2015stable, theillard2019sharp, bellotti2019rm, theillard2021vprm, blomquist2024stable, binswanger2025multiphase}) and do not introduce new algorithms. The appendix serves mainly to ensure completeness and reproducibility of the numerical framework.

\section{Computational Results} \label{sec:results}

\subsection{Advection Examples}
In this section, we use a set of analytic examples to compare the characteristic bending method with similar advection schemes (\textit{e.g.}, semi-Lagrangian, Coupled Level Set Reference Map, and Volume-Preserving Reference Map) in two dimensions. The examples used consist of a Gaussian distribution and a level set function subject to a constant angular velocity field and a level set function subject to an incompressible, but deformational flow. For each example, we examine the performance of the numerical schemes for the purely incompressible field as well as the incompressible field coupled with a compressible component of varying strength. Finally, we present a simple nonlinear example in the form of the incompressible Euler equations.

\subsubsection{Gaussian Advection} \label{sec:gauss_results}
%
The purpose of this example is to verify that our characteristic bending scheme, as well as the comparative methods, yields the correct order of accuracy when applied to the advection of an infinitely smooth scalar field. The benefit of using a scalar field defined by a Gaussian distribution is that we easily measure the mass loss by computing the first moment of the advected quantity. Furthermore, by using a rotational velocity field defined over a Cartesian grid, we eliminate any serendipitous aliasing\footnote{This is, for example, what happens when using the first-order upwinding scheme for a one-dimensional advection equation with a CFL of 1 defined on a uniform, periodic grid.} that could artificially enhance our results. Hence, the Gaussian Rotation test is a simple, but non-trivial testbed to compare our characteristic bending scheme against standard methods.

For this example, we consider the advection of the scalar function,
\begin{align}
    \phi(x,y) = \frac{1}{\sqrt{2 \sigma \pi}} \exp{\left ( - \frac{(x-0.5)^2 + y^2}{2 \sigma^2} \right ) },
\end{align}
in the domain, $\Omega = [-1,1]^2$. This function represents a radial Gaussian centered at $(0.5, 0)$ and we advect this function using the incompressible velocity field, $\mathbf{u}=(-y, x)$ from time $t=0$ to $t=2\pi$. We also consider adding a compressible component to the velocity field, in the form of artificial expansion, to simulate a scenario in which the advecting velocity is not entirely divergence free (\eg as a result of an approximation projection method). We scale this compressible component by the maximum refinement level of our mesh, $\Delta x_{min}$, with parameters, $\alpha$ and $\beta$, to represent either a first- or second-order deviation from the incompressible space. We center the compressible component at the origin, leading to the velocity field,
\begin{align}\label{eq:u_rotational}
    u(x, y, t) =
    \begin{cases}
        \begin{aligned}
            &-y \; + \; \alpha \Delta x_{min} \; x + \; \beta \left ( \Delta x_{min} \right )^2 \; x\\
            & \quad \; x \; + \; \alpha \Delta x_{min} \; y + \; \beta \left ( \Delta x_{min} \right )^2 \; y
        \end{aligned} & \text{for } t \in [0, 2\pi].
    \end{cases}
\end{align} 

We measure the accuracy of our numerical method using the $L^{\infty}$ norm
\begin{align}
    L^{\infty} = \max_\Omega \left ( \left | \phi(x,y,0) - \phi(x,y,2\pi) \right | \right )
\end{align}
and measure the relative mass loss, $M$, by computing,
\begin{align}
    M = \left | 1 - \frac{\int_\Omega \phi(x,y,2\pi) \; d\Omega}{\int_\Omega \phi(x,y,0) \; d\Omega} \right | .
\end{align}

The scalar field, $\phi$, and the velocity field, $\mathbf{u}$, are both infinitely smooth and the characteristics of the velocity field can be well resolved by the reference map based methods (\ie the Coupled Level Set Reference Map and the Volume Preserving Reference Map). We use a second-order Runge-Kutta method to compute the departure point and the quadratic WENO interpolation (see, \textit{e.g.} Equation~\eqref{eq:quadENO}) to interpolate the scalar, $\phi$, and the map, $\xi$, between grid points. Additionally, for this example we allow for extrapolation outside of the domain to remove any accuracy penalties from boundary clipping effects. Because the characteristics of this velocity field can be well resolved, the map, $\xi$, does not require any restarts.

We compare the results of our CB scheme with the SL, the CLSRM in \cite{bellotti2019rm}, and the VPRM in \cite{theillard2021vprm} using this example. While the CLSRM and VPRM methods were specifically developed in the level set context, they can naturally be applied to the advection of scalar fields. Additionally, we include results for our the augmented reference map characteristic bending method (RMCB). 

\begin{figure}[!tb]
    \centering
    \makebox[\linewidth][c]
    {
        \includegraphics[width=0.32\linewidth]{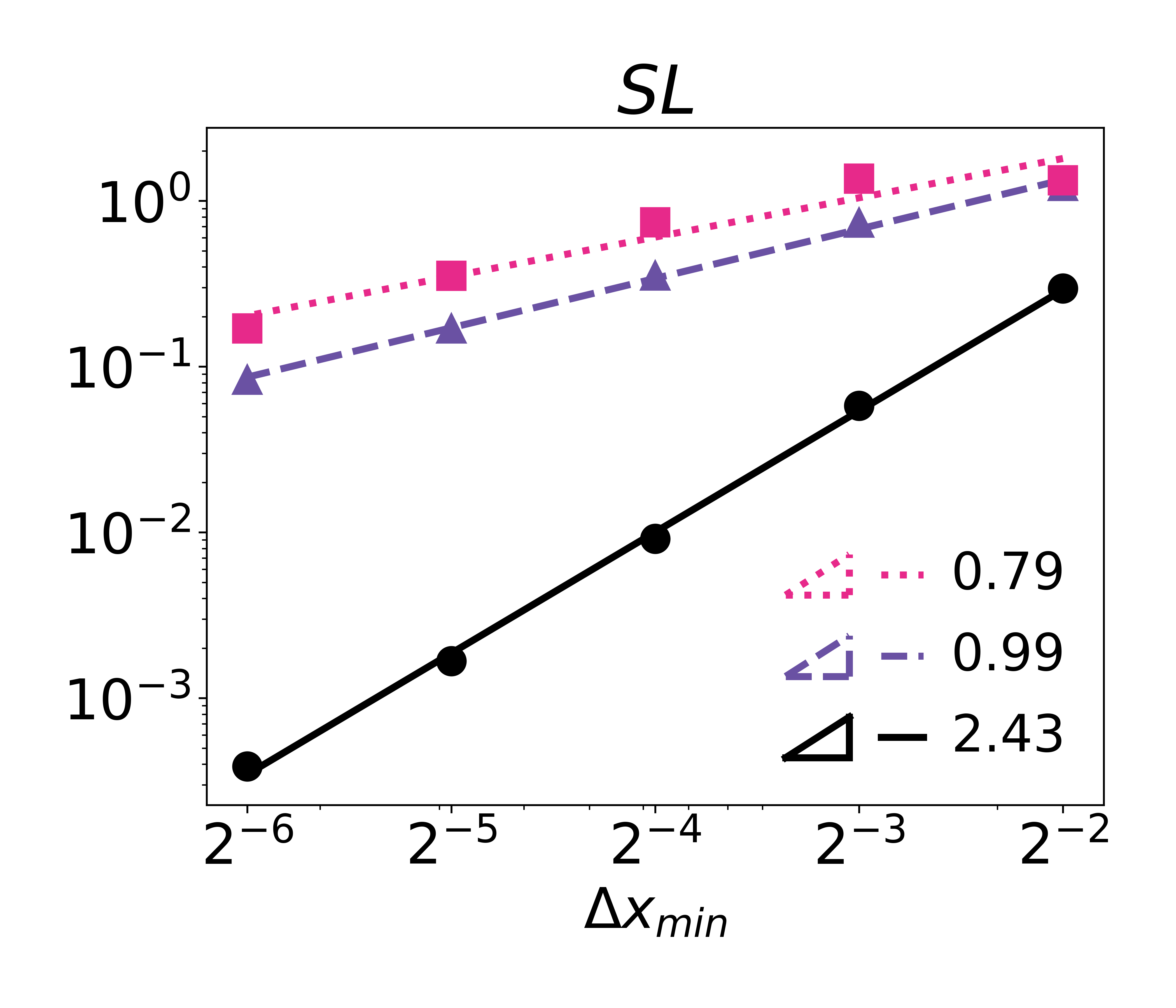}
        \hspace{0.03\linewidth}
        \includegraphics[width=0.32\linewidth]{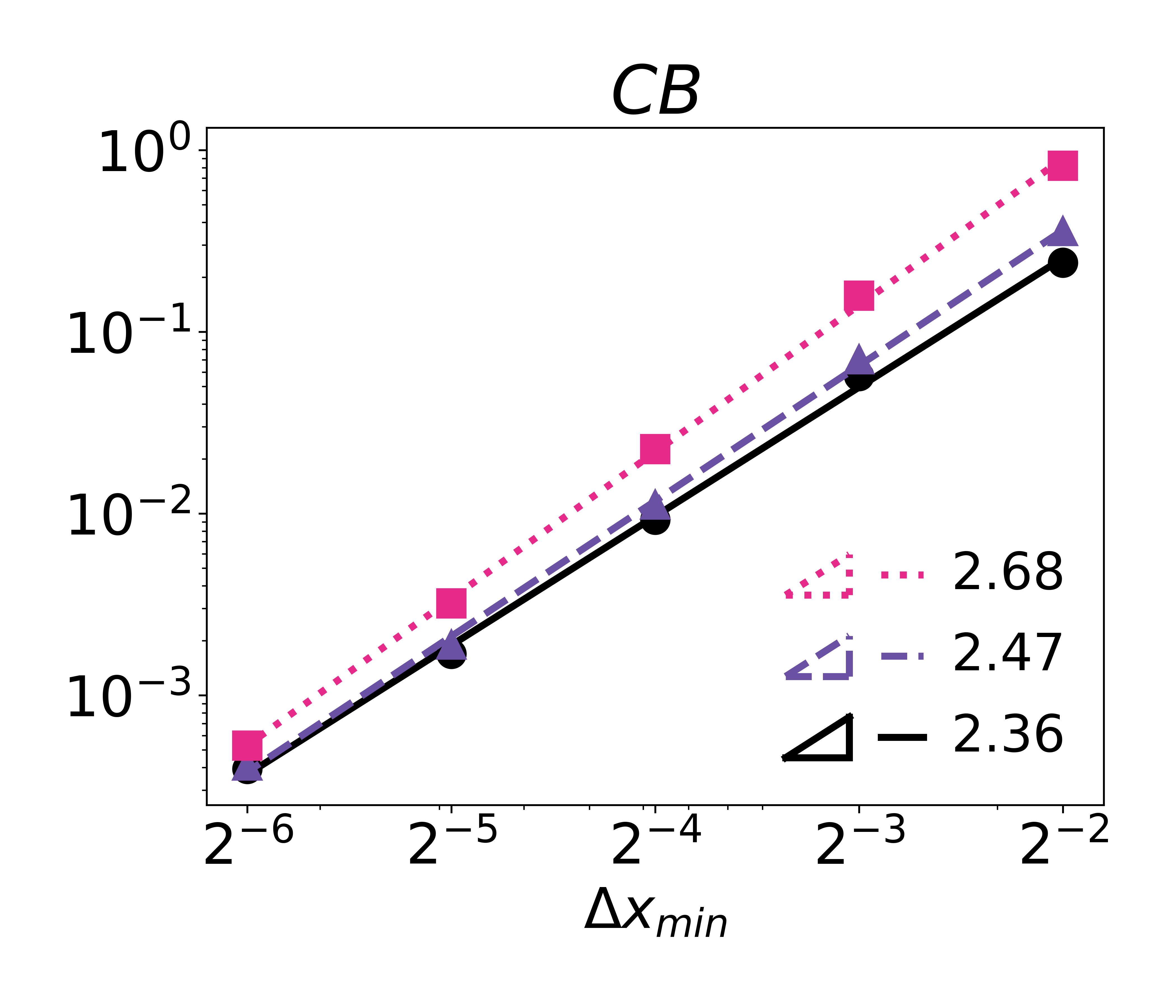}
    }
    \\[1em]
    \makebox[\linewidth][c]
    {
        \includegraphics[width=0.32\linewidth]{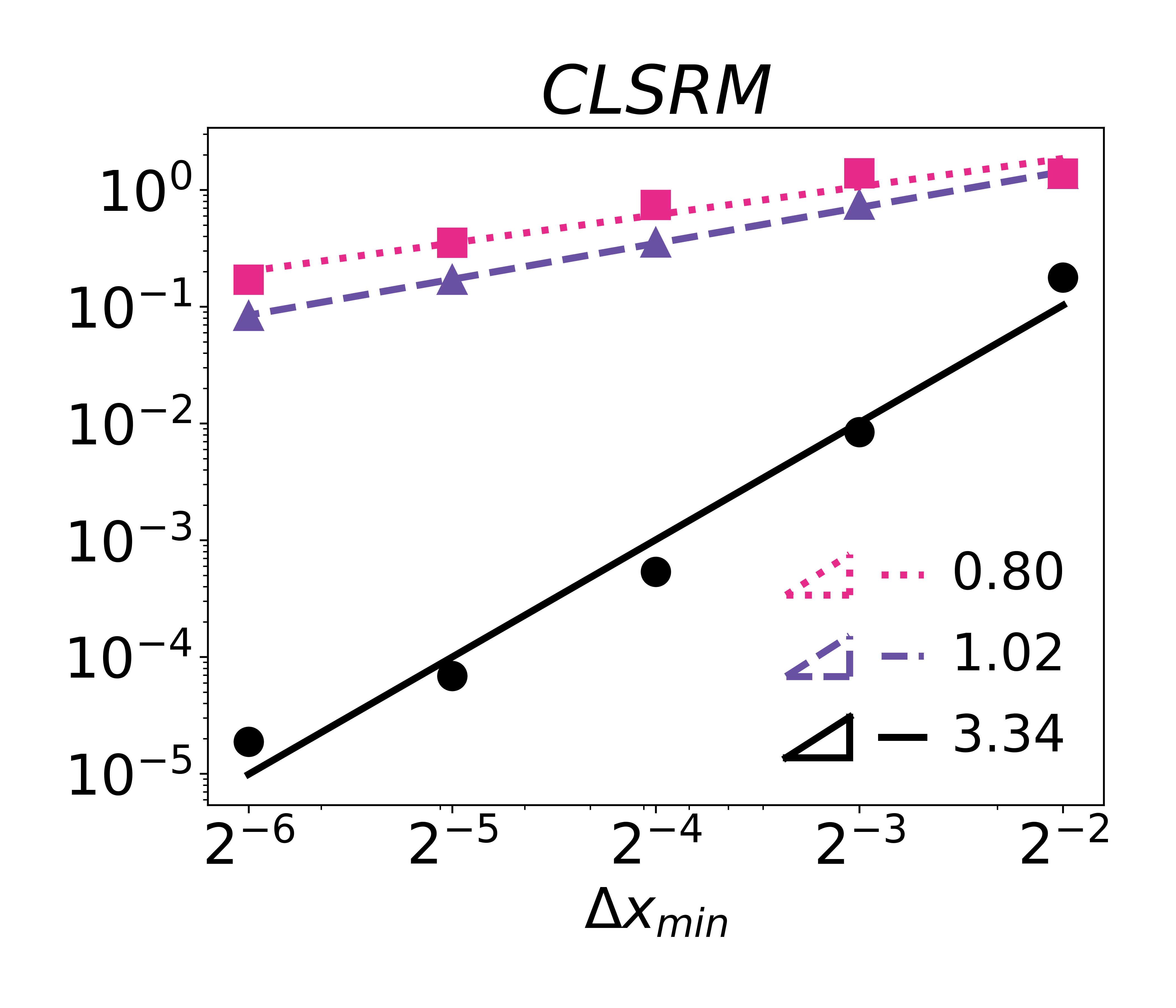}
        \hspace{0.03\linewidth}
        \includegraphics[width=0.32\linewidth]{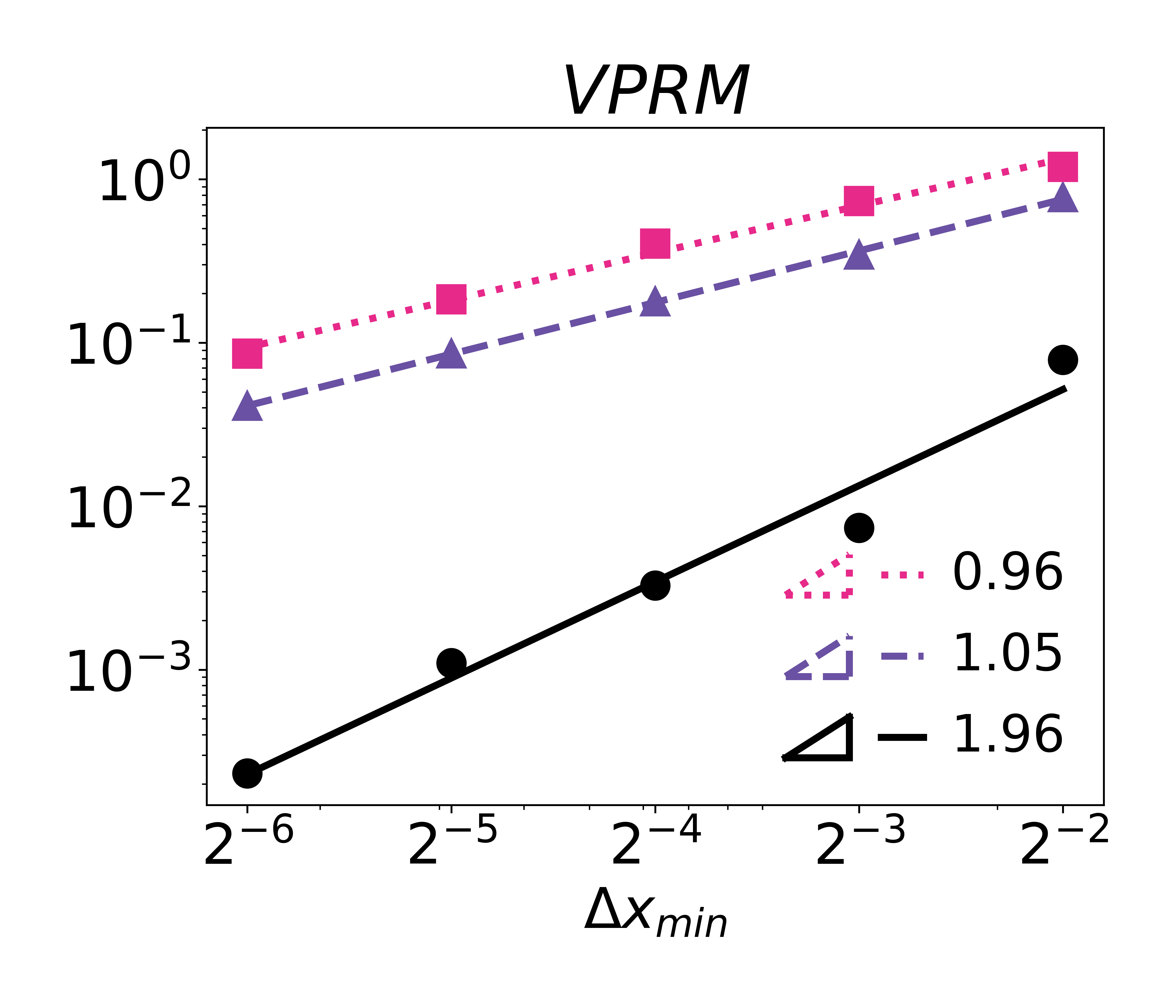}
        \hspace{0.03\linewidth}
        \includegraphics[width=0.32\linewidth]{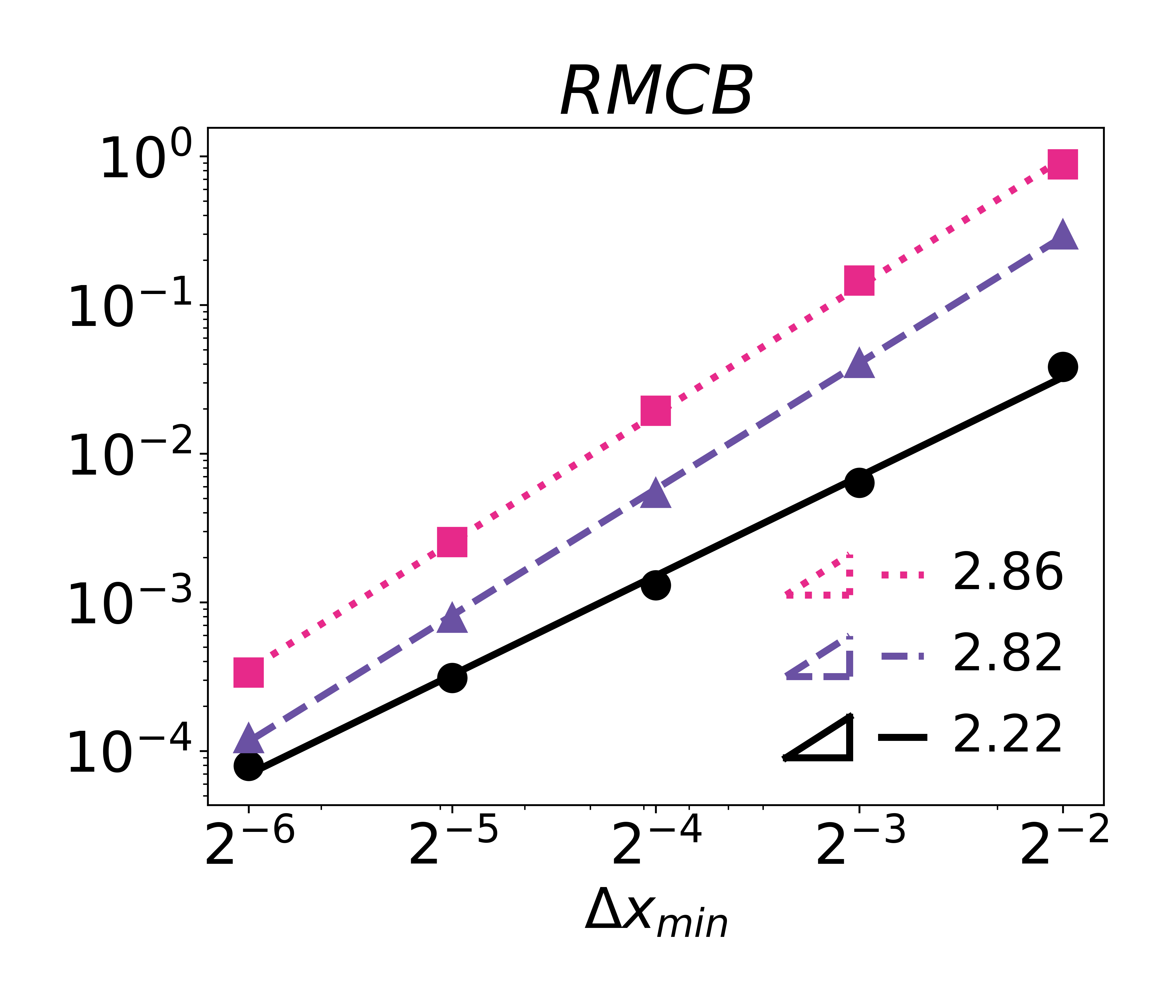}
    }
    \caption{Solution error $\|\phi-\phi_e\|_\infty$ for the Gaussian advection example with various level of artificial expansion added to the velocity field. \includegraphics[height=0.015\textwidth]{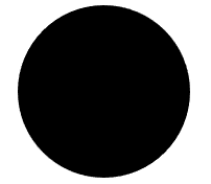} denotes the completely incompressible field, \includegraphics[height=0.015\textwidth]{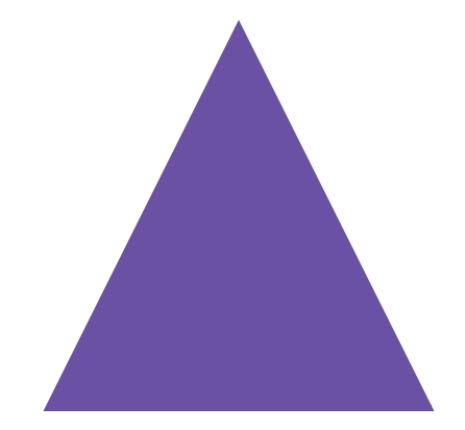} denotes the second-order expansion added to the incompressible velocity (\textit{i.e.} $\alpha=0, \; \beta=1$), and \includegraphics[height=0.015\textwidth]{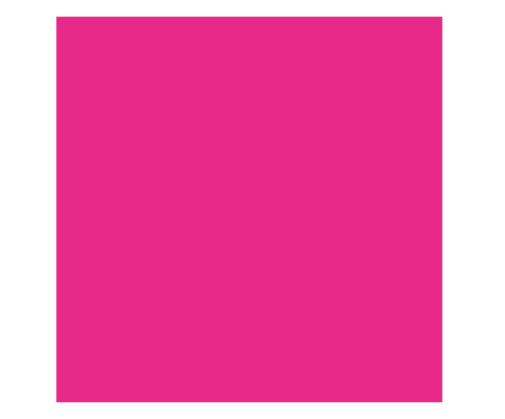} denotes the first-order expansion added to the analytic velocity field (\textit{i.e.} $\alpha=1, \; \beta=0$).}
    \label{fig:gauss_2d_intr}
\end{figure}

\begin{figure}[!htb]
    \centering
    \makebox[\linewidth][c]
    {
        \includegraphics[width=0.32\linewidth]{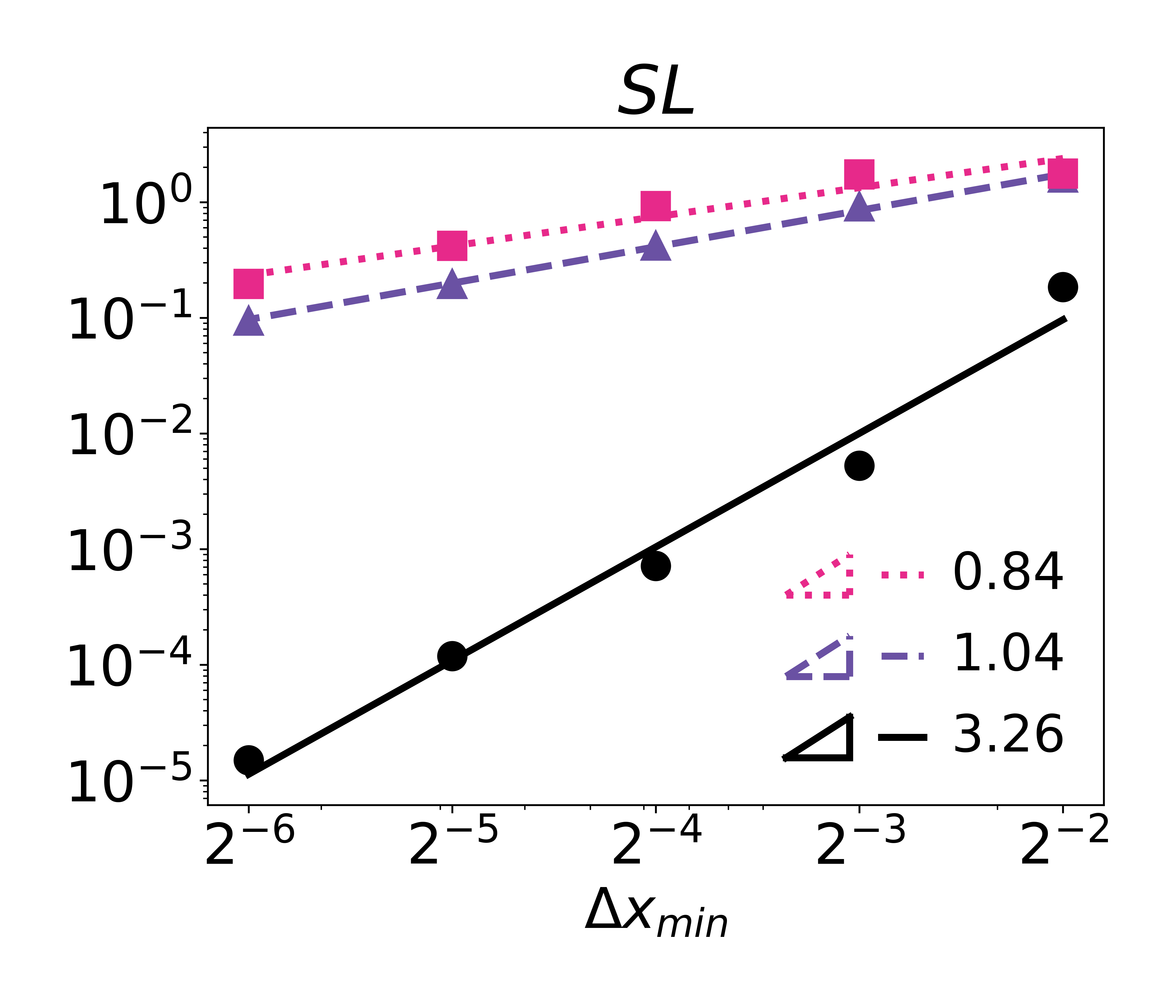}
        \hspace{0.03\linewidth}
        \includegraphics[width=0.32\linewidth]{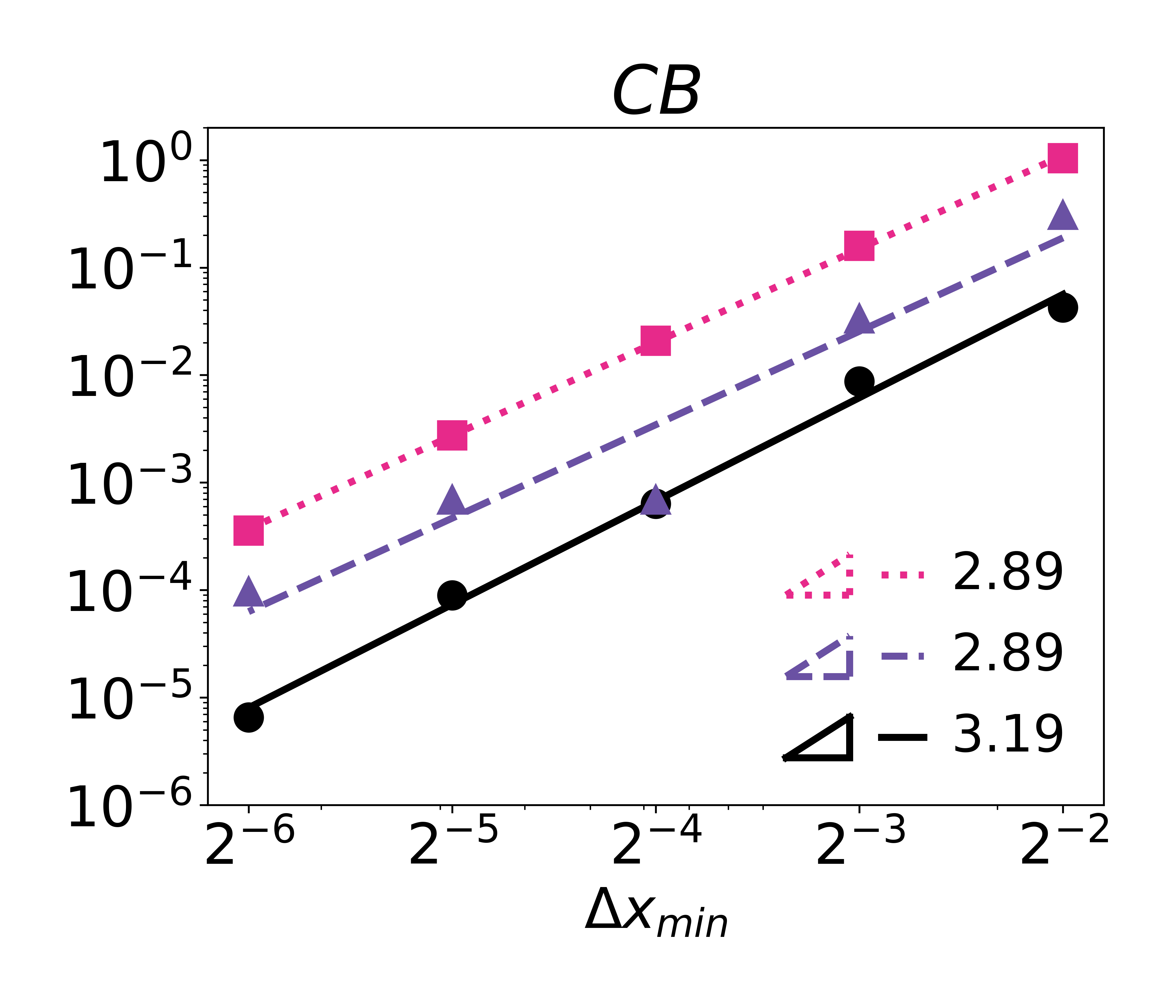}
    }
    \\[1em]
    \makebox[\linewidth][c]
    {
        \includegraphics[width=0.32\linewidth]{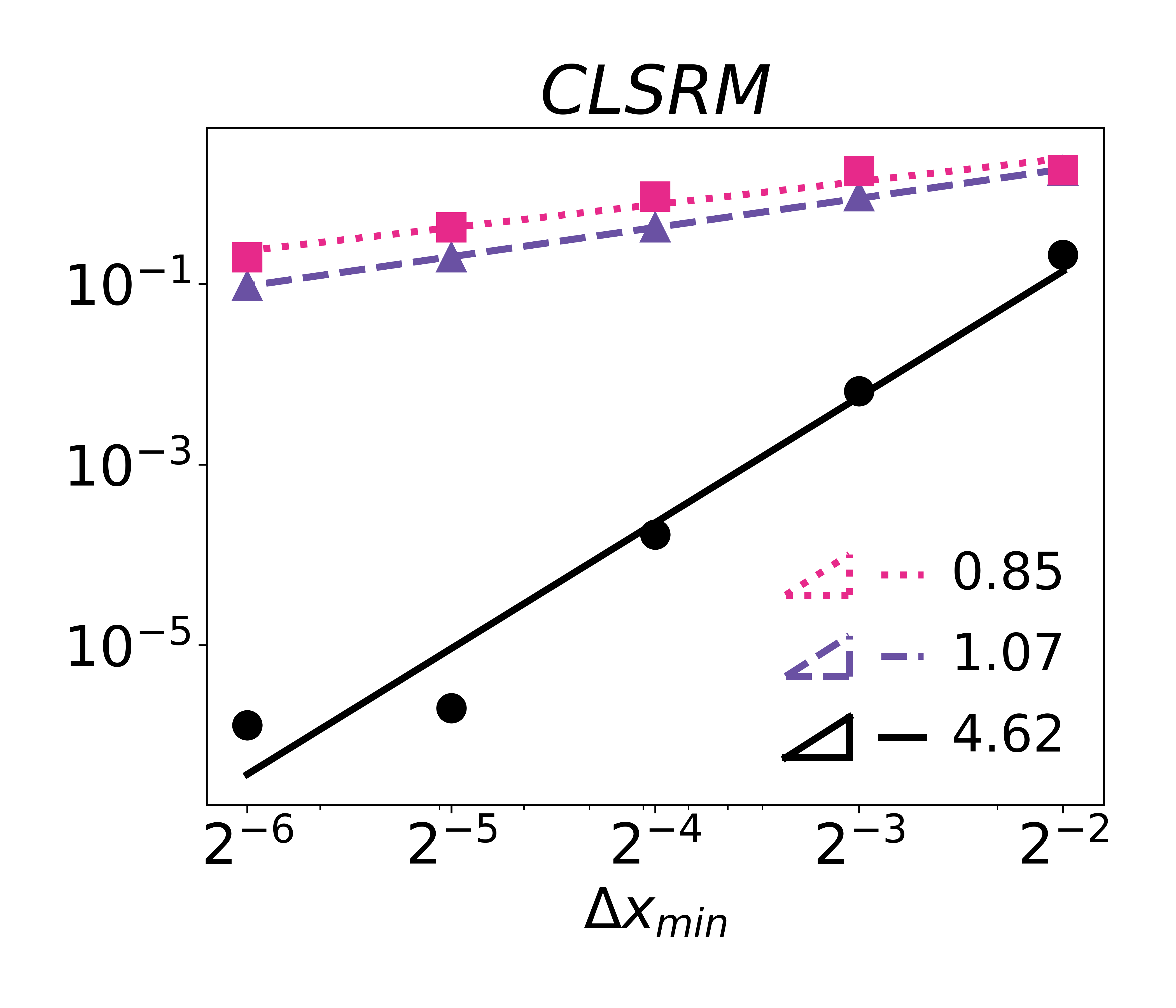}
        \hspace{0.03\linewidth}
        \includegraphics[width=0.32\linewidth]{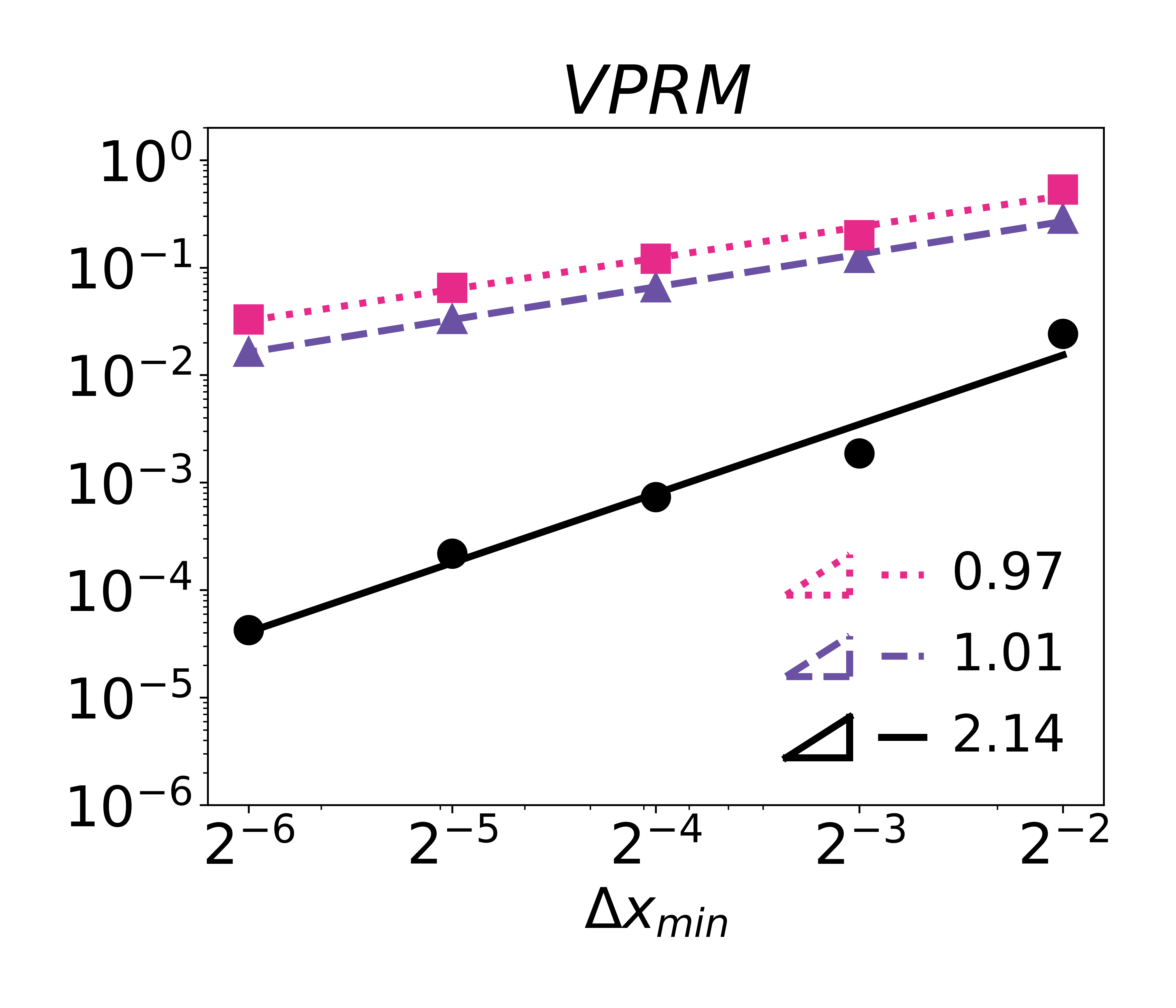}
        \hspace{0.03\linewidth}
        \includegraphics[width=0.32\linewidth]{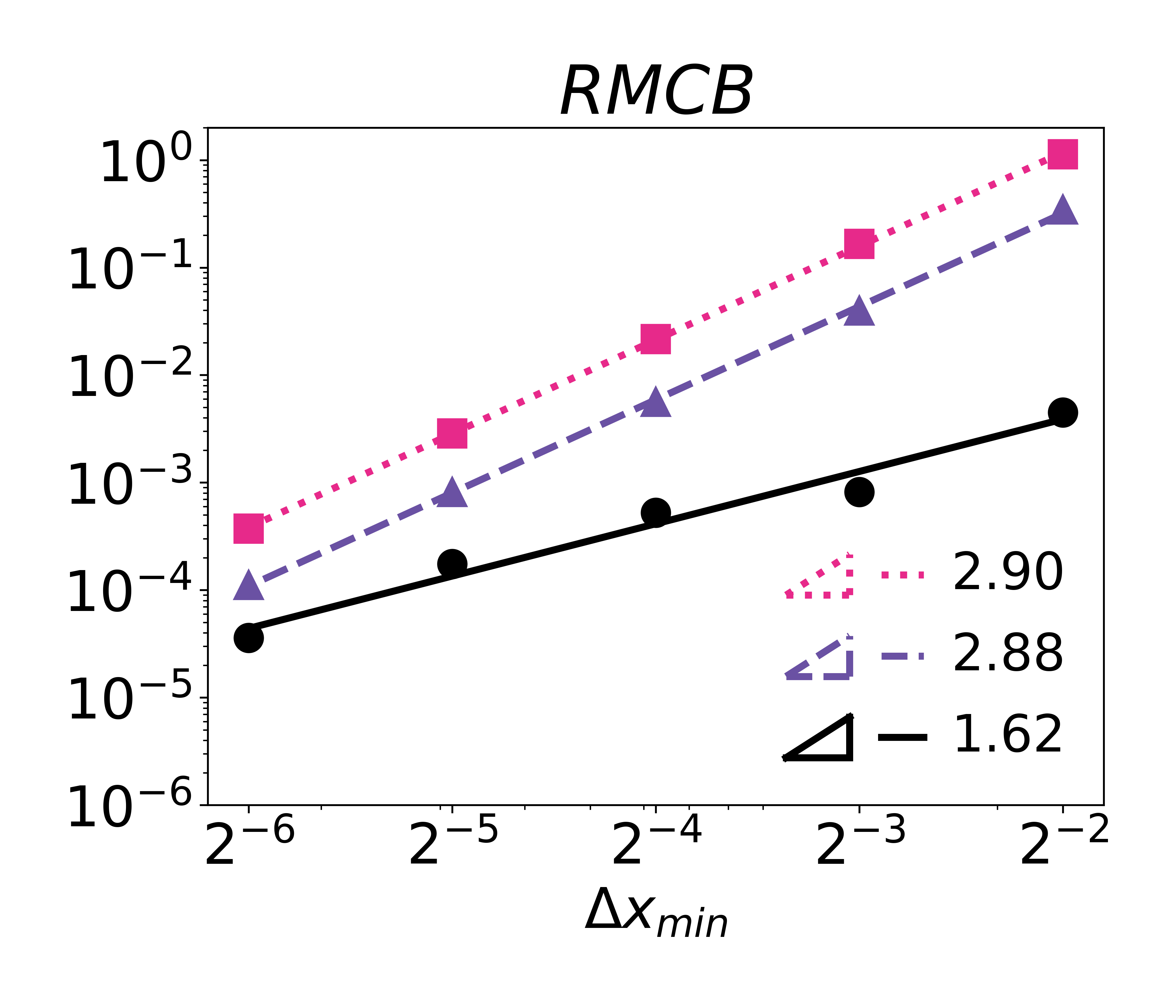}
    }
    \caption{Mass loss for the Gaussian advection example with various level of artificial expansion added to the velocity field. \includegraphics[height=0.015\textwidth]{figures/circle.pdf} denotes the completely incompressible field, \includegraphics[height=0.015\textwidth]{figures/purple_triangle.pdf} denotes the second-order expansion added to the incompressible velocity (\textit{i.e.} $\alpha=0, \; \beta=1$), and \includegraphics[height=0.015\textwidth]{figures/square_purple.pdf} denotes the first-order expansion added to the analytic velocity field (\textit{i.e.} $\alpha=1, \; \beta=0$).}
    \label{fig:gauss_2d_mass}
\end{figure}

We test each of these numerical methods on a uniform grid by setting the minimum and maximum refinement levels equal (we consider adaptive meshes in the subsequent examples). We test each scheme with no artificial expansion (\ie $\alpha=0, \beta=0$), a second-order deviation from the incompressible space (\ie $\alpha=0, \beta=1$), and a first-order deviation from the incompressible space (\ie $\alpha=1, \beta=0$). Each of these methods inherit the stability properties of the semi-Lagrangian scheme (see \cite{falcone1998slconvergence}) and we demonstrate this by using a CFL number of $5$. We present the results for SL, CLSRM, VPRM, CB, and RMCB in Figures \ref{fig:gauss_2d_intr} and \ref{fig:gauss_2d_mass} for the solution error and mass loss, respectively.

Starting with SL, we see that without any artificial expansion we see second-order convergence in the $L^\infty$ norm and third-order convergence in mass loss, $M$. When we introduce the artificial viscosity, however, the order of accuracy for the semi-Lagrangian experiences a significant drop. This is entirely expected as the SL method does not have any machinery to filter the added compressible modes. Similarly, we see the CLSRM perform exceptionally well with the analytically divergence-free velocity field and then suffer once the artificial expansion is added. Again, the CLSRM does not include any way to filter out compressible modes. The VPRM does perform slightly better than the previous methods, however, here we see the effect of projecting long-time map and the accumulation of errors. 

The results for the CB scheme present a much more robust handling of the artificial expansion. Without artificial expansion, both the SL and CB methods yield nearly identical results. CB continues to yield similar results even after the artificial expansion is added. Here, we do see an impact when the first order expansion is added at coarser levels of refinement. However, this impact decreases as the grid is refined. We see a similar dynamic with the RMCB and see that the error associated with the expansion decrease as the grid is refined.

In Figure~\ref{fig:gauss_2d_mass}, we compare the each method with respect to mass loss. The SL and CB methods yield very similar results when with no compressible modes in the advecting velocity. As the strength of the artificial expansion is increased, the SL method drops to first-order accuracy in the mass loss, whereas the CB method maintains second-order. For the VPRM and RMCB methods, we see similar results. Without artificial expansion, the methods perform similar. As the strength of the compressible modes are increase, the VPRM drops to a first-order accurate scheme whereas the RMCB remains second-order accurate.  

Next, we focus on advecting level set interfaces using adaptive quadtree meshes. This is the context that the CLSRM and VPRM methods were developed in and, thus, these examples serve as an illustration of the differences between these foundational schemes and the CB method. 

\subsubsection{Slotted Cylinder}
The purpose of this example is to explore how CB, in comparison to other methods, preserves sharp interface features. A canonical example problem for demonstrating this is the slotted cylinder of \cite{zalesak1979fully}, where the profile of a circle with a slot is embedded into an initial distribution function by setting the value of the distribution to 3 inside of the circle and 1 outside of the circle. This profile is then advected by a field with constant angular velocity for one full revolution around the axis of rotation. After one full revolution, we can measure the accuracy of the numerical scheme by comparing the final advected profile with the initial profile, making this a particularly effective example for visually demonstrating numerical diffusion.

\begin{figure}
    \centering
    \includegraphics[width=0.5\linewidth]{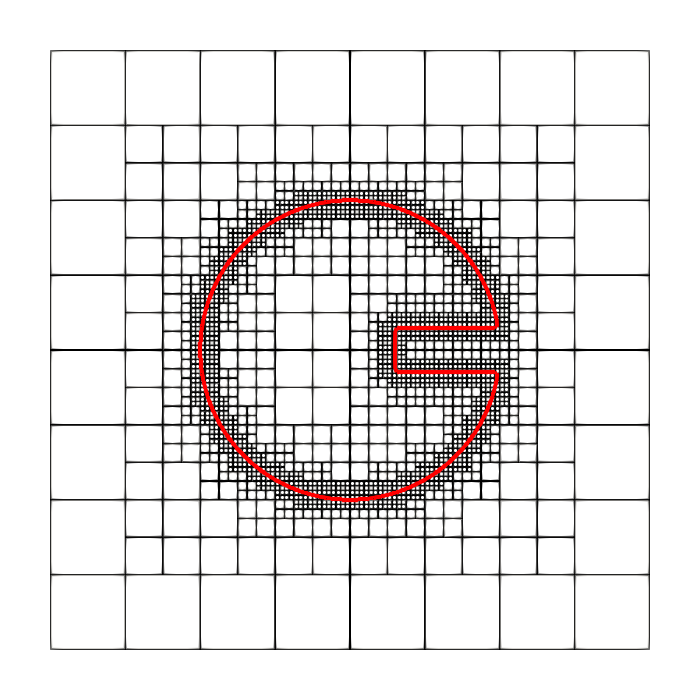}
    \caption{Visual representation of the initial profile (\textit{red}) for the Slotted Cylinder example using an adaptive quadtree mesh, with a minimum and maximum refinement level of 4 and 7, respectively. The circle is centered at the origin, $(0,0)$ with a radius of $R=0.5$. The slot has a width of $w=0.15$ and is placed at a height of $h=0.15$ from the origin. We note that while the level set representation is sharp, the rendering of the profile literally \textit{cuts corners} leading to a visual chamfer of the shape.}
    \label{fig:slotted_cylinder_init}
\end{figure}

A modern variation of this problem, and what we do herein, embeds the slotted cylinder profile using a level set function, centered in the computational domain. Specifically, we consider the advection of an interface defined by the zero-contour of the level set function shown in Figure \ref{fig:slotted_cylinder_init}. For this example, we use the domain $\Omega = [-1,1]^2$ and the velocity field, $\mathbf{u} = (-y, x)$. As with the previous example, we include a compressible component in the velocity field centered at the origin, which leads to the same velocity field defined by \eqref{eq:u_rotational}.

For this example, we use an adaptive quadtree mesh and refine the mesh near the interface (also shown in Figure \ref{fig:slotted_cylinder_init}). Specifically, we use the refinement criteria introduced in Section \ref{sec:amr} with the uniform band set to $8 \Delta x_{min}$\footnote{This has been a common practice for level set methods since the early work of Adalsteinsson and Sethian in \cite{adalsteinsson1999fast}.}. This value is based on a heuristic, but is loosely chosen to ensure that the foot of the characteristics originating from nodes near the interface lie in region of maximum refinement. Additionally, we set the span between the minimum and maximum refinement levels to be 4 (\ie minimum level 3, maximum level 7) and increment both the minimum and maximum refinement levels to compute the order of accuracy. Finally, we set the finest resolution for this example to a minimum level of 7 and a maximum level of 11. 

In this example, the interface of the slotted cylinder is embedded as the zero-contour of a signed distance function, $\phi$. In order to ensure that $\phi$ remains a signed distance function, $\phi$ must be periodically reinitialized. Following \cite{bellotti2019rm} and \cite{theillard2021vprm}, we reinitialize $\phi$ using the PDE-based approach proposed in \cite{sussman1994level} and solve the following axillary equation for $\phi$, 
\begin{align} \label{eq:PDE_reinit}
    \frac{\partial \phi}{\partial \tau} = sign(\phi) \left (|\nabla \phi | - 1 \right ) \qquad \forall x \in \Omega 
\end{align}
for a fictitious time, $\tau$, until steady state. To solve this reinitialization equation, we use a Total Variation Diminishing second-order Runge-Kutta (TVD-RK2) method \cite{shu1988efficient}. For a detailed review of the level set method, we refer the reader to the books \cite{osher2003lsbook} and \cite{sethian1999level}. 

One side effect of this systematic reinitialization procedure is that it tends to smooth sharp features of the level set function. As noted in \cite{sussman1999efficient} and \cite{sethian1999level}, iterating the reinitialization procedure has the effect of perturbing the zero-contour of the level set function, which can diffuse sharp interfaces and lead to mass loss. There are a number of alternative techniques that better preserve sharp features, such as the work in \cite{saye2014high}, but avoiding reinitialization altogether tends to be one of the best strategies. This, is one of the key benefits with reference map based schemes. Because the map does not require frequent restarts, we can maintain the signed distance nature of $\phi$ without frequent reinitialization because $\phi(\mathbf{x},t)$ is constructed directly from the composition of $\phi_0$ and the reference map $\xi$. 

For the work herein, we follow \cite{bellotti2019rm} and \cite{theillard2021vprm} (primarily for a fair comparison) and perform systematic reinitialation using the PDE-based scheme at every iteration of the SL and CB schemes. Similarly, we reinitialize the level set function every time we restart the map for the RM, VPRM and RMCB methods. We use the criteria presented in \cite{theillard2021vprm} when restarting the reference map, which monitors the bijectivity of the map and performs a restart when the columns of $\nabla \xi$ are nearly colinear. As with the previous example, the characteristics of the solid body rotation field (\textit{i.e.} $\mathbf{u}=(-y,x)$) are well resolved using the reference map approach and no restarting is required for the analytically divergence free field. When the artificial expansion terms are incorporated, however, this is no longer guaranteed. 

For each of these examples, we measure the $L^2$ error on the interface, which we define as follows,
\begin{align} \label{eq:interface_L2}
    L^2 = \left ( \int_{\Gamma} \left ( \phi(x,y,0) - \phi(x,y,2\pi) \right )^2 \; d \Gamma \right )^{1/2},
\end{align}
where $\Gamma$ is defined as the set of nodal values $(x_n,y_n)$ of $\phi$, such that $\phi(x_n,y_n) < \Delta x_{min}$ (\ie at least one of the neighboring nodes is across the interface). Finally, we compute the volume loss by integrating a constant density, $\rho=1$, over the volume defined as the interior of the interface (\ie $\mathbf{x} \in \Omega \; s.t. \; \phi(\mathbf{x},t) \le 0 $) as follows,
\begin{align} \label{eq:interface_ML}
    V = \left | 1 - \frac{\int_{V(2\pi)} \rho \; dV}{\int_{V(0)} \rho \; dV} \right |.
\end{align}

\begin{figure}[!htb]
    \centering
    \makebox[\linewidth][c]
    {
        \includegraphics[width=0.32\linewidth]{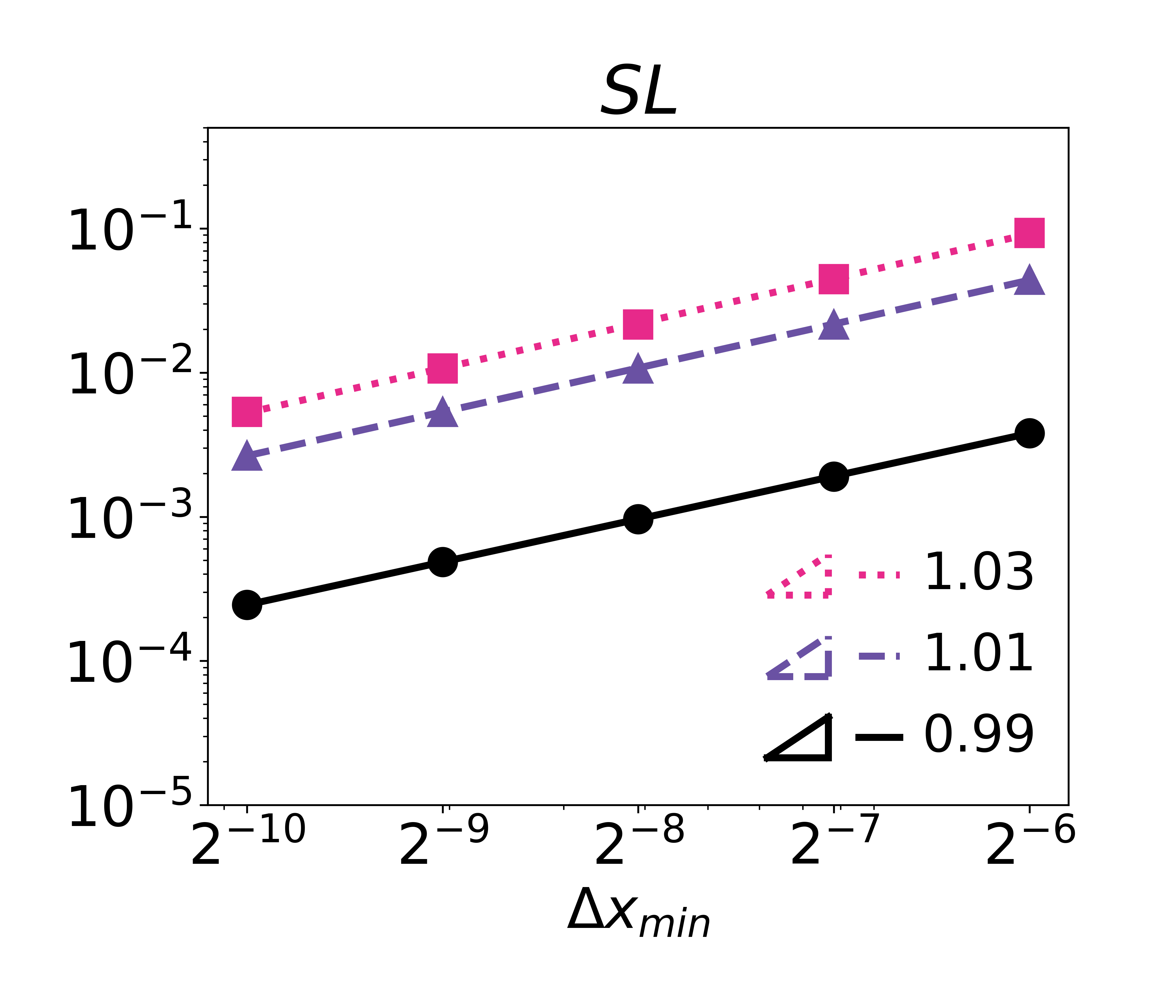}
        \hspace{0.03\linewidth}
        \includegraphics[width=0.32\linewidth]{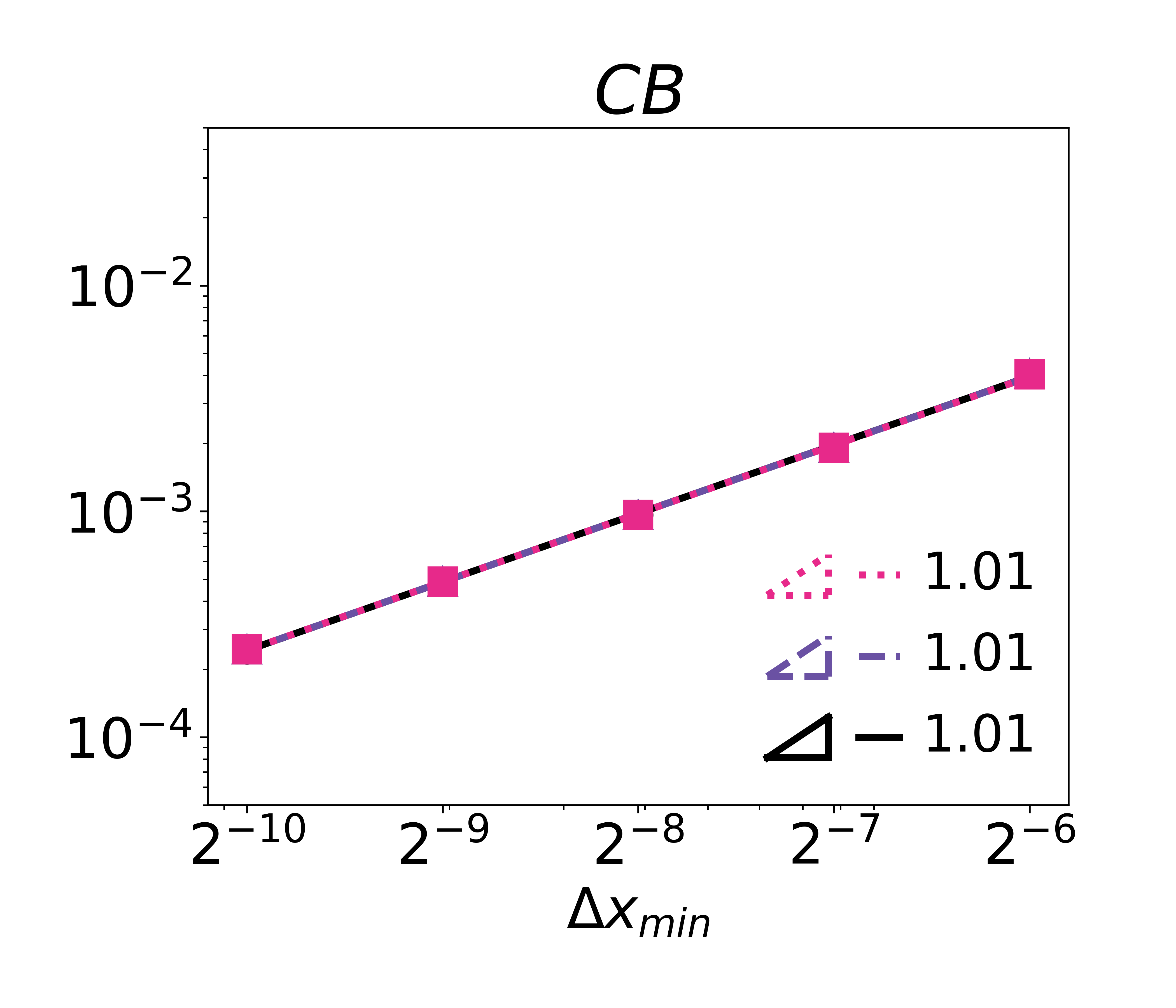}
    }
    \\[1em]
    \makebox[\linewidth][c]
    {
        \includegraphics[width=0.32\linewidth]{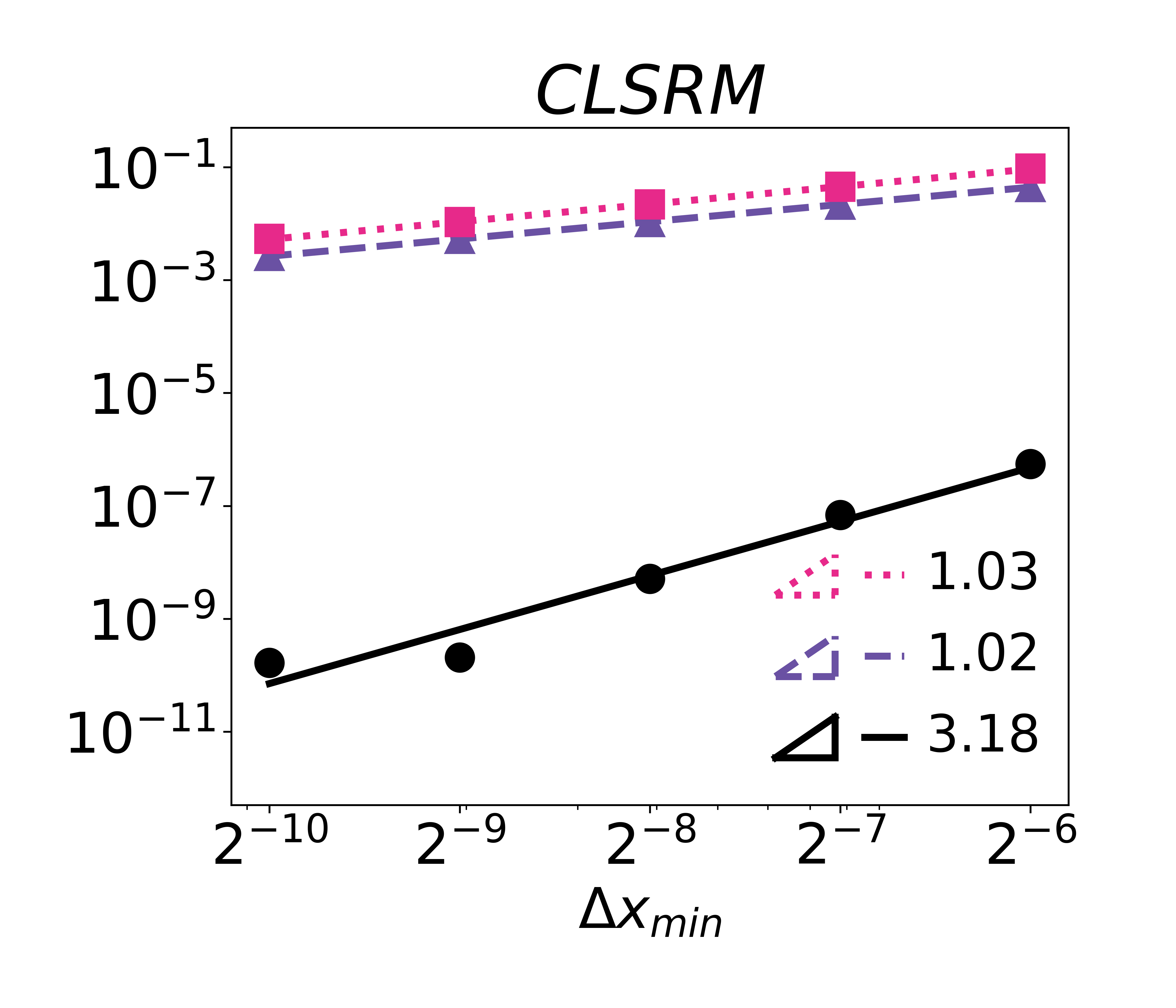}
        \hspace{0.03\linewidth}
        \includegraphics[width=0.32\linewidth]{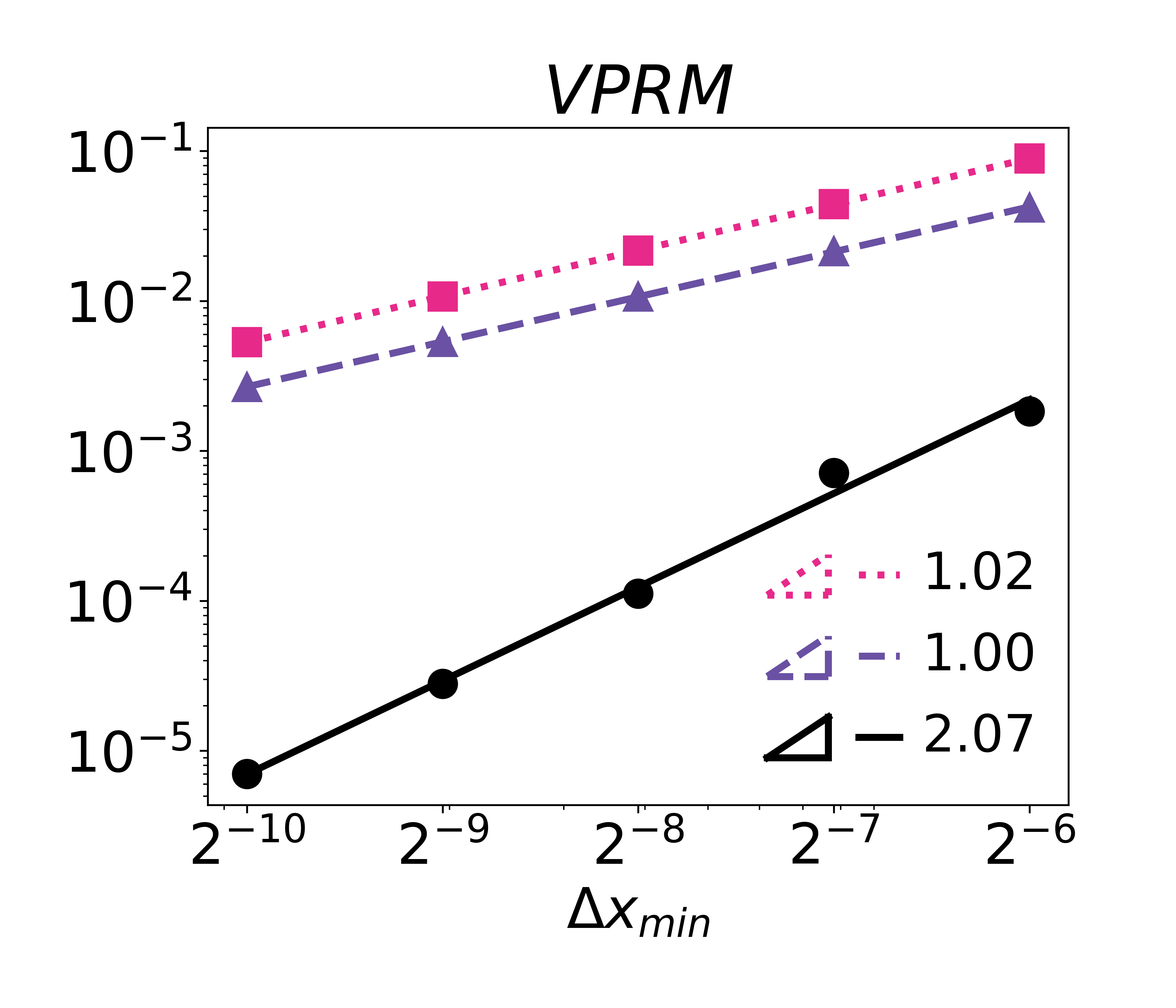}
        \hspace{0.03\linewidth}
        \includegraphics[width=0.32\linewidth]{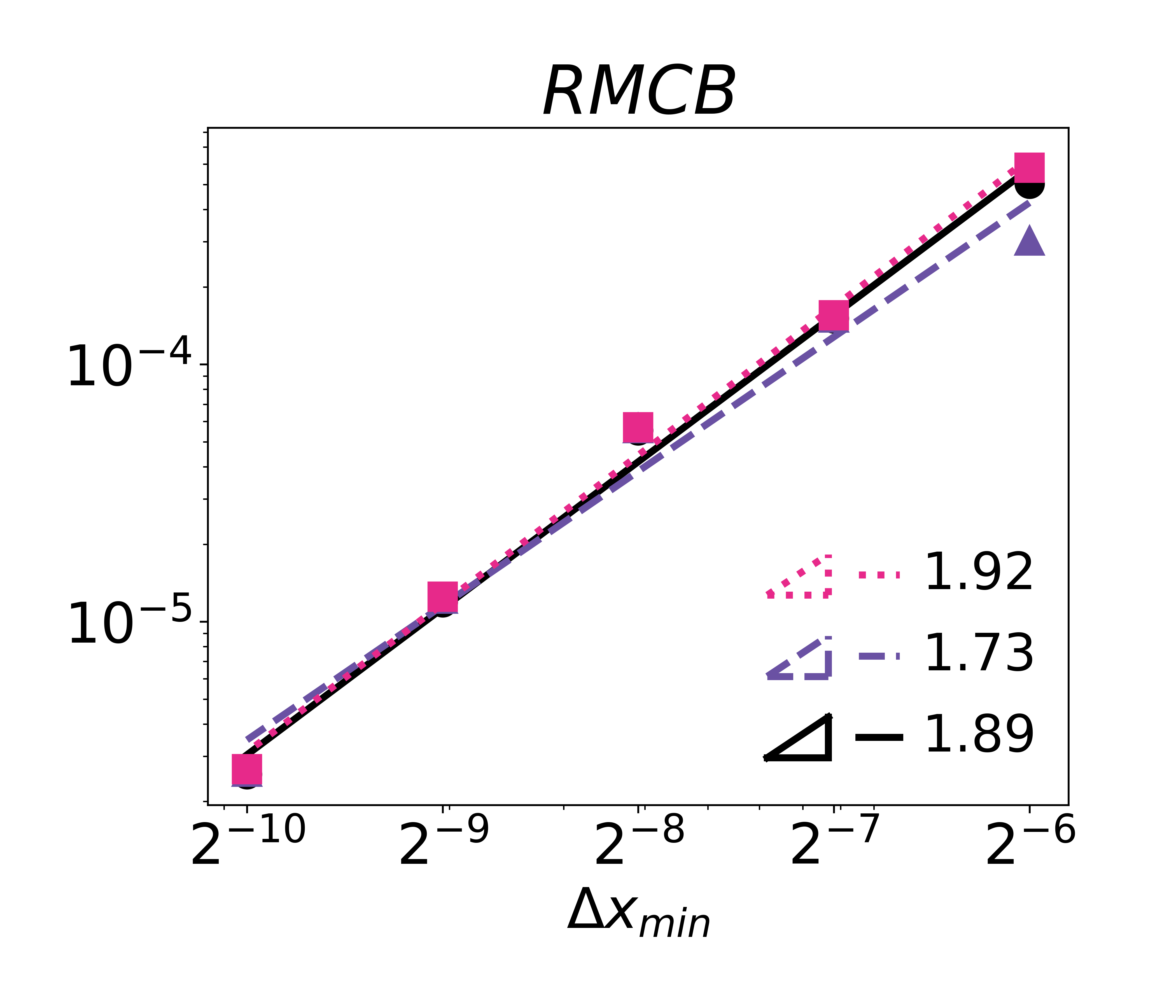}
    }
    \caption{Solution error $\|\phi-\phi_e\|_2$ for the Slotted Circle example with various level of artificial expansion added to the velocity field. \includegraphics[height=0.015\textwidth]{figures/circle.pdf} denotes the completely incompressible field, \includegraphics[height=0.015\textwidth]{figures/purple_triangle.pdf} denotes the second-order expansion added to the incompressible velocity (\textit{i.e.} $\alpha=0, \; \beta=1$), and \includegraphics[height=0.015\textwidth]{figures/square_purple.pdf} denotes the first-order expansion added to the analytic velocity field (\textit{i.e.} $\alpha=1, \; \beta=0$).}
    \label{fig:slot_2d_intr}
\end{figure}

\begin{figure}[!htb]
    \centering
    \makebox[\linewidth][c]
    {
        \includegraphics[width=0.32\linewidth]{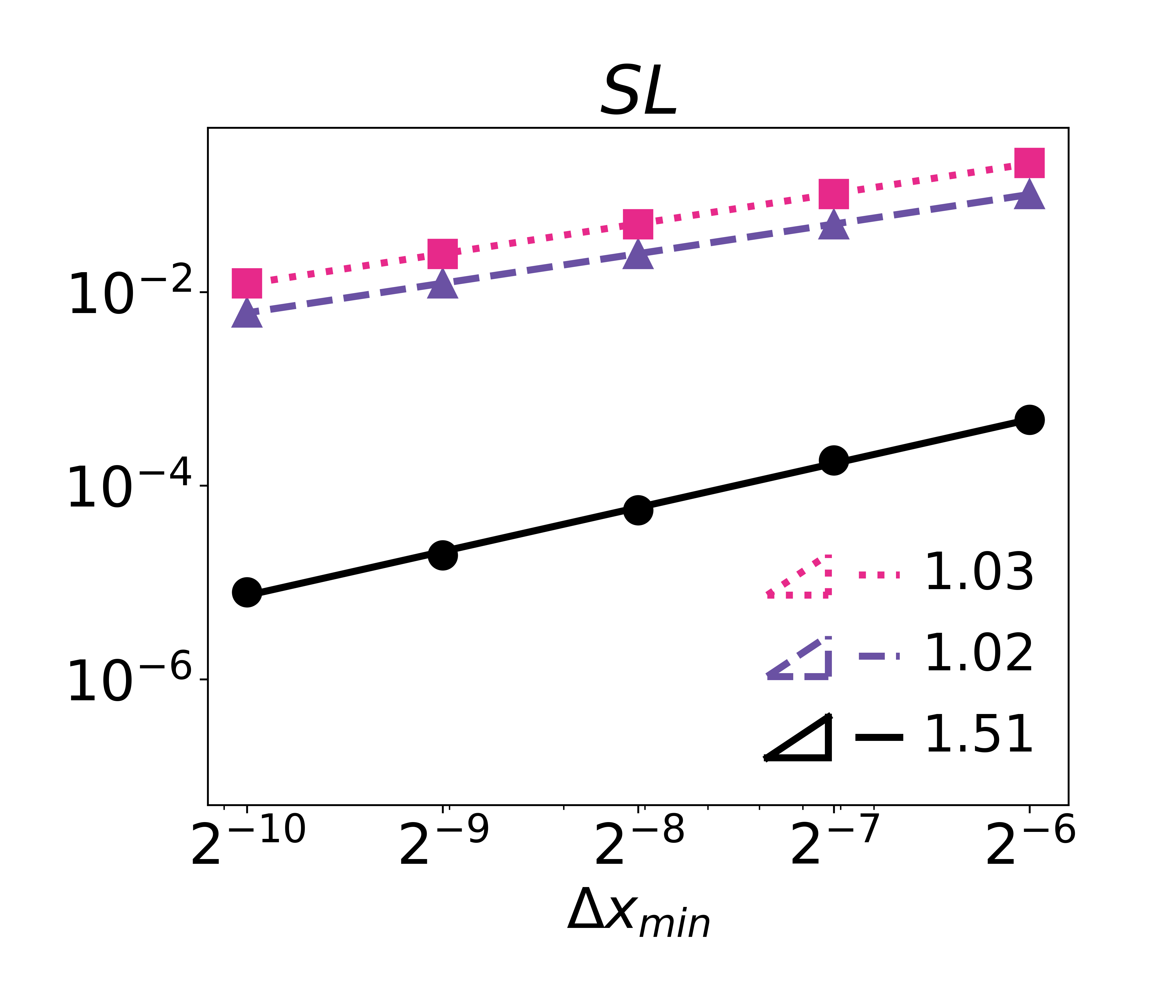}
        \hspace{0.03\linewidth}
        \includegraphics[width=0.32\linewidth]{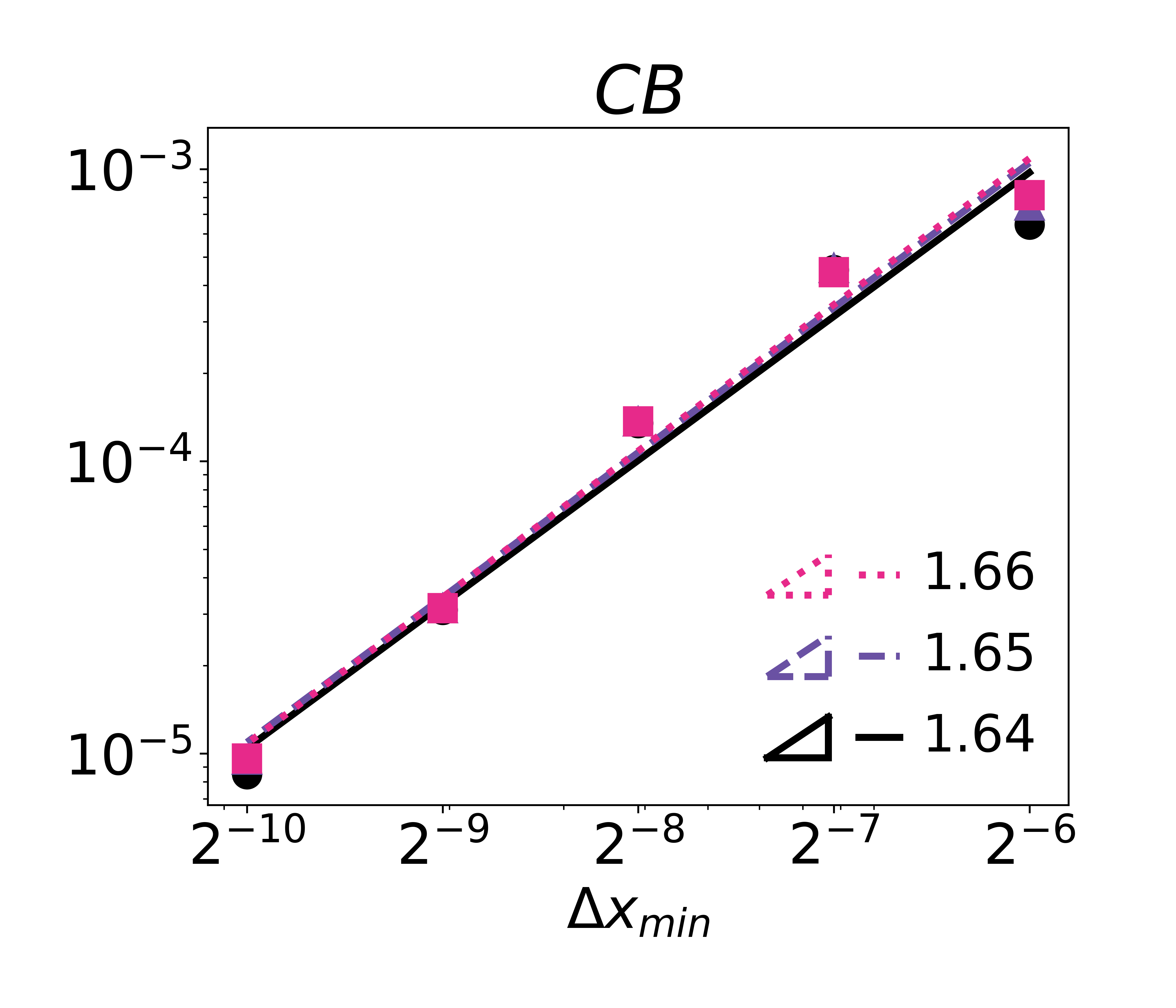}
    }
    \\[1em]
    \makebox[\linewidth][c]
    {
        \includegraphics[width=0.32\linewidth]{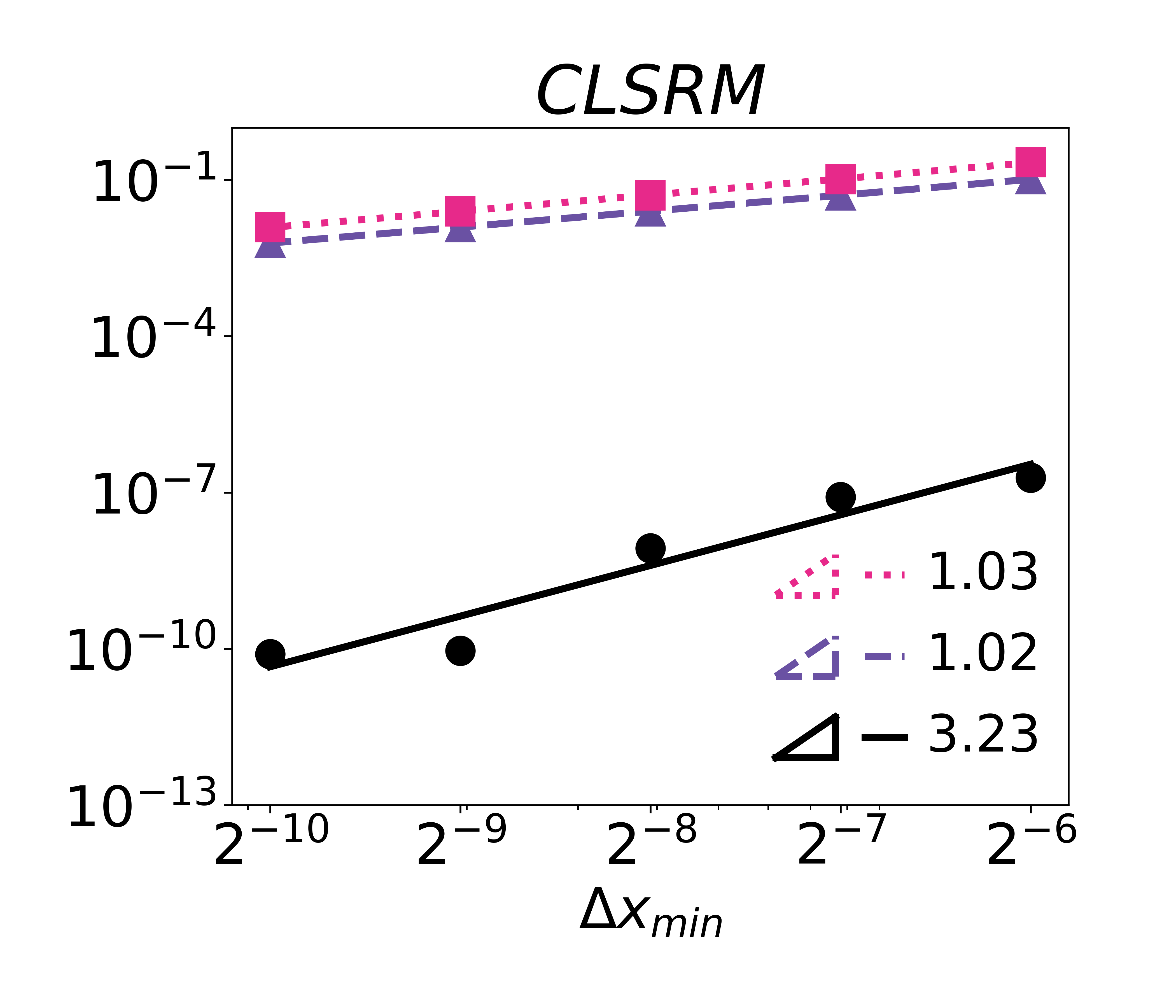}
        \hspace{0.03\linewidth}
        \includegraphics[width=0.32\linewidth]{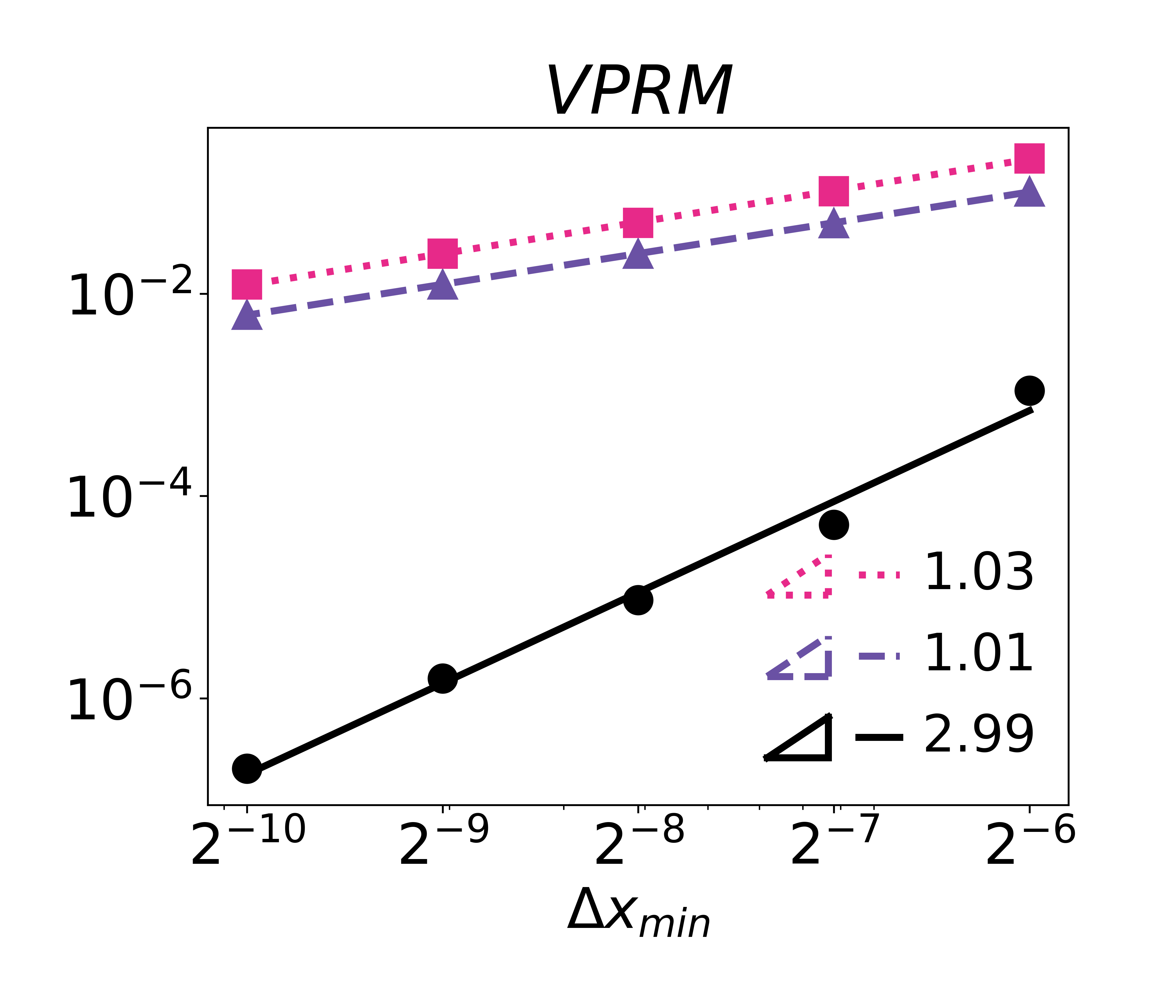}
        \hspace{0.03\linewidth}
        \includegraphics[width=0.32\linewidth]{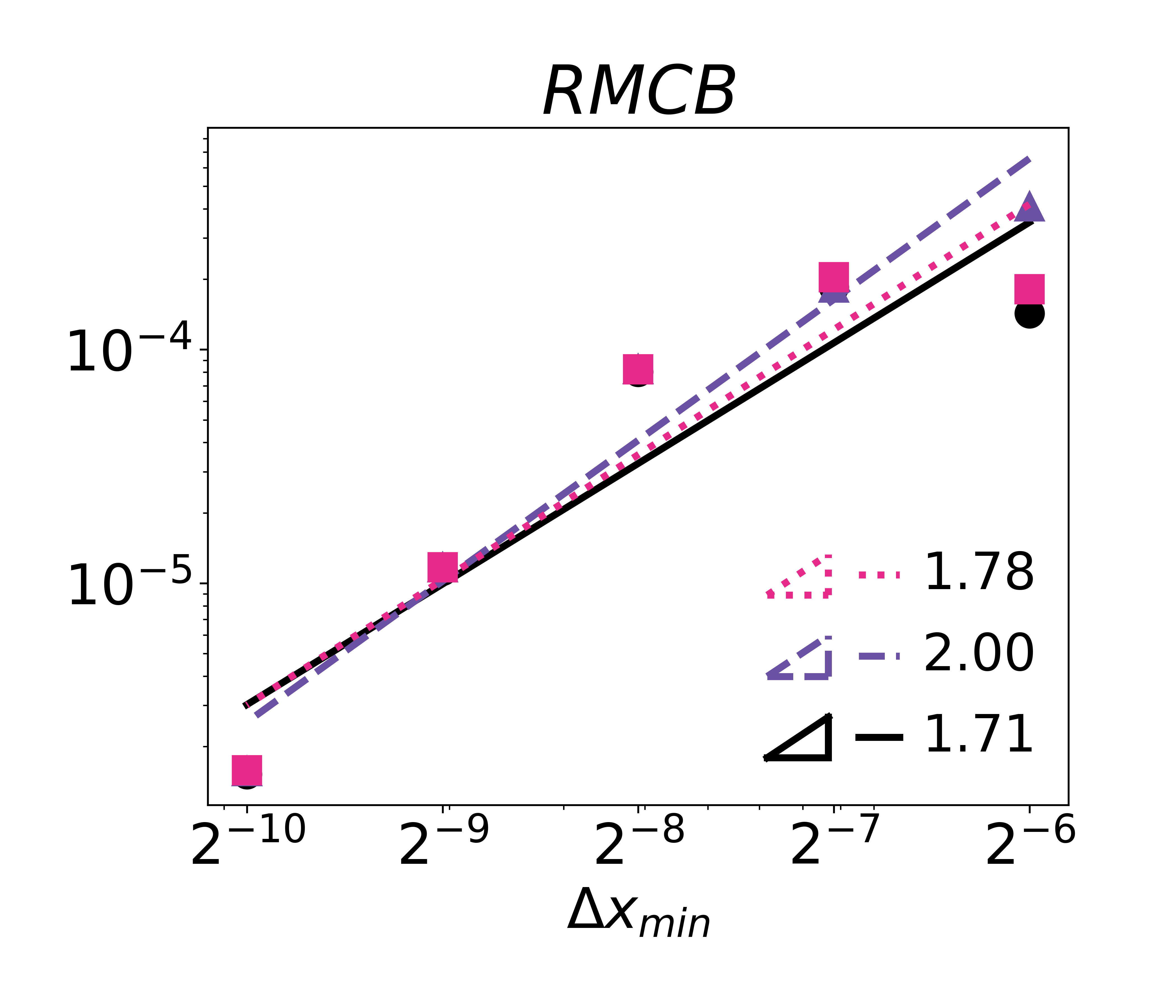}
    }
    \caption{Volume loss for the Slotted Circle example with various levels of artificial expansion added to the velocity field. \includegraphics[height=0.015\textwidth]{figures/circle.pdf} denotes the completely incompressible field, \includegraphics[height=0.015\textwidth]{figures/purple_triangle.pdf} denotes the second-order expansion added to the incompressible velocity (\textit{i.e.} $\alpha=0, \; \beta=1$), and \includegraphics[height=0.015\textwidth]{figures/square_purple.pdf} denotes the first-order expansion added to the analytic velocity field (\textit{i.e.} $alpha=1, \; \beta=0$).}
    \label{fig:slot_2d_vol}
\end{figure}

\begin{figure}[!htb]
    \centering
    \includegraphics[width=0.85\linewidth]{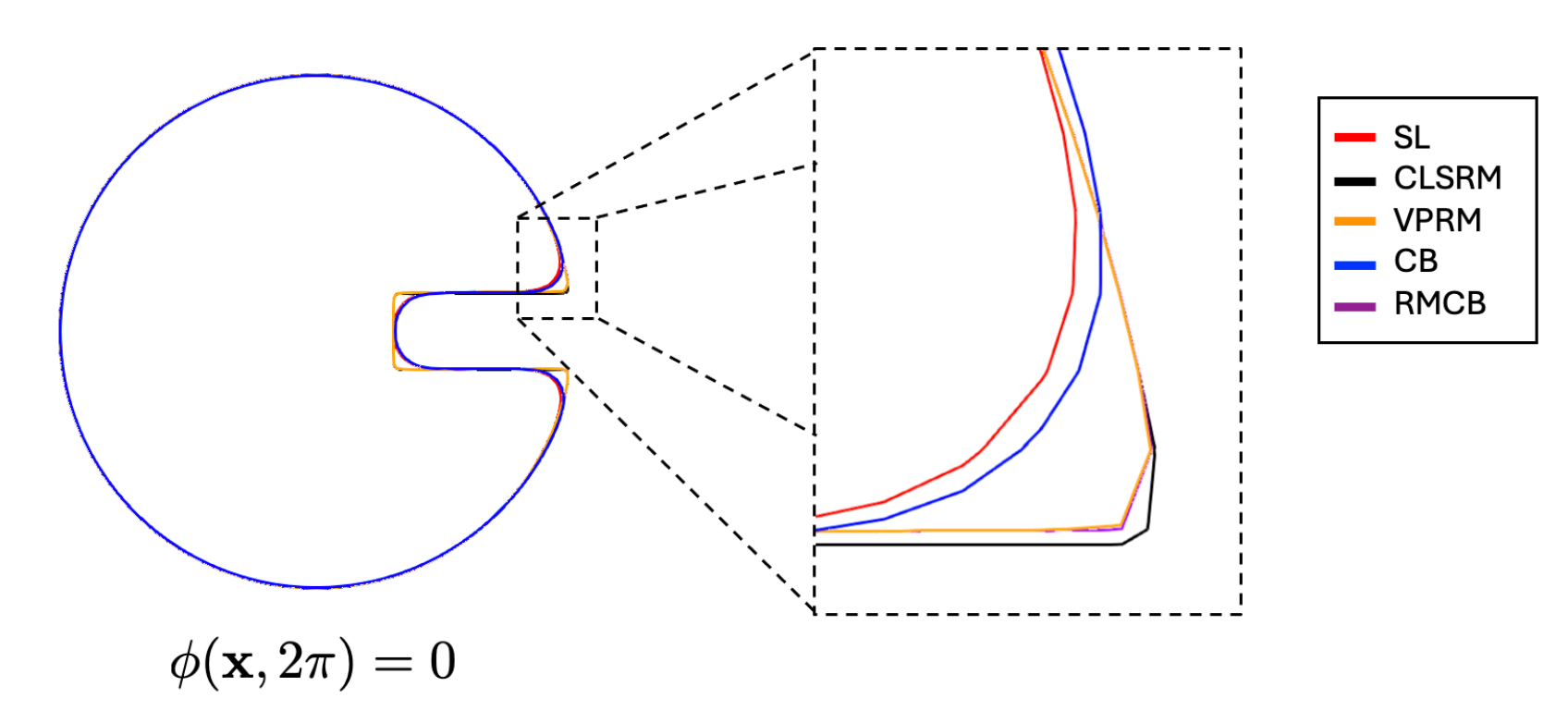}
    \caption{Comparison of rounded corners in the Slotted Circle example using each of the different methods. Results are shown using a quadtree grid with a minimum and maximum level of refinement of 3 and 7, respectively. The CLSRM method (black) is visually identical to the exact solution, $\phi(\mathbf{x},0)=0$.}
    \label{fig:slot_2d_zoom}
\end{figure}

\begin{figure}[!htb]
    \centering
    \includegraphics[width=0.95\linewidth]{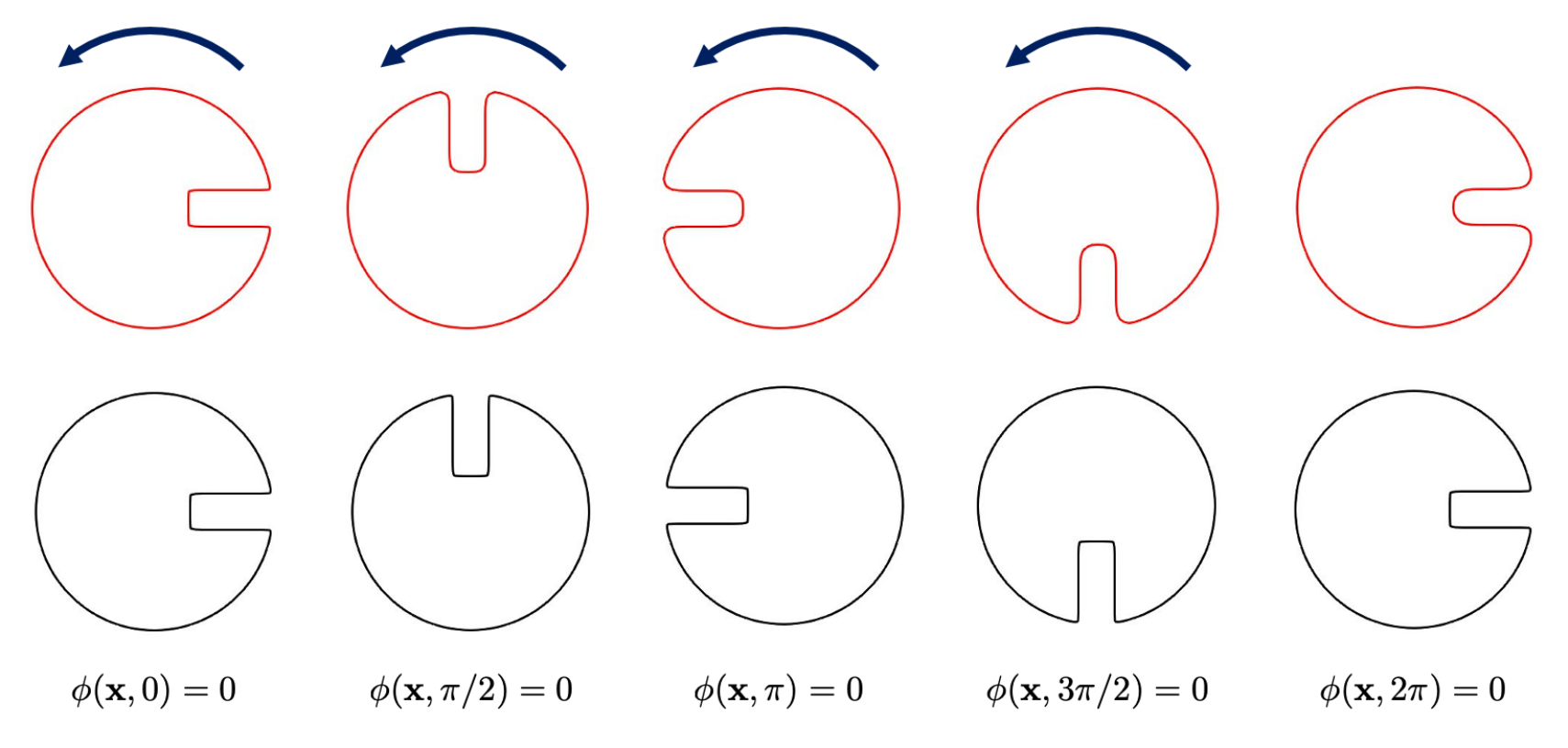}
    \caption{Evolution of the interface for the slotted circle example. The SL method (red) rounds sharp features of the interface as a result of interpolation and PDE-based reinitialization. The CLSRM method (black) avoids rounding sharp features by reconstructing the interface using the initial configuration and avoiding frequent reinitialization.}
    \label{fig:slot_2d_slvrm}
\end{figure}

We present the numerical results for each of the tested methods, using a CFL of $5$, in Figures \ref{fig:slot_2d_intr} and \ref{fig:slot_2d_vol} for interface error and volume loss, respectively. We immediately see that the SL method yields first-order accuracy for the interface error and between first- and second-order accuracy in volume loss. The decrease in accuracy for the SL method is directly related to the frequent interpolation of $\phi$ and the reinitialization of the interface at each time step. In Figures \ref{fig:slot_2d_zoom} and \ref{fig:slot_2d_slvrm}, we show how these effects result in a rounding effect at the corners of the slotted circle. The CLSRM avoids the accumulation of these errors by reconstructing the interface at every time step from the initial configuration avoiding the need for reinitialization. Unfortunately, the high-order accuracy of the CSLRM is significantly degraded when the artificial expansion is applied to the velocity field. Again, the CLSRM method does not have any machinery to filter compressible modes. This creates both error due to following the modified characteristics as well as necessitating frequent remaps as $\nabla \xi$ frequently triggers the colinear threshold.  

The VPRM performs slightly worse than the CLSRM for the incompressible case (\textit{i.e.} $\alpha=\beta=0$). This highlights one of the key findings from \cite{theillard2021vprm}, where it was shown that the VPRM can result in degraded performance compared with the CSLRM for analytically divergence-free velocities. Essentially, the added volume-preserving projection can perturb the interface. Here, we use a shell size equal to the uniform band (\textit{i.e.} $\phi < \| 8 \Delta x_{min} \|$) based on the commentary in \cite{theillard2021vprm}, but this parameter can have a significant impact on the overall performance of the method (as noted in \cite{theillard2021vprm}). For example, in the tests conducted here we see that increasing the shell size reduces the accuracy of the method for analytically divergence-free velocities, but improves performance (marginally) when artificial expansion is added to the advecting velocity. Because this parameter is typically chosen based on the assumption that the velocity is incompressible, we present results using the ideal shell size for the divergence-free condition.

\begin{figure}[!htb]
    \centering
    \includegraphics[width=0.95\linewidth]{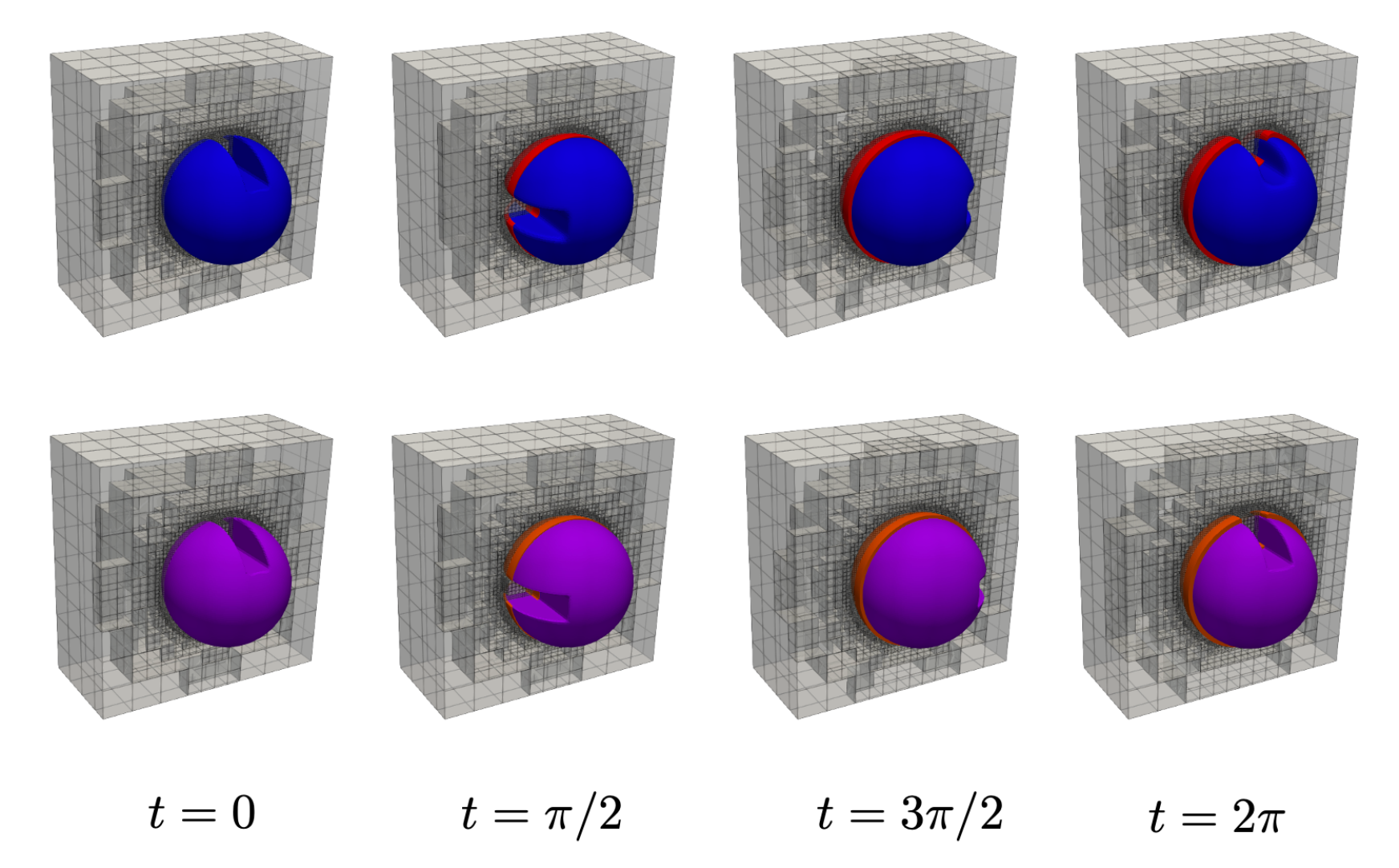}
    \caption{Evolution of the interface for a three-dimensional version of the slotted circle example using $\alpha=1, \; \beta = 0$. A comparison of the SL (red) and CB (blue) methods are shown on the top and a comparison of the VPRM (orange) and RMCB (purple) methods are shown on the bottom at 4 different times. For this example, we see that the CB and RMCB methods retain their shape as the SL and VRPM methods expand. Additionally, we see that the SL and CB methods smooth the corners, whereas the reference map-based methods remain sharp.}
    \label{fig:3d_ball_comparison}
\end{figure}

The results for the CB and RMCB methods demonstrate a similar invariance as in the Gaussian example. The CB method yields similar results to the SL method for the divergence-free condition, including the rounding of sharp features (see Figure \ref{fig:slot_2d_zoom}). When we incorporate the artificial expansion, the CB method yields nearly identical results to the divergence-free solution. Similarly, the RMCB method yields nearly identical results with and without the addition of artificial expansion. When looking at the volume loss error in Figure \ref{fig:slot_2d_vol}, we see similar results. The methods using CB retain performance even in the presence of artificial expansion, whereas the other methods degrade in performance. A three-dimensional example of the slotted cylinder is shown in Figure~\ref{fig:3d_ball_comparison}.

\subsubsection*{A Note on Level Set Reinitialization}

As mentioned above, it is possible to use a different reinitialization approach for the level set method. Here, we briefly demonstrate this possibility using the reinitialization scheme proposed in \cite{saye2014high} for the SL and CB methods\footnote{We only demonstrate the impact of reinitialization with the SL and CB schemes as the reference map-based methods, for this particular example, do not require any reinitialization.}. 

For this example, we use coarse uniform grids (to magnify error) and compare the interface and volume error between the PDE-based reinitialization and the algoim-based reinitialization of \cite{saye2014high}. For algoim, we use a third order reinitialization scheme to match our third-order PDE-based scheme and perform the reinitialization procedure at every iteration (as done with the PDE-based method). Here, we also use a slightly different metric to measure the volume loss. Specifically, we measure the difference between the volume of the level set function at the final time and the exact volume of the slotted circle. This removes potential discretization artifacts resulting from the coarse representation of the level set interface and focuses on the impact of reinitialization on the exact location of the interface. 

\begin{figure}[!htb]
    \centering
    \makebox[\linewidth][c]
    {
        \includegraphics[width=0.45\linewidth]{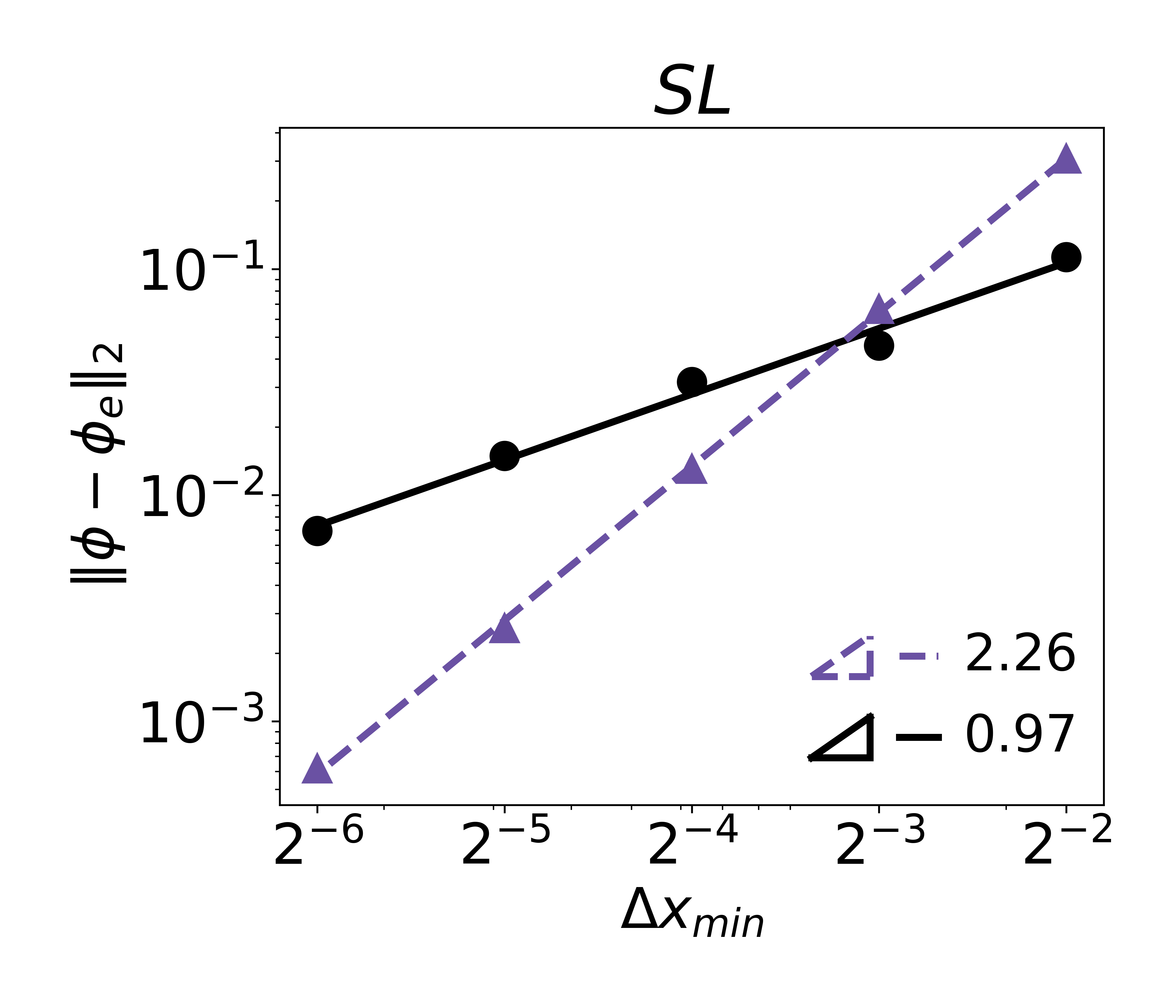}
        \hspace{0.03\linewidth}
        \includegraphics[width=0.45\linewidth]{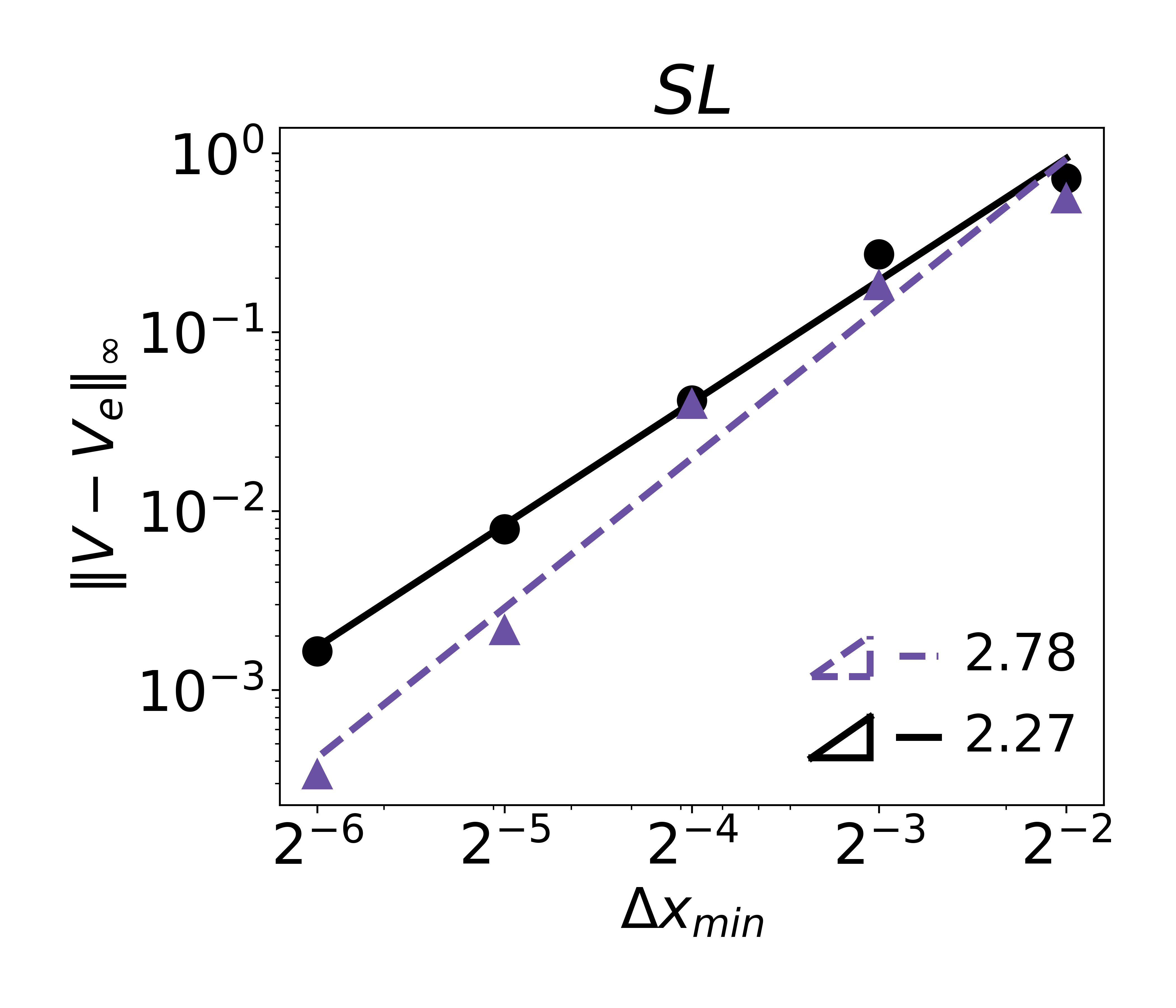}
    }
    \\[1em]
    \makebox[\linewidth][c]
    {
        \includegraphics[width=0.45\linewidth]{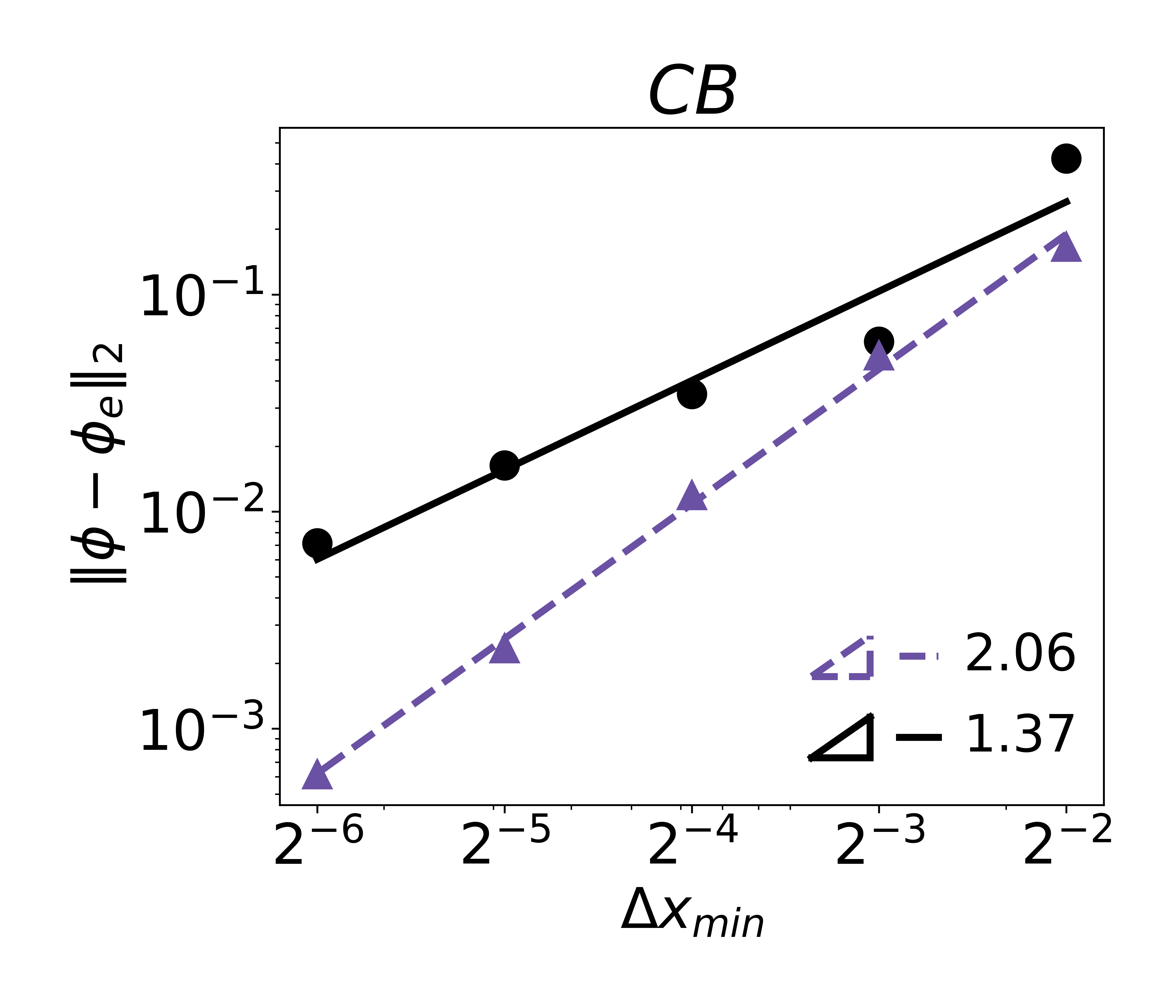}
        \hspace{0.03\linewidth}
        \includegraphics[width=0.45\linewidth]{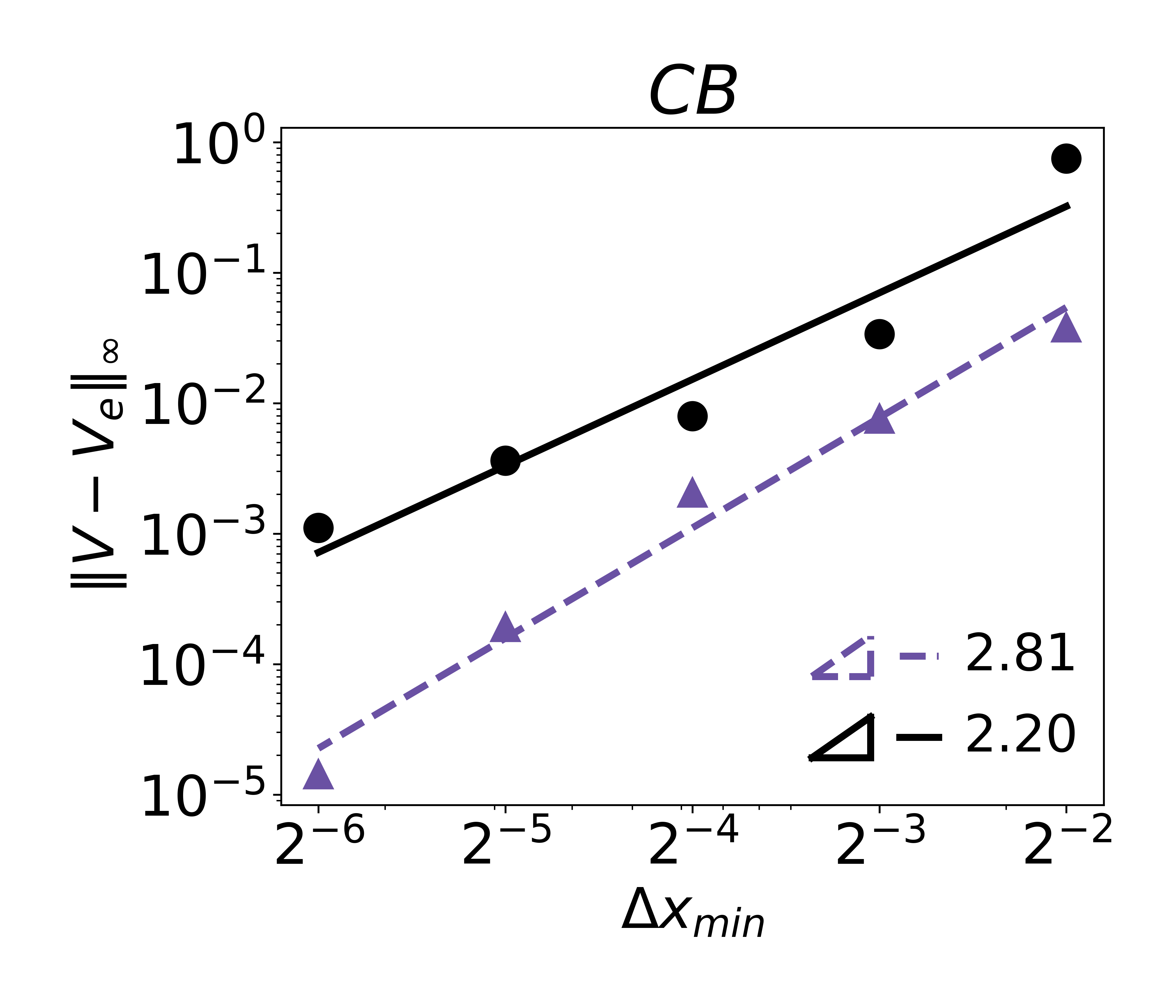}
    }
    \caption{Comparison of the PDE-based reinitialization (\includegraphics[height=0.015\textwidth]{figures/circle.pdf}) and algoim reinitialization (\includegraphics[height=0.015\textwidth]{figures/purple_triangle.pdf}) in \cite{saye2014high} for the level set function in the Slotted Circle example. Results for the solution error are shown on the left and results for volume loss are shown on the right.}
    \label{fig:slot_2d_algoim}
\end{figure}

In Figure \ref{fig:slot_2d_algoim}, we can see that the reinitialization scheme from \cite{saye2014high} leads to second-order accuracy in the $L^2$ norm, whereas the PDE-based reinitialization in \ref{eq:PDE_reinit} is first-order for the semi-Lagrangian and Characteristic bending schemes, which aligns with the results presented in \cite{saye2014high}. The interesting dynamic is that, when using algoim-based reinitialization, the volume error for the CB scheme is an order of magnitude lower than when using the PDE-based reinitialization. From these results, we can conclude that the CB method and volume-preservation in general will not correct errors resulting from reinitialization. The method, however, can see auxiliary benefits from using high-order reinitialization schemes. For the remainder of the results presented herein, we continue to use the PDE-based reinitialization scheme to compare with results from the literature. Additionally, for smooth surfaces (which are what we expect to deal with in later applications), the PDE-based reinitialization does indeed see third-order accuracy, as noted in \cite{sussman2003second}. 

\subsubsection{Deformational Flow}
The purpose of the single vortex example is to explore how the characteristic bending method behaves when subject to a strong deformational flow. This problem was first introduced in \cite{leveque1996singlevortex}\footnote{We also note that the incompressible field used in this example has been used in a variety of other studies (notably in \cite{bell1989second}). However, the example in \cite{leveque1996singlevortex} seems to be the first example of reversing the direction of the flow to test conservation and interface preservation properties.} and is a simple method for measuring the accuracy and robustness of numerical schemes for scalar advection. This example is particularly useful for measuring the distortion of an interface and the conservation properties (or lack thereof) for level set methods. As such, we use this example to highlight the benefits and limitations of the CB scheme.

We consider the advection of an interface defined by level set function
\begin{align}
    \phi(x,y,0) = \sqrt{ \left (x-0.5 \right )^2 + \left (y-0.75 \right )^2} - 0.2,
\end{align}
in the domain $\Omega = [0, 1]^2$. This level set function represents a circle of radius $R=0.2$ centered at $(0.5, 0.75)$ and we advect this interface using the incompressible velocity field,
\begin{align}
    \mathbf{u} = 
    \begin{cases}
        -\sin(2 \pi x)^2 \; \sin(2 \pi y) \\
         \quad \sin(2 \pi x) \; \; \sin(2 \pi y)^2,
    \end{cases}
\end{align}
from time $t=0$ until $t=\pi$. At time $t=\pi$, we reverse the direction of the velocity field by flipping the sign of each component and continue advecting the profile until time $t=2\pi$. As this process is reversible, the initial profile should be recovered exactly at time $t=2\pi$ and we can measure the accuracy and conservation properties of our numerical method by measuring the error between the initial profile and computed result.

Similar to the previous two examples, we consider including a compressible component to velocity field. However, we note that sign of the compressible component does not change at time $t=\pi$ as the reversal of this component could potentially negate the effects of the adding artificial expansion. Additionally, we center the compressible component at $(0.5,0.5)$, leading to the velocity field
\begin{align}
    u(x, y, t) =
    \begin{cases}
        \begin{aligned}
            &-\sin(2 \pi \; x)^2 \sin(2 \pi \; y) \; \; + \alpha \Delta x_{min} \; (x-0.5) + \; \beta \left ( \Delta x_{min} \right )^2 \; (x-0.5)\\
            & \quad \sin(2 \pi \; x) \; \; \sin(2 \pi \; y)^2 + \alpha \Delta x_{min} \; (y-0.5) + \; \beta \left ( \Delta x_{min} \right )^2 \; (y-0.5)
        \end{aligned} & \text{for } t \in [0, \pi] \\
        \\
        \begin{aligned}
            & \quad \sin(2 \pi \; x)^2 \sin(2 \pi \; y) \; \; + \alpha \Delta x_{min} \; (x-0.5) + \; \beta \left ( \Delta x_{min} \right )^2 \; (x-0.5) \\
            & - \sin(2 \pi \; x) \; \; \sin(2 \pi \; y)^2 + \alpha \Delta x_{min} \; (y-0.5) + \; \beta \left ( \Delta x_{min} \right )^2 \; (y-0.5)
        \end{aligned} & \text{for } t \in (\pi, 2\pi].
    \end{cases}
\end{align}

This velocity field, with $\alpha = \beta = 0$, is entirely divergence-free and, analytically, the characteristic trajectories do not cross. This may lead one to believe that the CLSRM and VPRM methods would not require restarting. However, in practice, and as noted in \cite{bellotti2019rm} and \cite{theillard2021vprm}, the bijectivity of the map can be lost due to numerical errors for this deforming flow and the reference map requires frequent restarting. Following the previous example, we reinitialize the level set function every time we restart the reference map. The SL and the CB schemes reinitialize the level set function at each iteration. 

As in the previous example, we use an adaptive quadtree mesh for this example and use a uniform band equal to $8 \; \Delta x_{min}$. We also use the $L^2$ interface error defined in \eqref{eq:interface_L2} and the volume loss defined in \eqref{eq:interface_ML}. Additionally, we integrate a clipping scheme at the domain boundary and no longer allow for extrapolation outside of the domain.  

\begin{figure}[!htb]
    \centering
    \makebox[\linewidth][c]
    {
        \includegraphics[width=0.32\linewidth]{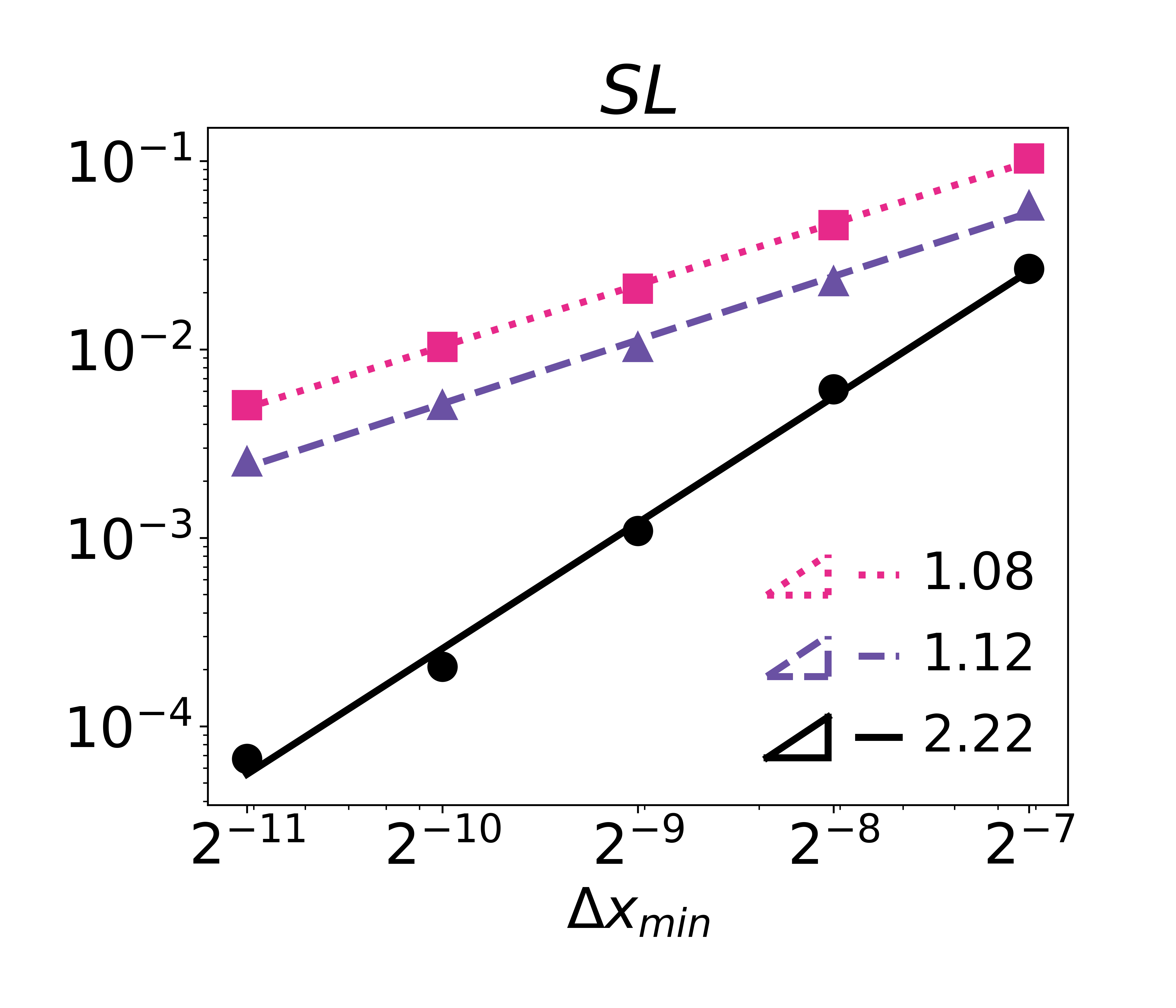}
        \hspace{0.03\linewidth}
        \includegraphics[width=0.32\linewidth]{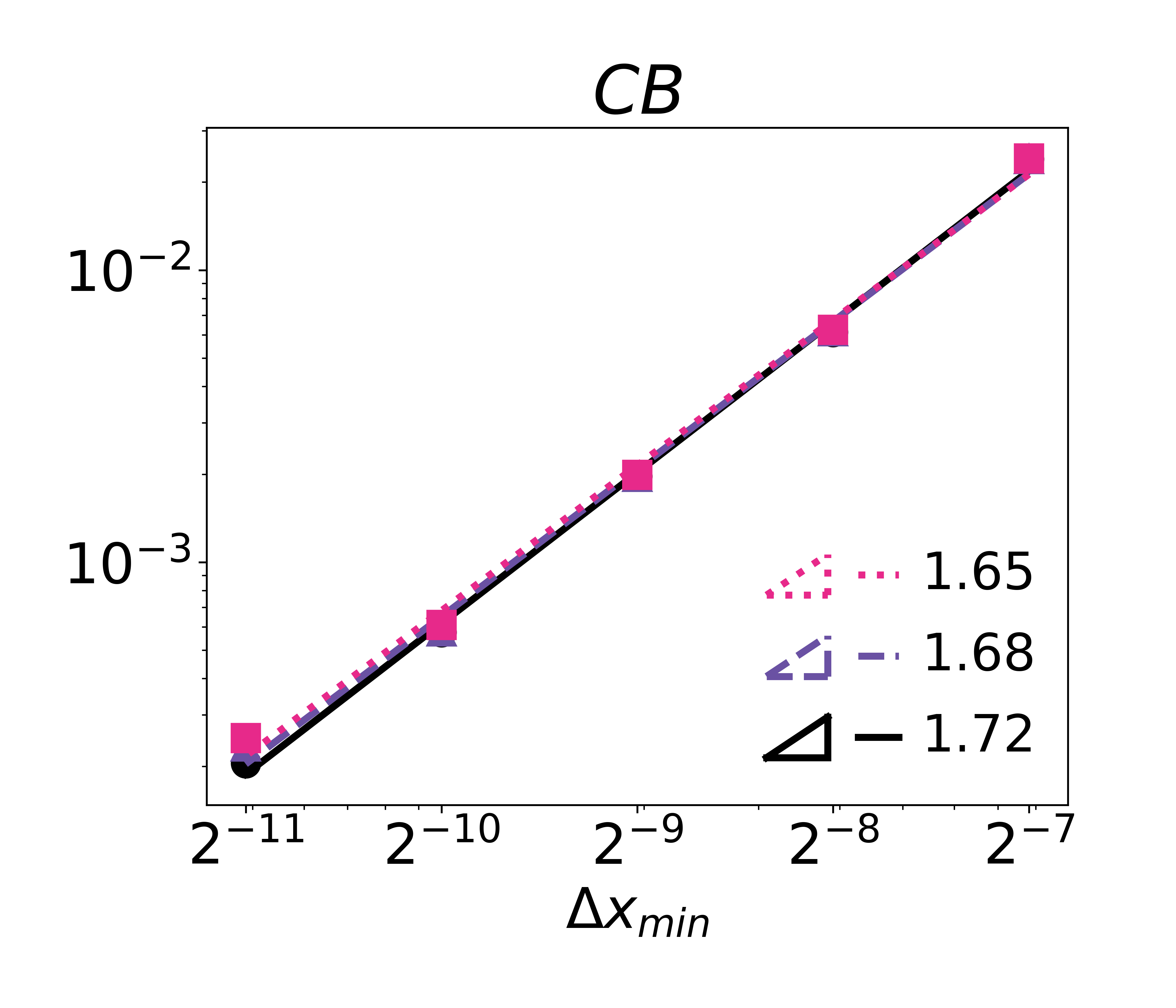}
    }
    \\[1em]
    \makebox[\linewidth][c]
    {
        \includegraphics[width=0.32\linewidth]{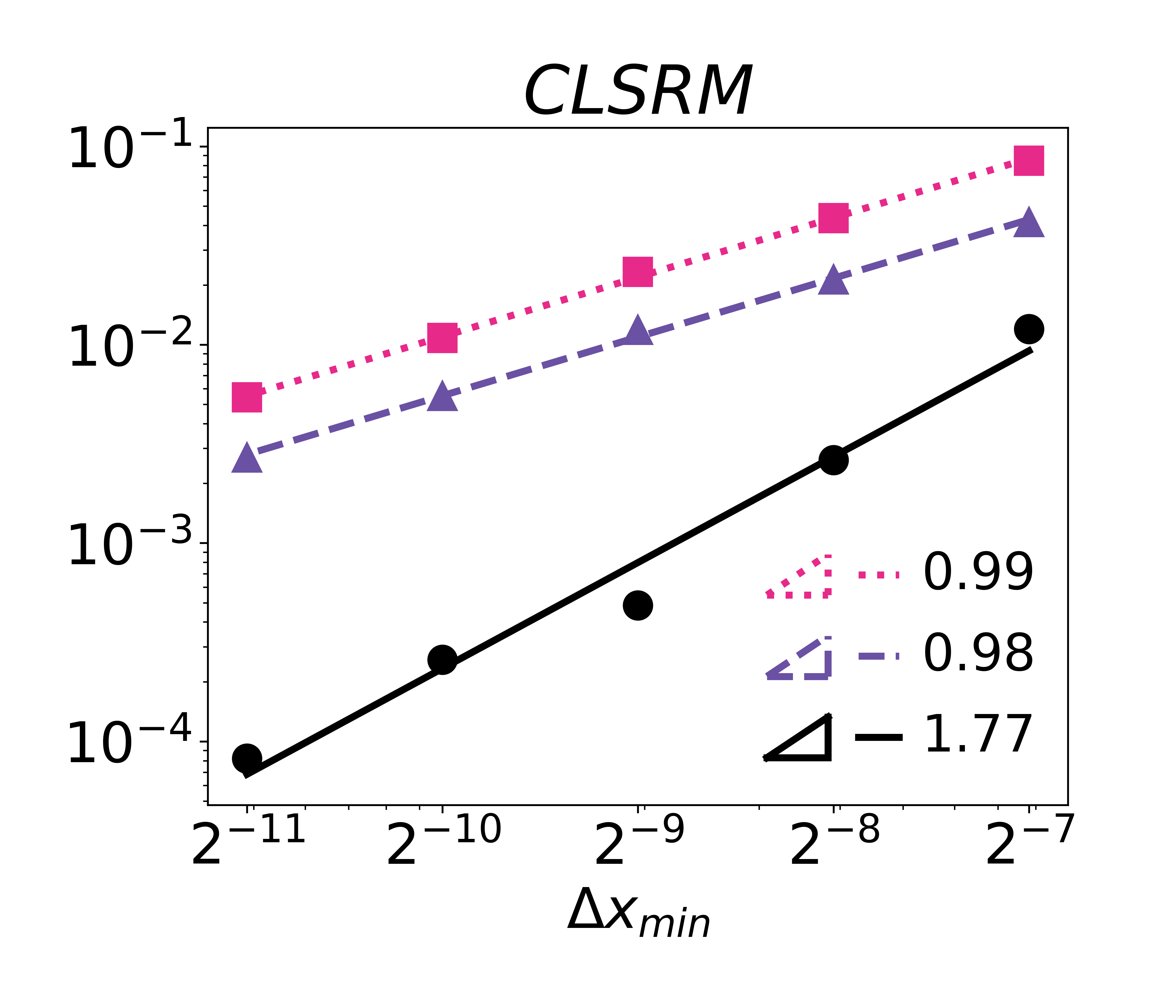}
        \hspace{0.03\linewidth}
        \includegraphics[width=0.32\linewidth]{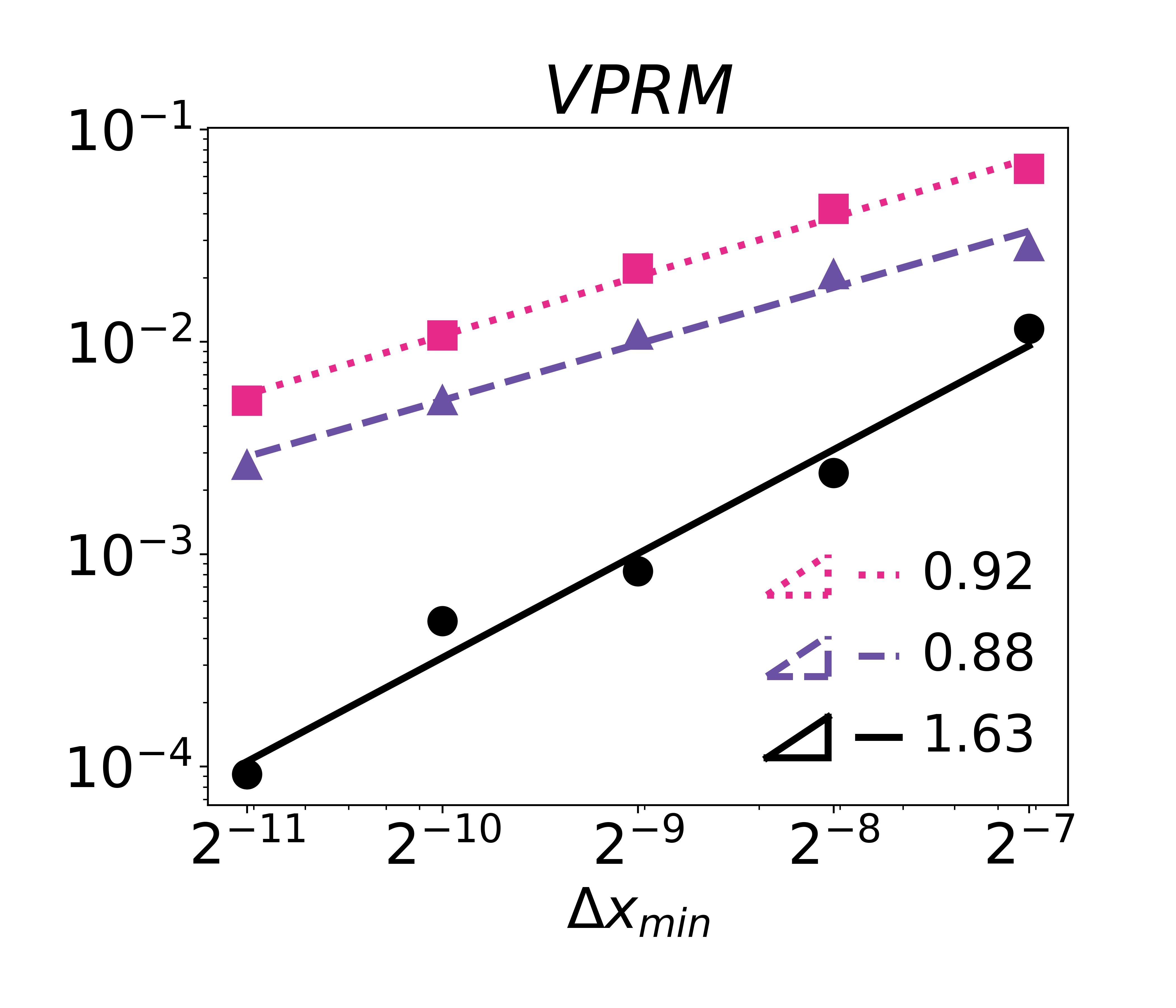}
        \hspace{0.03\linewidth}
        \includegraphics[width=0.32\linewidth]{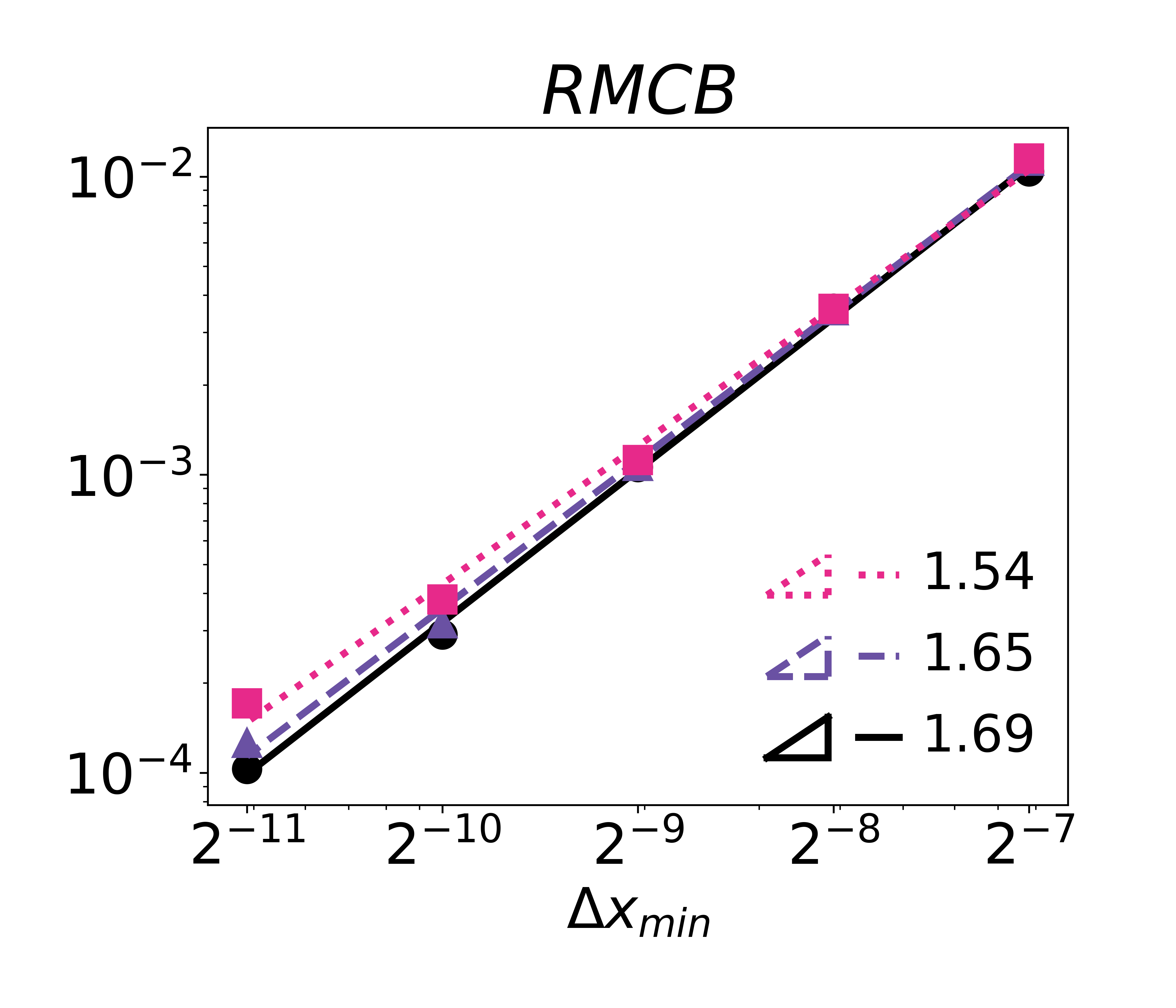}
    }
    \caption{Solution error $\|\phi-\phi_e\|_2$ for the Single Vortex example with various levels of artificial expansion added to the velocity field. \includegraphics[height=0.015\textwidth]{figures/circle.pdf} denotes the completely incompressible field, \includegraphics[height=0.015\textwidth]{figures/purple_triangle.pdf} denotes the second-order expansion added to the incompressible velocity (\textit{i.e.} $\alpha=0, \; \beta=1$), and \includegraphics[height=0.015\textwidth]{figures/square_purple.pdf} denotes the first-order expansion added to the analytic velocity field (\textit{i.e.} $\alpha=1, \; \beta=0$).}
    \label{fig:vortex_2d_intr}
\end{figure}

\begin{figure}[!htb]
    \centering
    \makebox[\linewidth][c]
    {
        \includegraphics[width=0.32\linewidth]{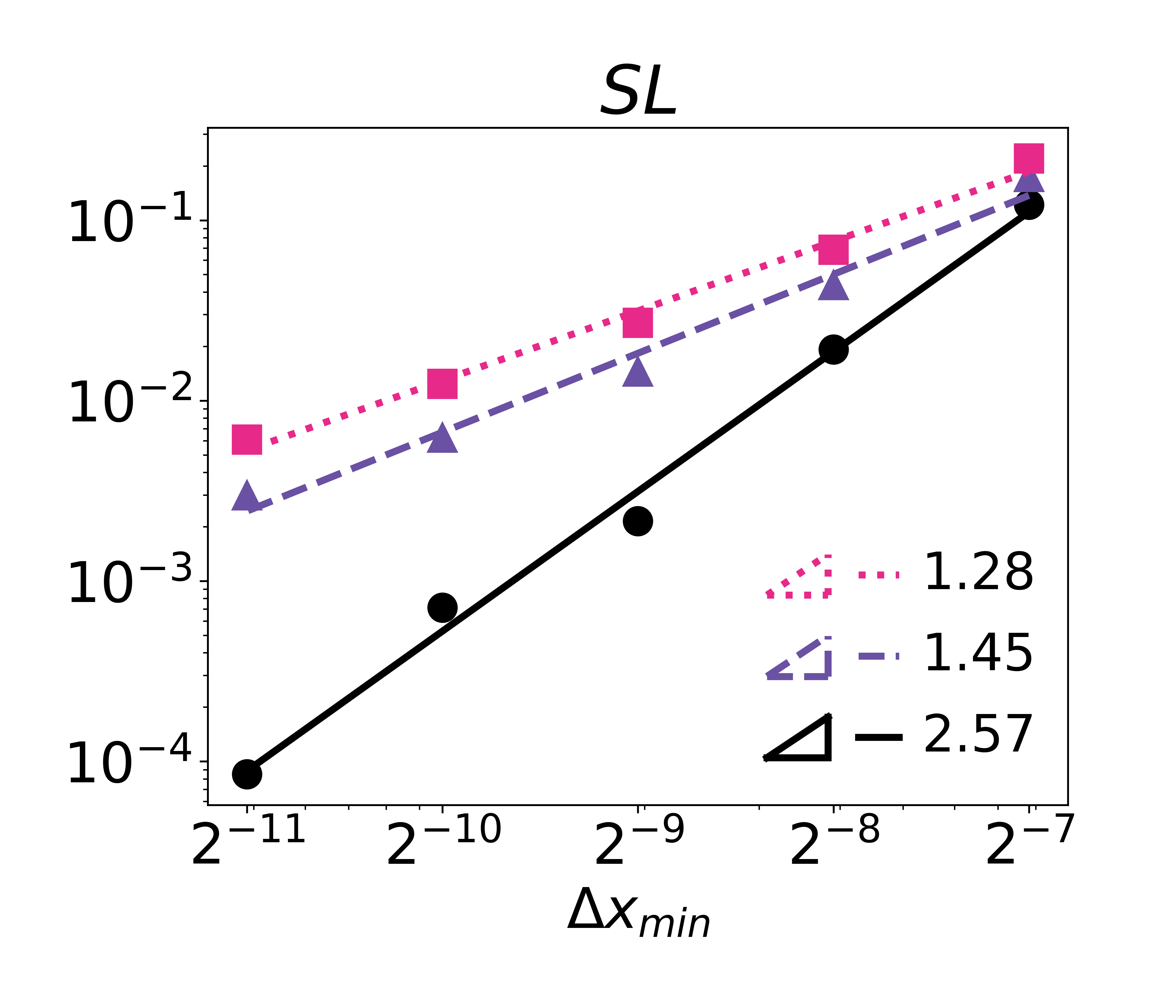}
        \hspace{0.03\linewidth}
        \includegraphics[width=0.32\linewidth]{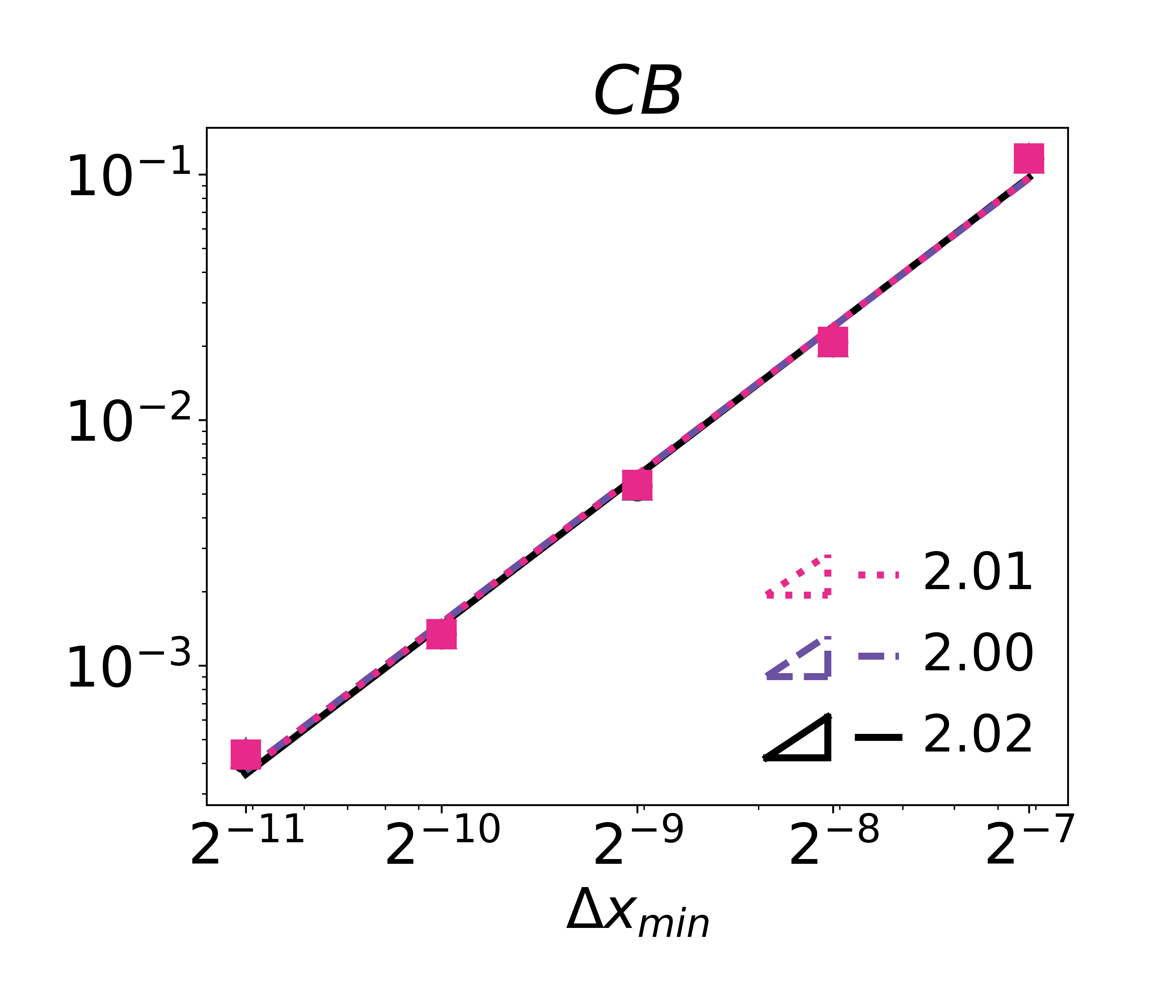}
    }
    \\[1em]
    \makebox[\linewidth][c]
    {
        \includegraphics[width=0.32\linewidth]{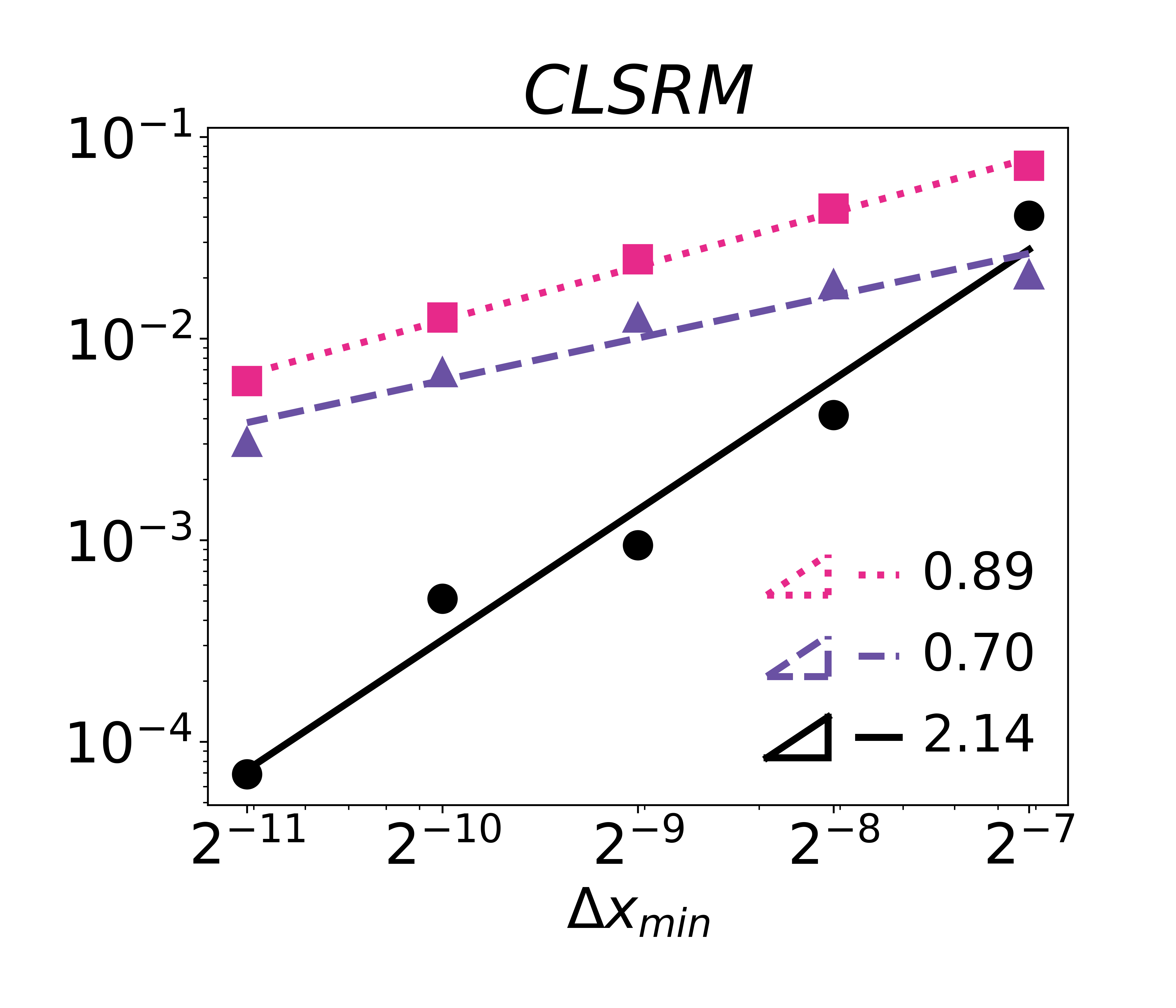}
        \hspace{0.03\linewidth}
        \includegraphics[width=0.32\linewidth]{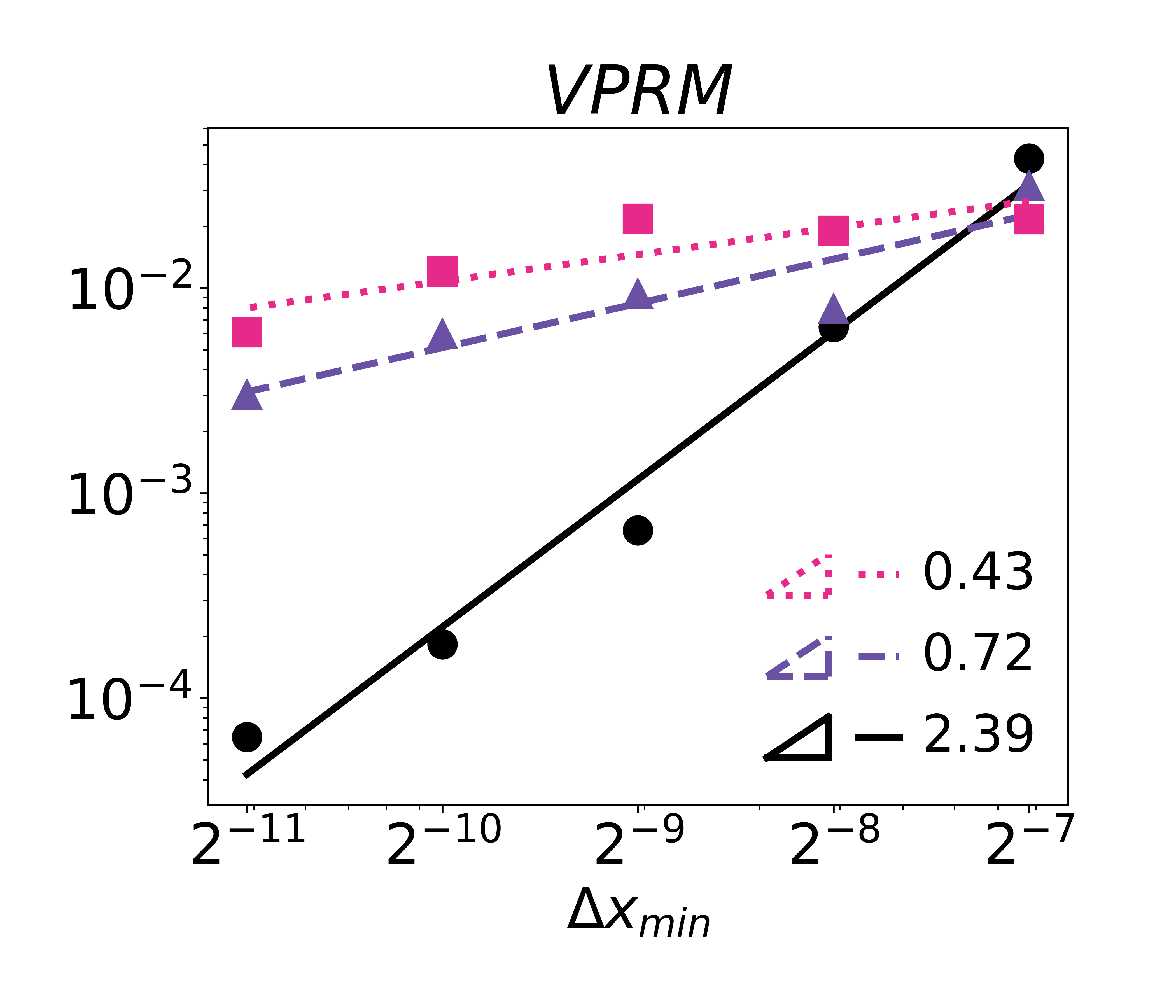}
        \hspace{0.03\linewidth}
        \includegraphics[width=0.32\linewidth]{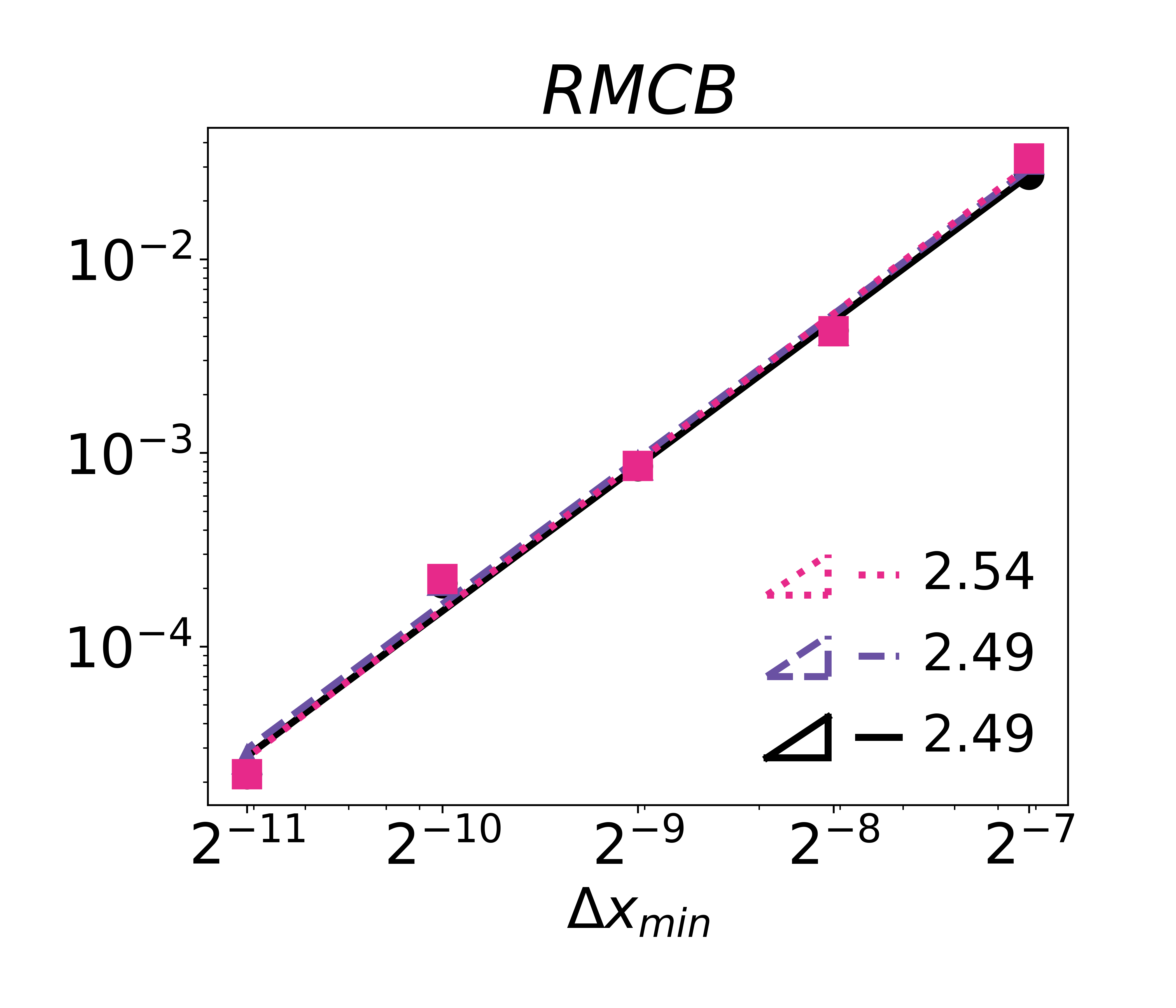}
    }
    \caption{Volume loss for the Single Vortex example with various level of artificial expansion added to the velocity field. \includegraphics[height=0.015\textwidth]{figures/circle.pdf} denotes the completely incompressible field, \includegraphics[height=0.015\textwidth]{figures/purple_triangle.pdf} denotes the second-order expansion added to the incompressible velocity (\textit{i.e.} $\alpha=0, \; \beta=1$), and \includegraphics[height=0.015\textwidth]{figures/square_purple.pdf} denotes the first-order expansion added to the analytic velocity field (\textit{i.e.} $\alpha=1, \; \beta=0$).}
    \label{fig:vortex_2d_vol}
\end{figure}

The results for the SL, CLSRM, VPRM, CB, and RMCB methods using a $\text{CFL}=5$ are shown in Figures \ref{fig:vortex_2d_intr} and \ref{fig:vortex_2d_vol}. As expected, we see between first- and second-order accuracy for the interface error for all of the methods tested. Furthermore, we see second-order convergence for the volume loss with each method. When we add artificial expansion to the velocity field, we observe the same trend noted in the previous two examples. The SL, CLSRM, and VPRM schemes all yield accuracy between first- and second-order, whereas the CB and RMCB schemes produce results nearly identical to the purely incompressible case.

\begin{figure}[!tb]
    \centering
    \includegraphics[width=0.75\linewidth]{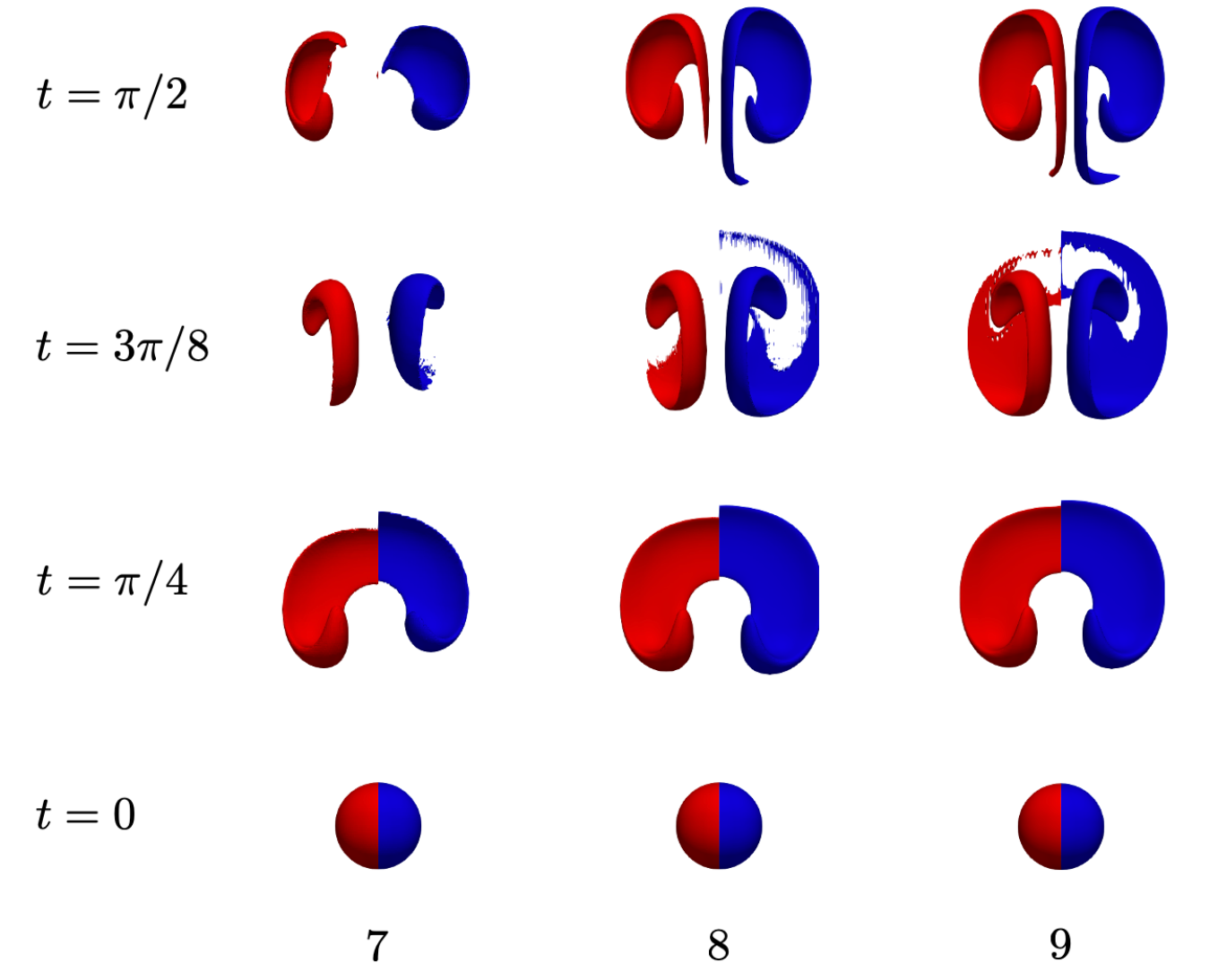}
    \caption{Evolution of the interface for a three-dimensional deformational flow proposed by Enright \textit{et al.} \cite{enright2005fast} using a first-order artificial compression added to the incompressible field (\textit{i.e.} $\alpha = -1, \; \beta = 0$). The SL (red) and CB (blue) methods are shown using at four time steps with increasing maximum levels of refinement. Here, we see visual improvements with the CB method with the coarse grid that diminish (visually) as the grid is refined.}
    \label{fig:single_vortex_3d_sl_v_cb}
\end{figure}

\begin{figure}[!tb]
    \centering
    \includegraphics[width=0.49\linewidth]{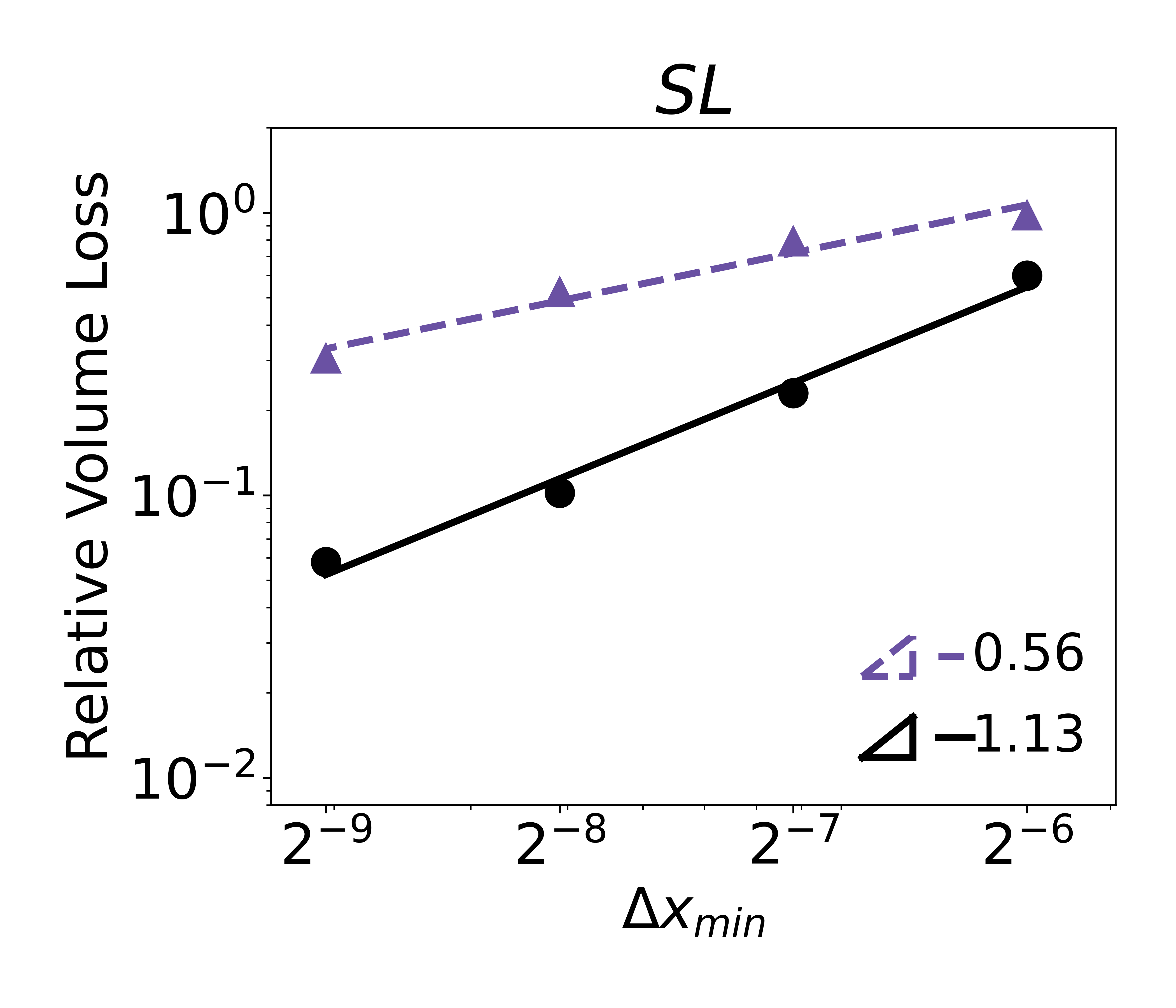}
    \hfill
    \includegraphics[width=0.49\linewidth]{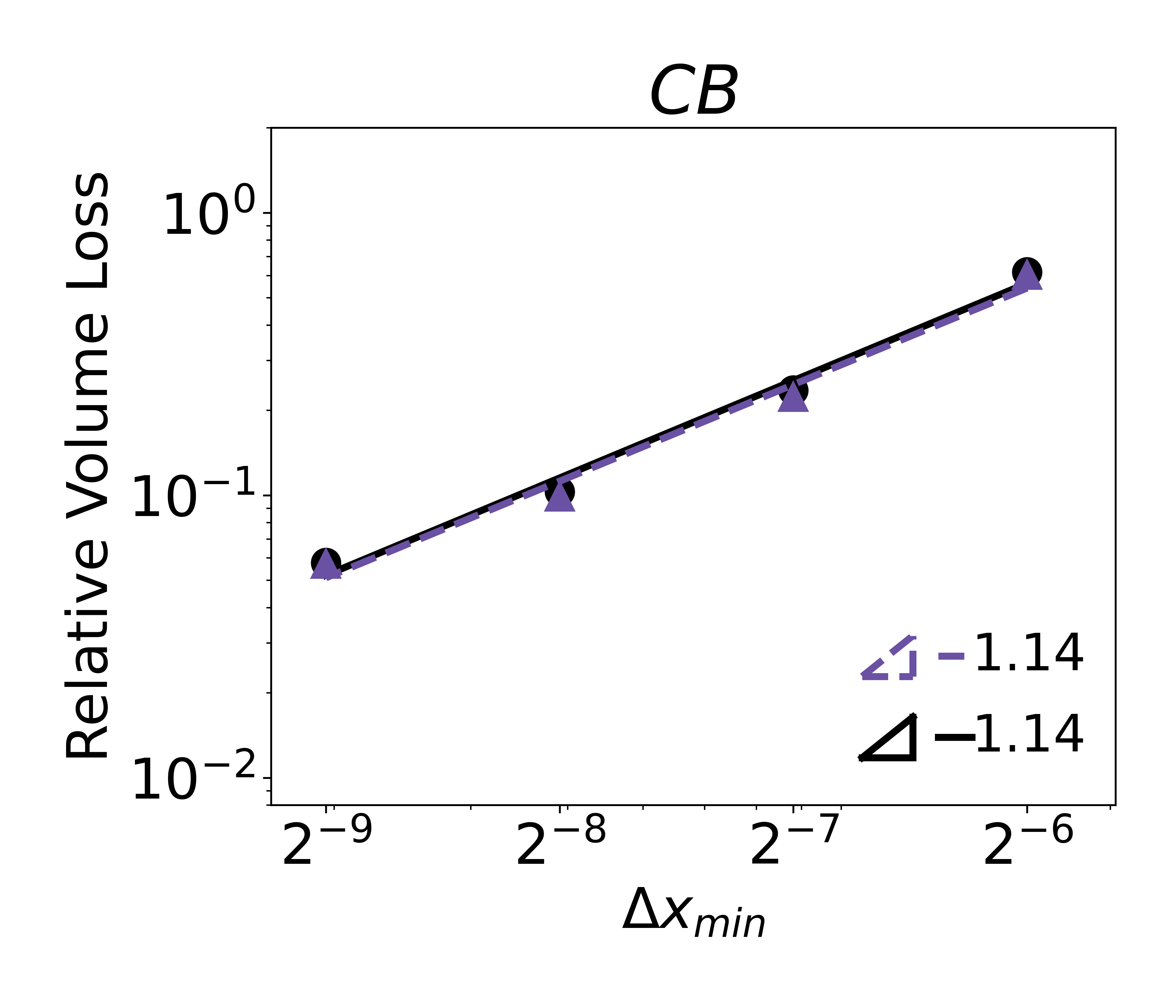}
    \caption{Volume loss for the three-dimensional deformational flow proposed by Enright \textit{et al.} \cite{enright2005fast} using various levels of artificial expansion added to the velocity field. \includegraphics[height=0.015\textwidth]{figures/circle.pdf} denotes the completely incompressible field and \includegraphics[height=0.015\textwidth]{figures/purple_triangle.pdf} denotes a first-order compression added to the incompressible velocity (\textit{i.e.} $\alpha=1, \; \beta=0$).}
    \label{fig:single_vortex_3d_sl_v_cb_vol}
\end{figure}

Figure~\ref{fig:single_vortex_3d_sl_v_cb} provides a qualitative comparison of the SL and CB methods using a three-dimensional example from \cite{bellotti2019rm}, originally proposed by Enright \textit{et al.} \cite{enright2005fast}\footnote{The reference map based methods (CLSRM, VPRM, and RMCB) exhibit minimal volume loss for this example and are extensively compared in \cite{bellotti2019rm} and \cite{theillard2021vprm}.}. Here, we advect the flow forward in time until the level set function loses a significant amount of mass, rather than reversing the velocity field as in the two-dimensional case. This visualization highlights the ability of the CB method to preserve the topology of the level set surface, particularly on coarse grids where the impact of the artificial compression\footnote{For this example, adding artificial expansion increases the mass of the level set function. So we have used compression to highlight the differences more clearly.} is more pronounced. In Figure~\ref{fig:single_vortex_3d_sl_v_cb_vol}, we show the corresponding loss in volume at the final time for both methods using $\alpha = 0, \text{ and } -1$. In contrast to the SL method, the CB scheme remains largely invariant with respect to the artificial expansion, demonstrating its robustness. Note that because the breakup of the interface produces a non-smooth problem, we do not expect second-order convergence. Nevertheless, these results clearly illustrate the advantages of the CB method in this three-dimensional, deformable flow.

\subsubsection{Incompressible Euler} \label{sec:inc_Euler}
%
For the final advection example, we consider a simple example of nonlinear advection in the form of the incompressible Euler equations,
\begin{align}
    \frac{\partial \mathbf{u}}{\partial t} + \mathbf{u} \cdot \nabla \mathbf{u} = -\nabla p. \label{eq:inc_Euler}
\end{align}
Specifically, we construct an analytic solution to \eqref{eq:inc_Euler} using the velocity field,
\begin{align}\label{eq:u_Euler}
    u(x, y, t) =
    \begin{cases}
        \begin{aligned}
            & \quad \sin(x) \cos(y) \\
            & -\cos(x) \sin(y)
        \end{aligned}
    \end{cases}
\end{align} 
and compute the corresponding pressure gradient, $\nabla p$. This yields a stationary problem, where the exact solution is simply the initial velocity field for all time. Numerically, we solve this system using the standard Lformulation by integrating the right-hand side, $-\nabla p$, along characteristic curves in the Lagrangian frame, following \cite{boyd2001, karniadakis2005spectral}.

The goal of this test is twofold. First, it provides a manufactured nonlinear problem in which to compare the B scheme with the standard SL method. Second, it highlights how the CB scheme behaves in a Navier–Stokes setting. This problem is set up as the advection step in a projection method and by iterating the advection step, we can observe how error accumulates and how the divergence of the velocity field is affected.

\begin{figure}[!htb]
    \centering
    \includegraphics[width=0.45\linewidth]{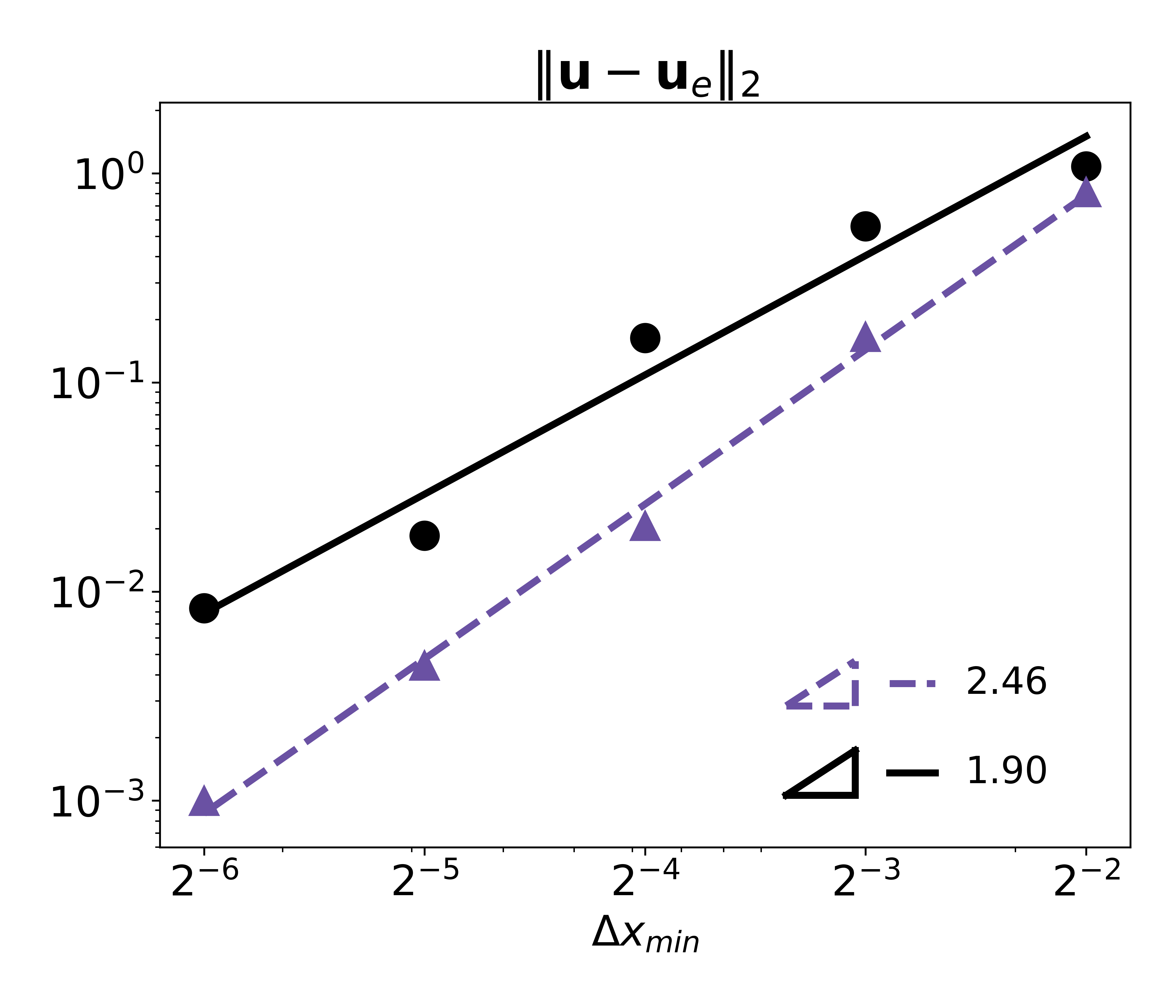}
    \hspace{0.03\linewidth}
    \includegraphics[width=0.45\linewidth]{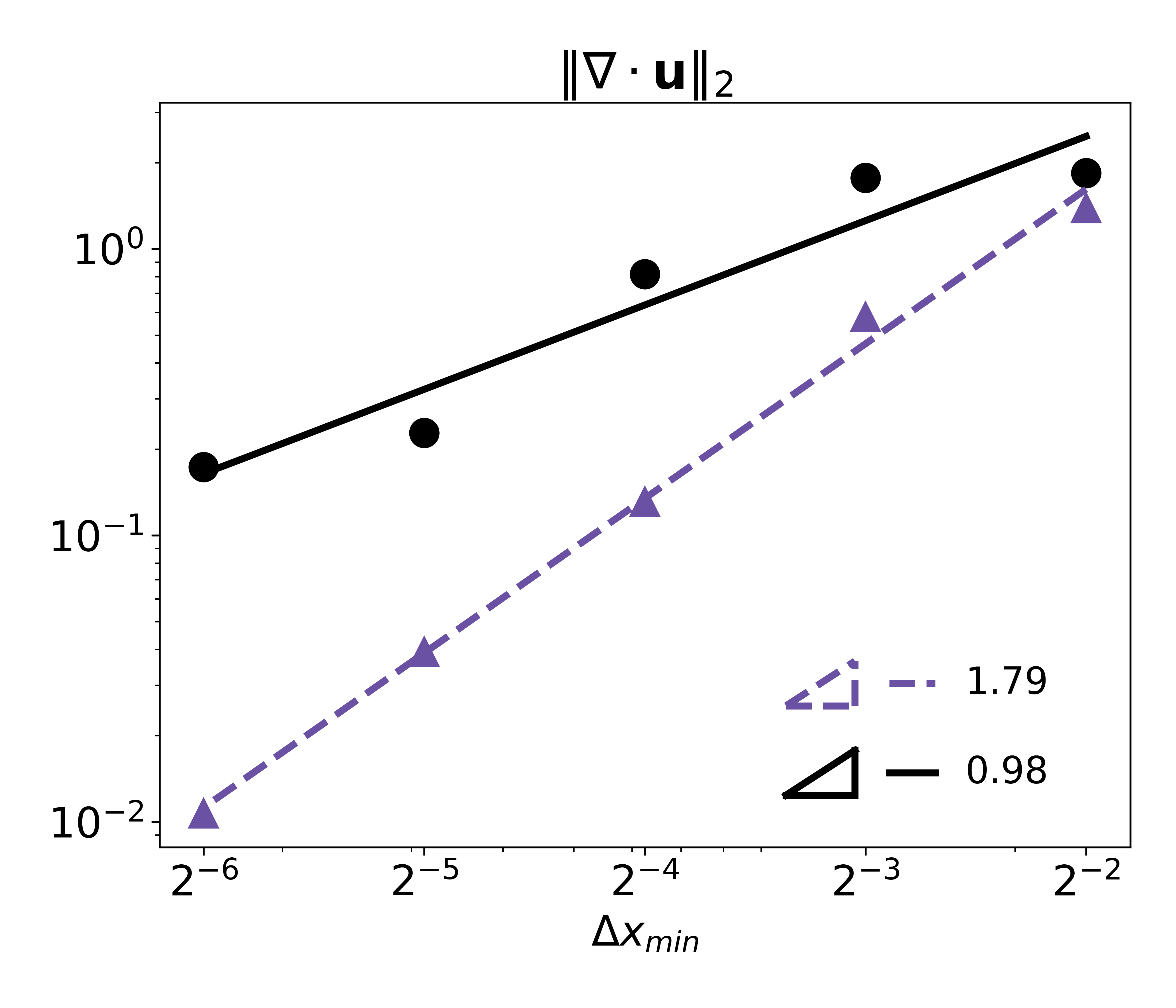}
    \caption{Comparison of the SL (\includegraphics[height=0.015\textwidth]{figures/circle.pdf}) and CB (\includegraphics[height=0.015\textwidth]{figures/purple_triangle.pdf}) methods for the incompressible Euler example. Velocity error is shown on the left and the divergence is shown on the right.}
    \label{fig:euler_2d}
\end{figure}

We run this problem on a uniform grid with refinement levels ranging from 3 to 7, and use CFL-based time stepping with $CFL = 3$. Recall that the semi-Lagrangian method is unconditionally stable for linear advection problems. However, this property does not automatically extend to nonlinear problems (see \cite{falcone1998slconvergence}) and, for this problem, large CFL numbers may cause the solution to blow up. Empirically, we found that both SL and CB remain stable with $CFL=3$ when combined with an RK2 backward time integrator and WENO-based quadratic interpolation.

The results are summarized in Figure \ref{fig:euler_2d} using a $CFL=3$. We report errors at a final time $t=8\pi$, corresponding to the point at which oscillations in the velocity error and divergence are stabilized. The SL method shows between first- and second-order convergence for the velocity field, while no consistent convergence is observed in the divergence. By contrast, the CB method exhibits clear second-order convergence in the velocity and nearly second-order convergence for the divergence. Notably, the CB method preserves incompressibility, which is consistent with its underlying volume-preserving formulation. 

We also observed during testing that the CB method remains stable with higher-order interpolation, while the SL scheme quickly becomes unstable and diverges from the exact solution. This effect is most pronounced with non-symmetric stencils (e.g. Taylor polynomials rather than bicubic or tricubic interpolation, as implemented in \cite{saye2014high}). In these cases, oscillations introduced by high-order interpolation appear to be smoothed by the volume-preserving correction, which aligns with the commentary in \cite{theillard2021vprm}. For comparability with prior studies, we restrict results here to second-order interpolation, but note this as a promising direction for future analysis.

Finally, we acknowledge that this test somewhat favors the CB method. Unlike the SL method, which has no mechanism to enforce incompressibility, the CB method effectively performs a divergence-free projection in the Lagrangian frame. That said, there are extensions of the SL method that address this limitation (e.g., conservative formulations such as \cite{lauritzen2010cslam, zerroukat2002slice}). Another possible solution is to reformulate the SL method in terms of the reference map framework and then use a Piola transform (see \textit{e.g.}, \cite{boffi2013mixed}) to ensure that the transformation preserves divergence-free fields. Nevertheless, the results clearly illustrate the benefit of the characteristic bending method for maintaining a divergence-free velocity field.

\subsection{Incompressible Multiphase Navier-Stokes} \label{sec:multiphase}
%
In this section, we test the characteristic bending scheme in the context of two-phase, incompressible flows. Specifically, we compare the CB implementation against the SL method in an analytic setting and then in a practical example of rising bubbles. We explain where the CB method can be incorporated into a multiphase projection method and show how these changes impact both the solution and the robustness of the modified solver. 

\subsubsection{Governing Equations}
The fluid velocity and pressure are modeled by the incompressible Navier-Stokes equations\footnote{Note, we have chosen to omit the $\pm$ on the fluid variables for the clarity of presentation.},
\begin{align}
    \begin{rcases}
        \begin{aligned}
            \rho_i \left (\frac{\partial \mathbf{u}}{\partial t} + \mathbf{u} \cdot \nabla \mathbf{u} \right) & = -\nabla p + \mu_i \Delta \mathbf{u} + \mathbf{f} \quad \\
            \nabla \cdot \mathbf{u} & = 0
        \end{aligned}
    \end{rcases}
    \text{ in } \Omega_i \text{,} \label{eq:momentum}
\end{align}
subject to the interfacial jump conditions,
\begin{align}
    \begin{rcases}
        \begin{aligned}
            [ \mathbf{u} ] &= 0  \quad \\
            [ \sigma \cdot \mathbf{n} ] &= \gamma \kappa \mathbf{n} + \mathbf{q} \quad
        \end{aligned}
    \end{rcases}
    \text{ on } \Gamma \text{,} \label{eq:velocity_jump}
\end{align}
where,
\begin{align}
    \sigma = -p \ \mathbf{I} + \mu \left (\nabla \mathbf{u} + \nabla \mathbf{u}^T \right )
\end{align}
is the stress tensor, $\kappa$ is the curvature of the interface, and $\gamma$ is the surface tension. The jump operator, $[ \; \cdot \; ]$, is defined as $[\chi] = \chi^+ - \chi^-$ and represents the jump of the quantity $\chi$ across the interface $\Gamma$. We use $\mathbf{f}$ and $\mathbf{q}$ to capture any additional forcing terms acting on the fluid or at the interface, respectively. Finally, we use the standard convection where the $-$ superscript represents quantities in the interior of the interface and the $+$ superscript represents quantities in the exterior of the interface.

\subsubsection{Overview of the Navier-Stokes Solver}
%
In order to solve the two-phase, incompressible Navier-Stokes equations, we use the solver developed in \cite{binswanger2025multiphase}. This solver employs a collocated, node-based framework on adaptive quadtree (2D) and octree (3D) grids, where all the fluid variables are stored at the grid nodes. This choice of variable arrangement greatly simplifies the design of data structures and algorithms for the Navier-Stokes solver, which makes it an ideal testbed for the characteristic bending advection schemes. 

We modify this particular solver by replacing the SL method used in the SLBDF scheme with the CB method and by replacing the VPRM scheme used to advance the fluid interface with the RMCB scheme. The SLBDF scheme is used to advanced the momentum equation by discretizing the equation along the characteristics of the velocity field, treating the nonlinear advection term explicitly, and then uses a backward differentiation formula to handle the diffusive terms implicitly. The explicit treatment of the advection term is done entirely using the SL method and thus can be replaced with the CB scheme. The evolution of the interface in \cite{binswanger2025multiphase} is done using the VPRM approach for which the RMCB method can be used as in the previous advection examples.

\subsubsection{Analytic Vortex}
%
We verify the convergence for the full two-phase solver using the standard, analytic vortex problem (see \cite{theillard2019sharp, binswanger2025multiphase}). We solve the complete incompressible, two-phase Navier-Stokes equations, Eq. \eqref{eq:momentum} - \eqref{eq:velocity_jump}, given the exact analytic solution, 
\begin{align}
    \mathbf{u}^{\pm}(x,y,t) & =
    \begin{cases}
        \quad \sin (x) \cos (y) \cos (t) \\
        - \cos (x) \sin (y) \cos (t) 
    \end{cases} \label{eq:avort_exact_vel} \\
    p^{\pm}(x,y,t) & = \; 0 \text{,}
\end{align}
with the interface, $\Gamma$, defined by the level set function,
\begin{align}
    \phi (x,y) = 0.1 - \sin (x) \sin (y) \text{.}
\end{align}
We then set the forcing terms, $\mathbf{f}^\pm(x,y,t)$ and $\mathbf{q}(x,y,t)$, as
\begin{align}
    \mathbf{f}^\pm(x,y,t) = & \rho^\pm \left ( \frac{\partial \mathbf{u}}{\partial t} + \mathbf{u} \cdot \nabla \mathbf{u} \right ) - \mu^\pm \Delta \mathbf{u} \\
    \mathbf{q}(x,y,t) = & - \gamma \nabla \cdot \frac{\nabla \phi}{|\nabla \phi|} + \left [ \mu \nabla \mathbf{u} \cdot \frac{\nabla \phi}{|\nabla \phi|} \right ] \text{.}
\end{align}
Finally, we set the remaining parameters as,
\begin{align}
    \mu^+, \rho^+ = 1, \qquad \mu^-, \rho^- = 10, \qquad 
    \gamma = 0.1, \qquad \Omega = \left [ -\frac{\pi}{3}, \frac{4\pi}{3} \right ]^2, \qquad t_{\text{final}} = \pi .
\end{align}

We prescribe no-slip boundary conditions on the walls of the computational domain, $\Omega$, and use an adaptive quadtree mesh with a span of 4 between the minimum and maximum refinement levels. For this particular example, we refine only near the interface using the criteria specified in Section \ref{sec:amr} (\ie no vorticity-based refinement is used). 

\begin{figure}[!tb]
    \centering
    \makebox[\linewidth][c]
    {
        \includegraphics[width=0.32\linewidth]{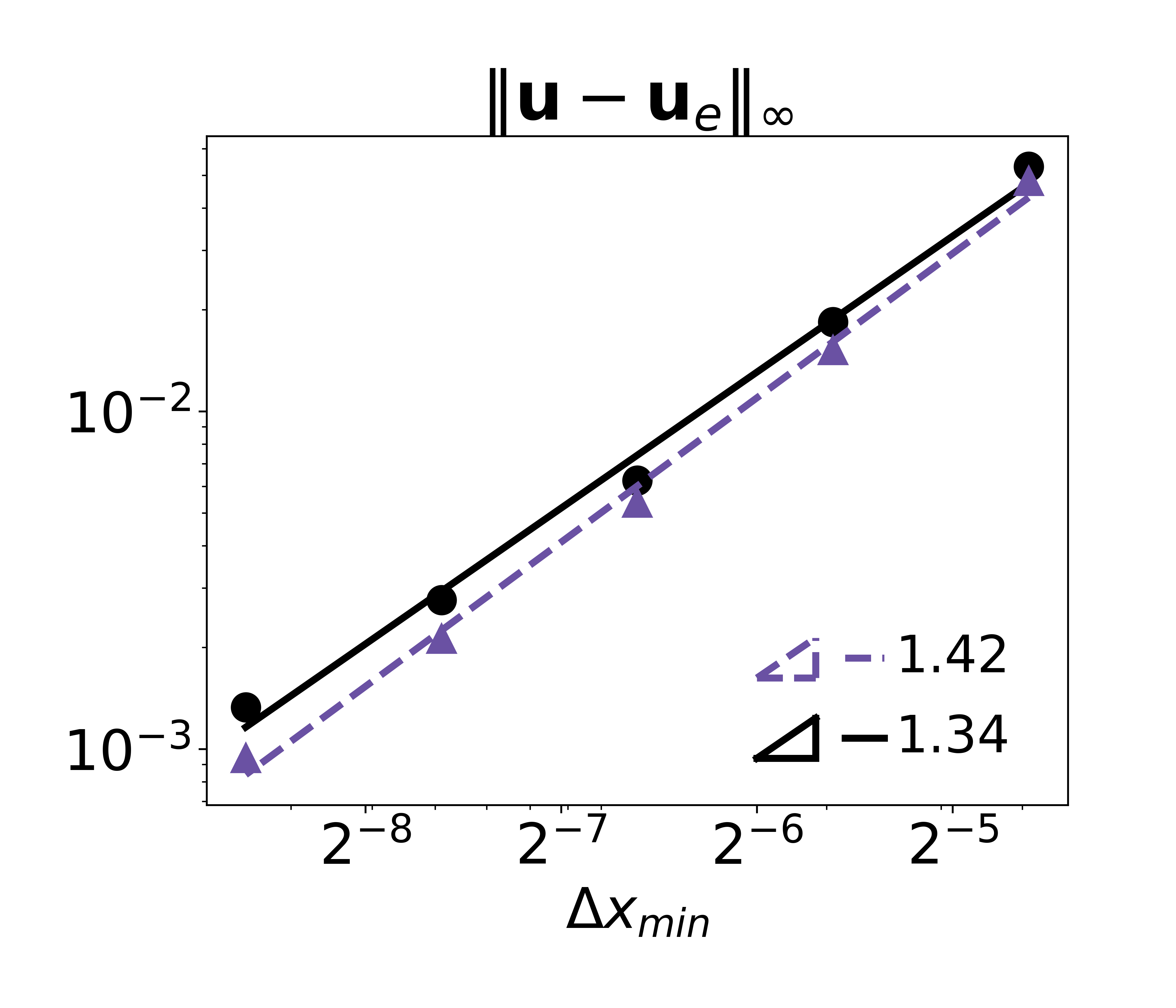}
        \hspace{0.03\linewidth}
        \includegraphics[width=0.32\linewidth]{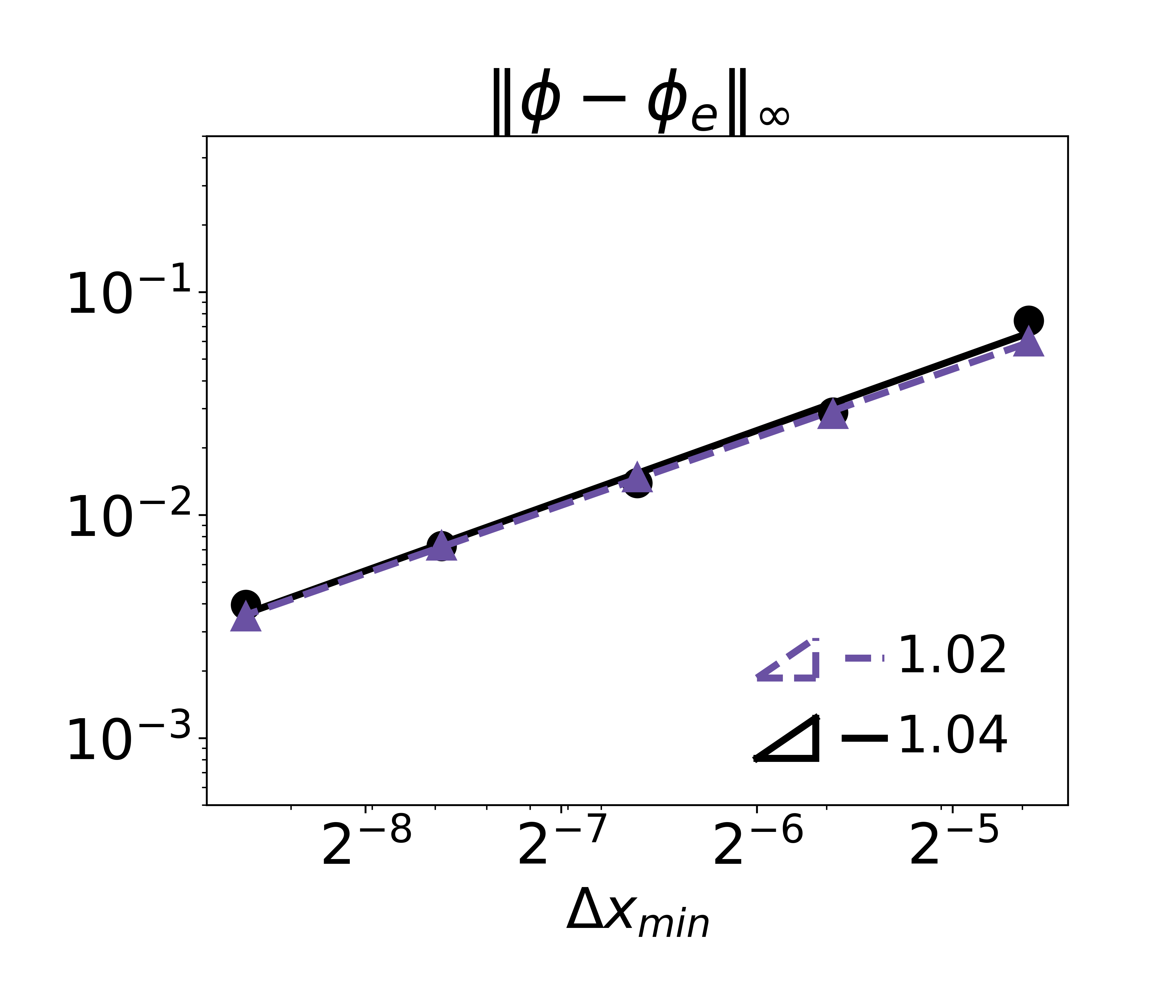}
        \hspace{0.03\linewidth}
        \includegraphics[width=0.32\linewidth]{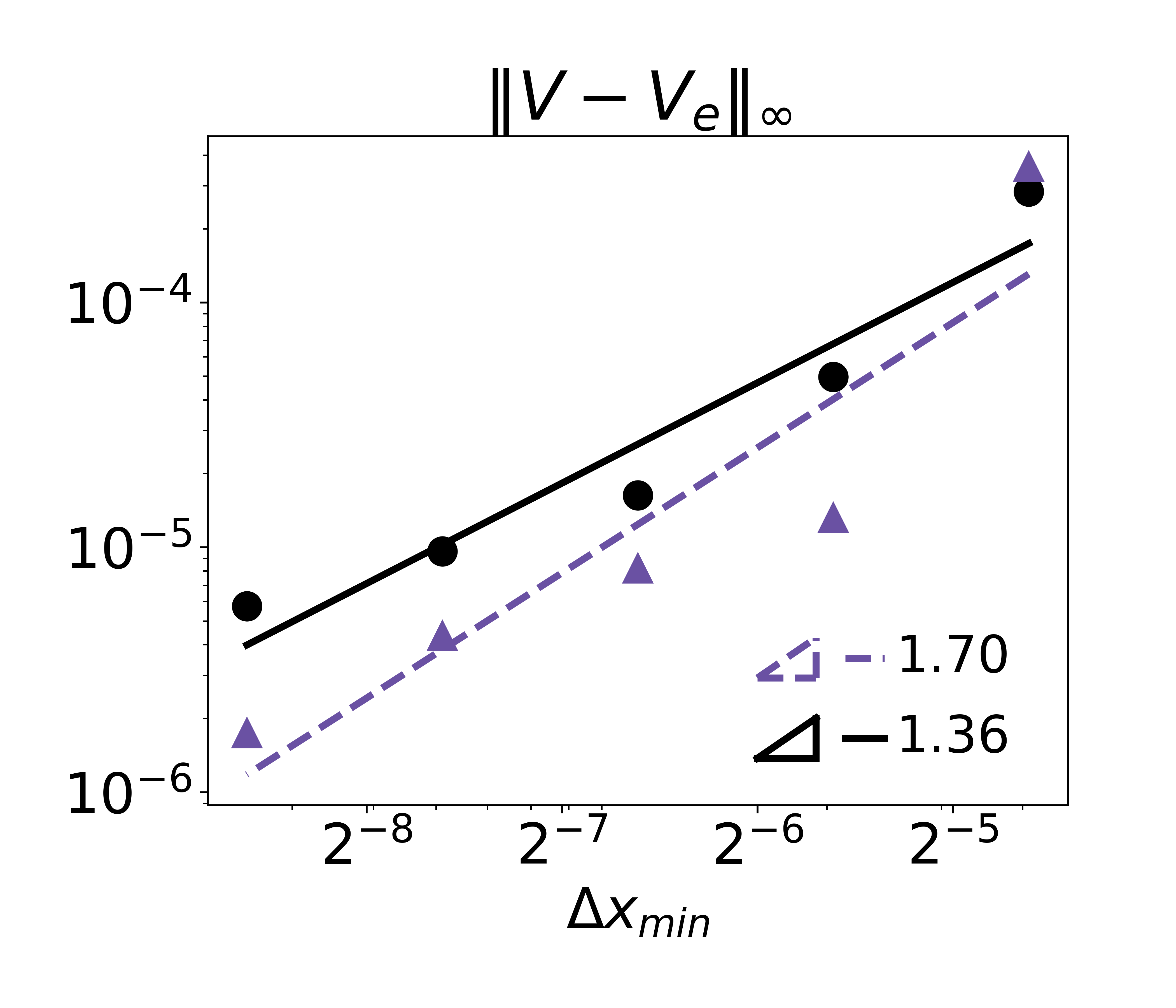}
    }
    \caption{Comparison of the SL with VPRM (\includegraphics[height=0.015\textwidth]{figures/circle.pdf}) and CB with RMCB (\includegraphics[height=0.015\textwidth]{figures/purple_triangle.pdf}) methods for the two-phase, analytic vortex example. Velocity error is shown on the left and the interface error is shown on the right. For this example, the SL method is coupled with the VPRM for the advection of the interface. For the CB scheme, the RMCB is used to advect the interface.}
    \label{fig:avortex_2d}
\end{figure}

In Figure~\ref{fig:avortex_2d}, we present the results using the exact formulation in \cite{binswanger2025multiphase} (SL/VPRM) and our modified version of the solver using the combination of CB and RMCB. We see that our modified solver yields a slight improvement in the accuracy of the solution and the interface position. This is expected as the interface for this particular example does not undergo large deformations and the velocity field should remain largely free of compressible modes. We do see an improvement in the volume loss, which is due to the subtle treatment of the correction in RMCB. However, the key importance of these results is that the usage of the CB/RMCB modification improves the solver in \cite{binswanger2025multiphase}.

\subsubsection{Rising Bubbles}
%
For the final example, we consider rising bubbles in two-dimensions. These examples are inspired by the experimental work in \cite{bhaga1981bubbles} and we consider Cases \textit{a}, \textit{e}, and \textit{f} to serve as a qualitative demonstration the differences between each of the different advection schemes presented herein (e.g., SL, RM, VPRM, CB, and RMCB). We consider an initially circular bubble suspended in another denser and more viscous fluid, where the density difference between the two fluids induces a buoyancy force causing the bubble to rise and deform. The dynamics of these rising bubbles are described by the non-dimensional Morton ($Mo$), E\"otv\"os ($Eo$), and Reynolds ($Re$) numbers, defined as
\begin{align*}
    Mo = \frac{g (\mu^{-})^4 }{\rho^{-} \gamma^{3}} \text{,} 
    \quad 
    Eo = \frac{g d^2 \rho^{-}}{\gamma} \text{,} 
    \quad 
    \text{and} 
    \quad 
    Re = \frac{\rho^- U d}{\mu^-} \text{,}
\end{align*}
where $U$ is the asymptotic rising velocity measured at the tip of the bubble, $g$ is the acceleration due to gravity, and $d$ is the initial diameter of the undeformed bubble. Cases $a$, $e$ and $f$ in \cite{bhaga1981bubbles} are defined using these dimensionless numbers and we summarize these parameters in Table \ref{table:rising_bubbles_params}. 

\begin{figure*}[!h]
\centering
\captionof{table}{Dimensionless parameters, Mo, Eo, and Re, for the Bhaga-Weber Cases $a$, $e$ and $f$.}
\begin{tabular}{|c|ccc|} 
        \hline
        Case &   $a$ &  $e$      & $f$     \\ \hline
        $Mo$ &   711 &  4.63e-3  & 8.20e-4 \\
        $Eo$ &   8.67 &  115      & 237     \\
        $Re$ &   7.80e-2 &  94       & 259     \\
        \hline
\end{tabular}
\label{table:rising_bubbles_params}
\end{figure*}

In order to simplify the presentation of our results, we set the rising velocity to $U=1$ and the initial bubble diameter to $d=4$, and then define the remaining simulation parameters as,
\begin{align*}
    \rho^- = 1 \text{,} \quad \frac{\rho^-}{\rho^+} = 10^3 \text{,} \quad \mu^- = \frac{\rho^-}{Re} \text{,} \quad \frac{\mu^-}{\mu^+} = 10^2 \text{,} \quad \gamma = \frac{(\mu^-)^2}{\rho^-}\sqrt{\frac{Eo}{Mo}} \text{,} \quad \text{and} \quad g = \frac{(Mo) \rho^- \gamma^3}{(\mu^-)^4} \text{.}
\end{align*}
We use the computational domain $\Omega = [-8, 24]^{2}$ and initialize the bubble at $\mathbf{x} = (0, 0)$. We chose the distance from the walls to minimize boundary effects and define no-slip boundary conditions at all but the top wall. On the top wall (\eg $\mathbf{x}=(x, 24$)), we impose a no-flux boundary condition. For each case, we run the simulation until the bubble reaches its asymptotic shape and rising velocity or until the interface ruptures. Similar to the previous studies, we use a minimum refinement level of 4 and adjust the maximum level of refinement to highlight differences between the advection schemes. 

\begin{figure}[!htb]
    \centering
    \includegraphics[width=0.95\linewidth]{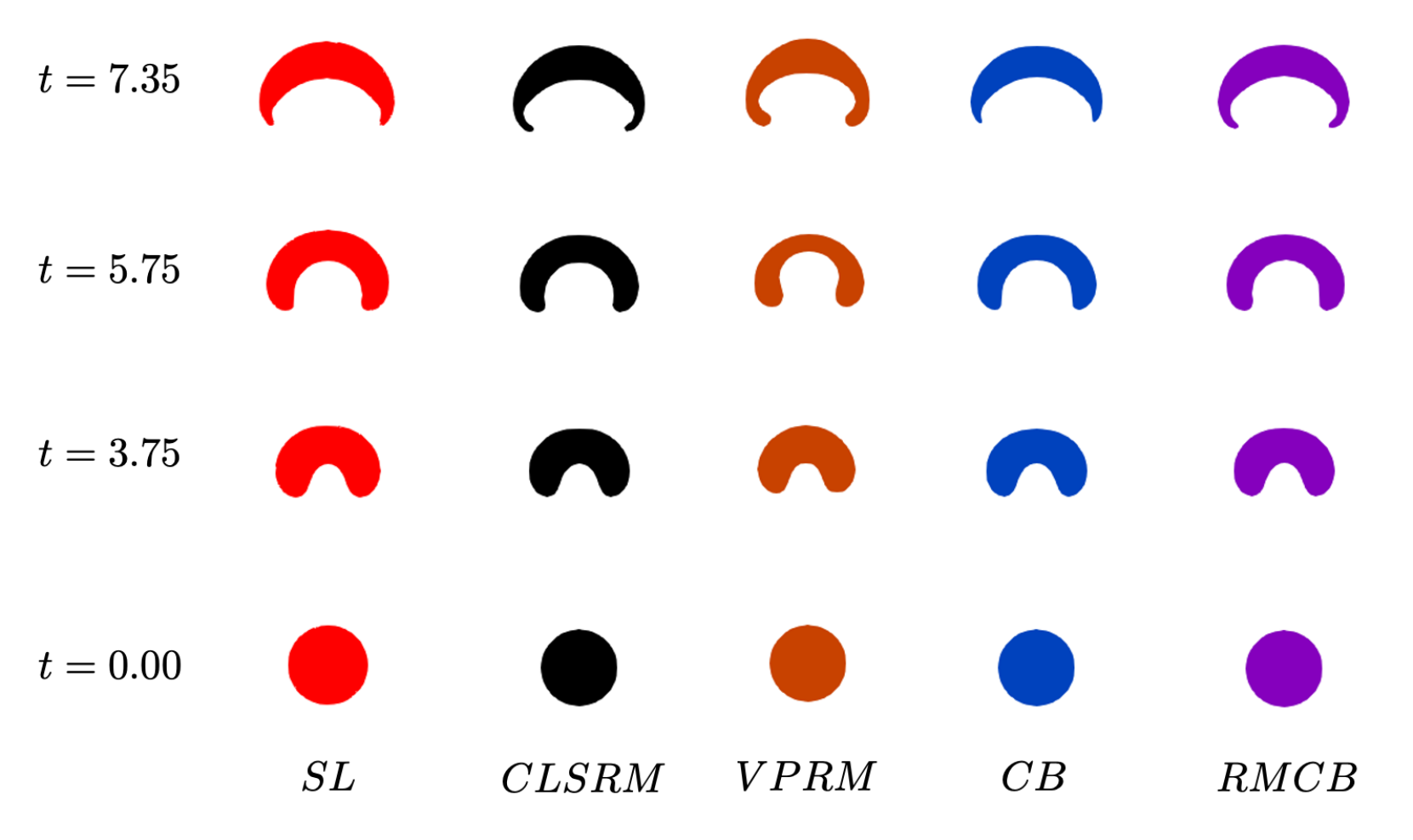}
    \caption{Evolution of a rising bubble in two-dimensions for Bhaga-Weber Case e using different advection schemes. Results shown for a minimum and maximum refinement level of 3 and 8, respectively.}
    \label{fig:rising_bubble_case_e}
\end{figure}

\begin{figure}[!tb]
    \centering
    \includegraphics[width=0.95\linewidth]{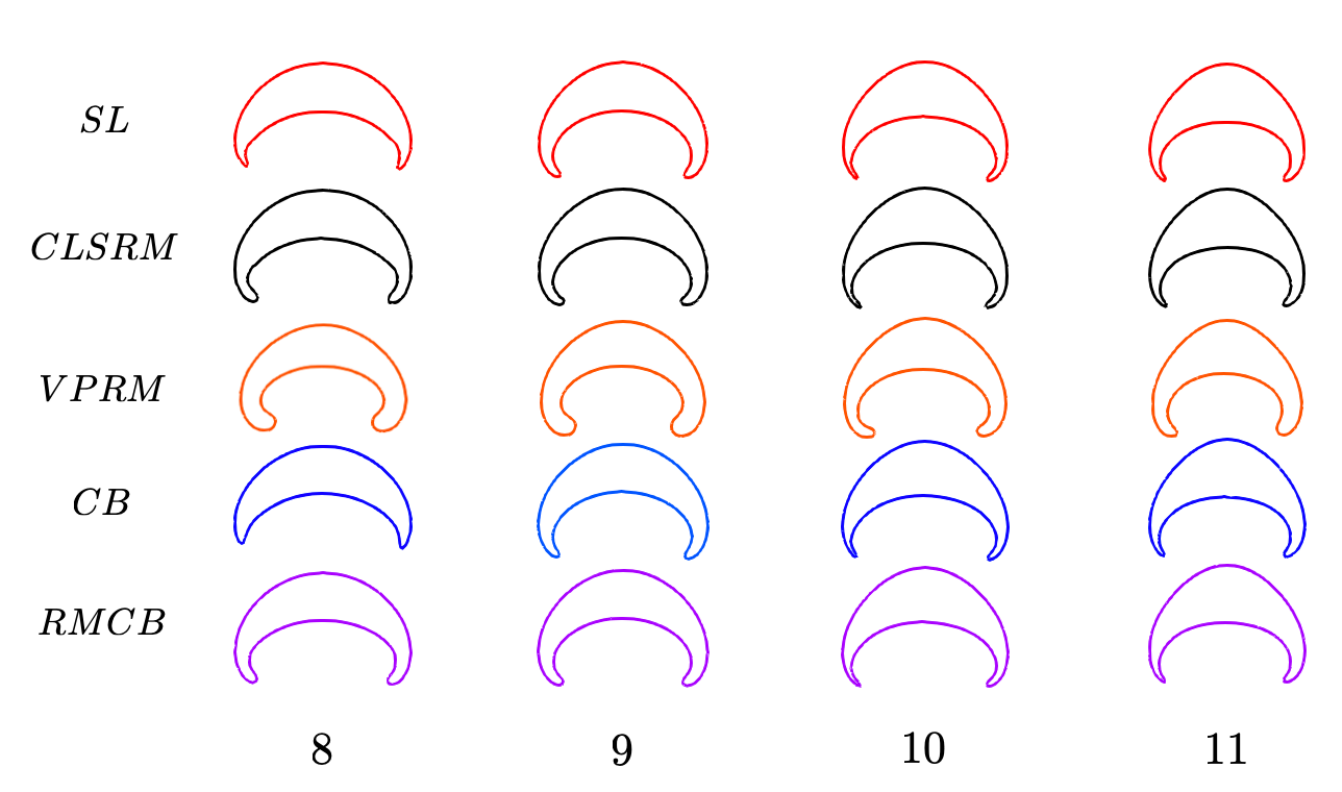}
    \caption{Convergence of the final interface profile for the Rising Bubble example using the parameters from Bhaga-Weber Case e. Interface profiles for each of the advection schemes (SL, RM, VPRM, CB, and RMCB) are shown at maximum refinement levels of 8, 9, 10, and 11. Here, we can see that interface profiles converge as the grid is refined and the difference between each of the methods is largely eliminated.}
    \label{fig:rising_bubble_case_e_interface}
\end{figure}

In Figure~\ref{fig:rising_bubble_case_e}, we present the bubble shape for Bhaga-Weber Case e using a maximum refinement level of 8. This is a relatively coarse grid (\textit{e.g.}, compared to the grids used in \cite{binswanger2025multiphase}), but this resolution highlights the major differences for each of the advection schemes. The most prominent difference for these simulations is with the skirt of the bubble (trailing edge). With the SL advection, we can see the effects of reinitialization with slightly truncated corners, similar to the effects present in the slotted circle example. This effect is slightly less pronounced with the CB method, but both the RM and RMCB schemes well-preserve the sharp corners. If we focus on the VPRM scheme, we can clearly see that the volume-preserving projection results in significant deformation of the bubble shape. We note that these discrepancies largely disappear as the refinement of the grid increases and we show how the interface profiles converge for this example in Figure~\ref{fig:rising_bubble_case_e_interface}. 

\begin{figure}[!tb]
    \centering
    \includegraphics[width=0.49\linewidth]{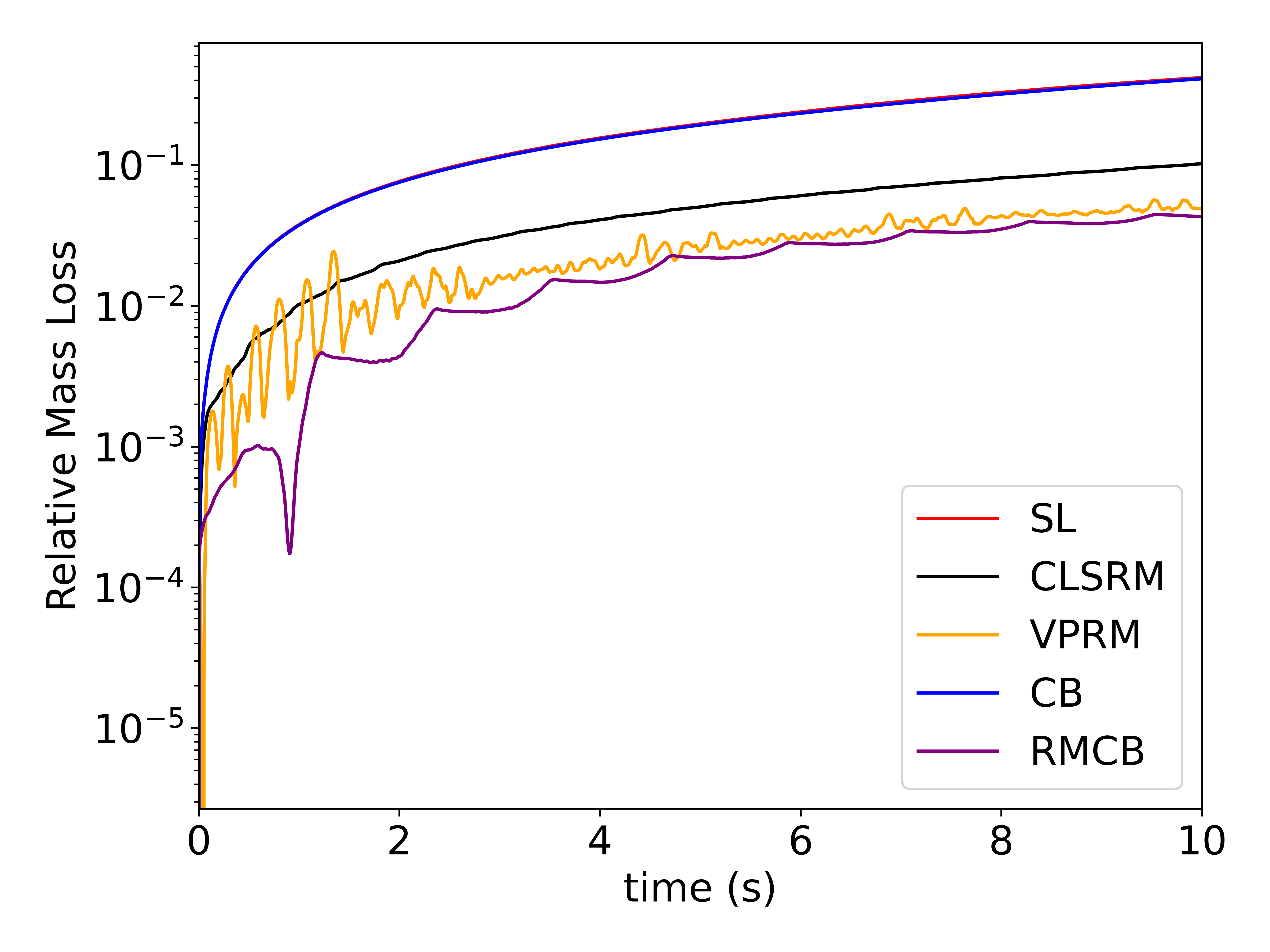}
    \hfill
    \includegraphics[width=0.49\linewidth]{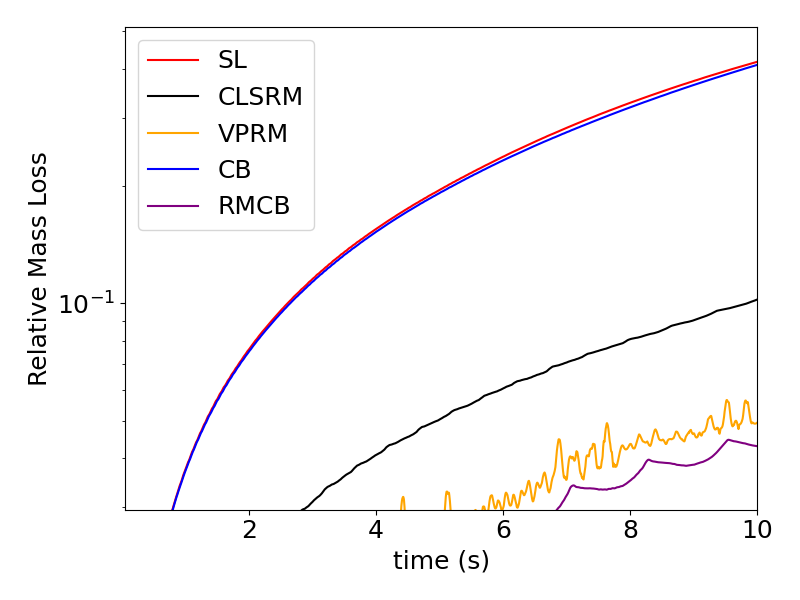}
    \caption{Relative volume loss for each of the different advection schemes (SL, RM, VPRM, CB, and RMCB) with a maximum refinement level of 8 for Case \textit{a} of \cite{bhaga1981bubbles}. The right panel provides an enlarged view of the left figure, specifically highlighting the subtle differences between the SL and CB methods.}
    \label{fig:rising_bubble_case_a}
\end{figure}

Figure~\ref{fig:rising_bubble_case_a} presents the relative mass loss for each advection scheme using Case \textit{a} from \cite{bhaga1981bubbles}. We highlight the mass loss using Case $a$, because this example exhibits only minor interface deformation and effectively showcases the benefits of the reference map-based methods. Additionally, the parameters of Case $a$ closely align those presented in the mass loss tests presented in previous studies (\textit{e.g.}, \cite{theillard2021vprm}). 

We observe that the SL and CB methods produce nearly identical results, with the CB scheme yielding a slight reduction in mass loss. A more pronounced distinction emerges when comparing the VPRM and RMCB methods. The RMCB scheme offers a clear improvement in relative mass loss and generates noticeably smoother solutions. Comparisons among the SL, CLSRM, and VPRM schemes reveal trends consistent with those reported in \cite{theillard2021vprm}. This example therefore helps to situate the CB and RMCB methods within the existing literature.

\begin{figure}[!tb]
    \centering
    \includegraphics[width=0.95\linewidth]{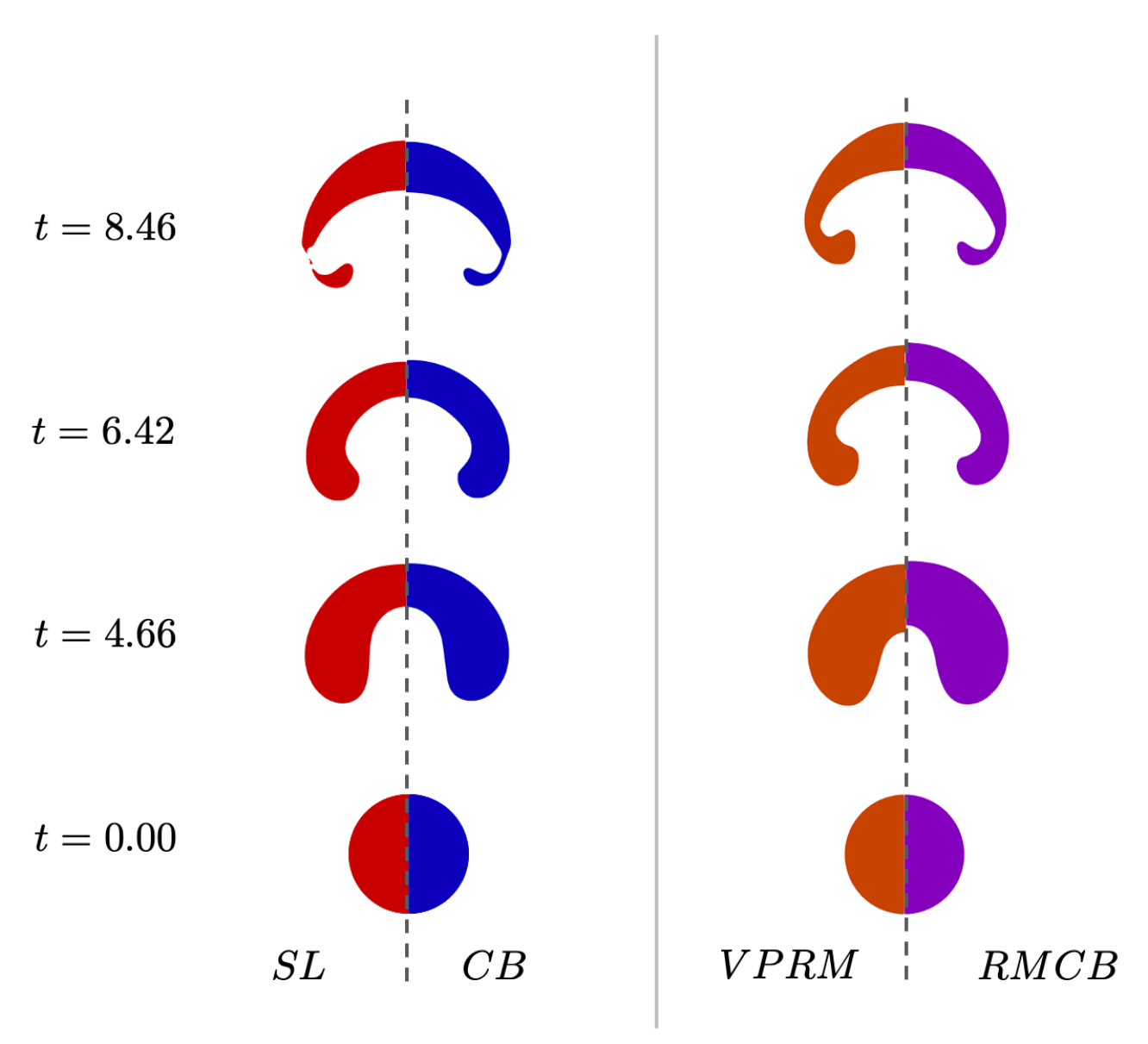}
    \caption{Comparison of characteristic bending advection with the traditional semi-Lagrangian method and the Volume-Preserving Reference Map \cite{theillard2021vprm} for Bhaga-Weber case f in two-dimensions. Results are shown for a minimum and maximum level of 4 and 9, respectively.}
    \label{fig:rising_bubble_case_f}
\end{figure}

In Figure~\ref{fig:rising_bubble_case_f}, we present a comparison of the SL, CB, VPRM, and RMCB schemes for Bhaga-Weber Case $f$ using a maximum refinement level of 9. As noted in \cite{binswanger2025multiphase}, Case $f$ presents some challenges for the multiphase solver and we can highlight the early rupture of the interface using this two-dimensional example. Comparing the SL and CB schemes, we can see that the CB method preserves the interface shape longer than the standard SL method. We do expect this improvement as the CB is equipped with the volume-preserving projection. Focusing on the comparison of the VPRM and RMCB methods, we see that the RMCB implementation results in a significant improvement in retaining the shape of the bubble, whereas the VPRM method deforms the bubble profile. We also see that the VPRM moves mass into the concave locations of interface thickening the skirt. As with the previous example, these differences decrease as the mesh is refined and the shapes of the interface converge. 

\section{Conclusions} \label{sec:conclusions}
%
The characteristic bending (CB) method extends the foundational ideas of the Coupled Level Set Reference Map (CLSRM) \cite{bellotti2019rm} and the Volume-Preserving Reference Map (VPRM) \cite{theillard2021vprm} to efficiently and accurately solve advection problems in incompressible flows. By reformulating the volume-preserving projection of \cite{theillard2021vprm} to operate on maps close to the identity, we obtain a generalized algorithm that improves stability and robustness across a range of test cases. Our results demonstrate that the CB method maintains accuracy even in the presence of artificial compressible modes and performs particularly well in preserving interface geometry for highly deformational flows.

Beyond scalar advection, the CB framework also performs well for nonlinear systems. In the incompressible Euler and multiphase Navier–Stokes examples, the advecting velocity fields contain only the errors introduced by the underlying spatial and temporal discretizations. Nevertheless, the CB method yields noticeable improvements. In the Euler case, it enhances accuracy and better preserves the divergence-free structure of the velocity field. In the Navier–Stokes simulations, the RMCB approach (the CLSRM method of \cite{bellotti2019rm} augmented with CB) reduces mass loss while maintaining a sharp interface. This is a notable improvement over the VPRM approach in \cite{theillard2021vprm}, which excels at preserving mass but can distort the interface. Although these improvements diminish as the grid is refined, the CB method consistently produces more robust and geometrically faithful results.

There are several promising research directions for this work. A natural next step is to explore more complex advection problems, including additional nonlinear cases and further examples involving the incompressible Euler equations, such as magnetohydrodynamic (MHD) flows as in \cite{yin2024characteristic}. In the multiphase context, it would be valuable to examine how the CB method influences the dynamics of interacting interfaces, for instance, in simulations of merging bubbles, and to assess its performance in flows with variable density or strong stratification (\textit{e.g.}, \cite{mandel2020retention}). Finally, an exciting direction is to investigate whether characteristic bending can be applied to compressible flows, where the reference map framework might be leveraged to bend characteristics around shocks and rarefactions in a physically consistent way.

We conclude by highlighting that the CB method presented herein is a general approach to solving transport problems with incompressible fields. Throughout this work, we emphasized the similarities of our approach to earlier methods and highlight the key reasons why our formulation is well-suited to filtering out artificial or spurious compressible modes in the advecting velocity. By recasting the classic semi-Lagrangian method in a reference map setting, we are able to leverage a volume-preserving projection to bend the characteristics of the velocity field, yielding an accurate and robust advection scheme. We show how the CB method can be used as a drop in replacement for traditional semi-Lagrangian schemes and how CB can be used to augment existing reference map-based formulations, demonstrating that CB is a flexible approach for constructing accurate and stable methods for incompressible advection. CB provides a unified perspective that builds on previous work and offers a practical foundation for developing more accurate and robust algorithms for incompressible flows.

\section{Acknowledgments}
This material is based upon work supported by the National Science Foundation under Grant No. DMS-1840265.

\section*{CRediT author statement }  
{\bf Matthew Blomquist}: Conceptualization, Formal analysis, Investigation, Methodology, Software, Validation, Visualization, Writing - Original Draft, Writing - Review $\&$ Editing. {\bf Stéphane Gaudreault}: Conceptualization, Writing - Original Draft, Writing - Review $\&$ Editing. {\bf Maxime Theillard}: Conceptualization, Supervision, Formal analysis, Methodology, Software, Project administration, Writing - Original Draft, Writing - Review $\&$ Editing.

\appendix
\section{Numerical Implementation}
\subsection{Data Structures}
\begin{figure}[!htb]
\centering
\includegraphics[width=0.85\linewidth]{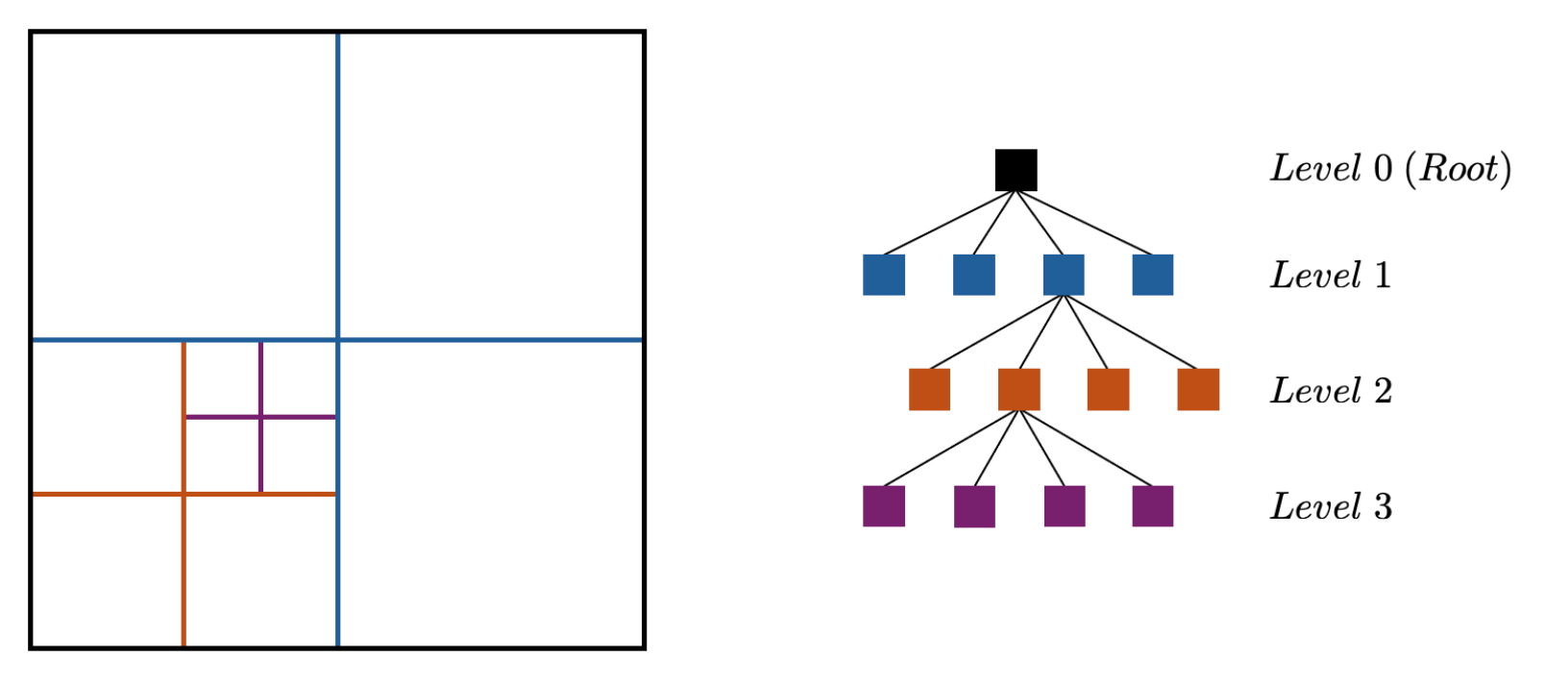}
\caption{Quadtree data structure for adaptive grid representation. The root cell covers the full domain, and refinement is applied recursively.}
\label{fig:quadtree_grid}
\end{figure}

%
We discretize the computational domain using adaptive quadtree (2D) and octree (3D) grids. The root cell represents the entire domain, and refinement proceeds recursively: each parent cell is subdivided into four (in 2D) or eight (in 3D) children, as illustrated in Figure \ref{fig:quadtree_grid}.

We employ non-graded grids, meaning that no restriction is imposed on the relative refinement levels of neighboring cells. This flexibility introduces the standard issue of hanging nodes (or T-junctions), where a node lacks a direct neighbor along one coordinate direction. For finite-difference discretizations, this creates difficulties in constructing standard stencils.

\begin{figure}[!ht]
\centering
\includegraphics[width = .75\textwidth]{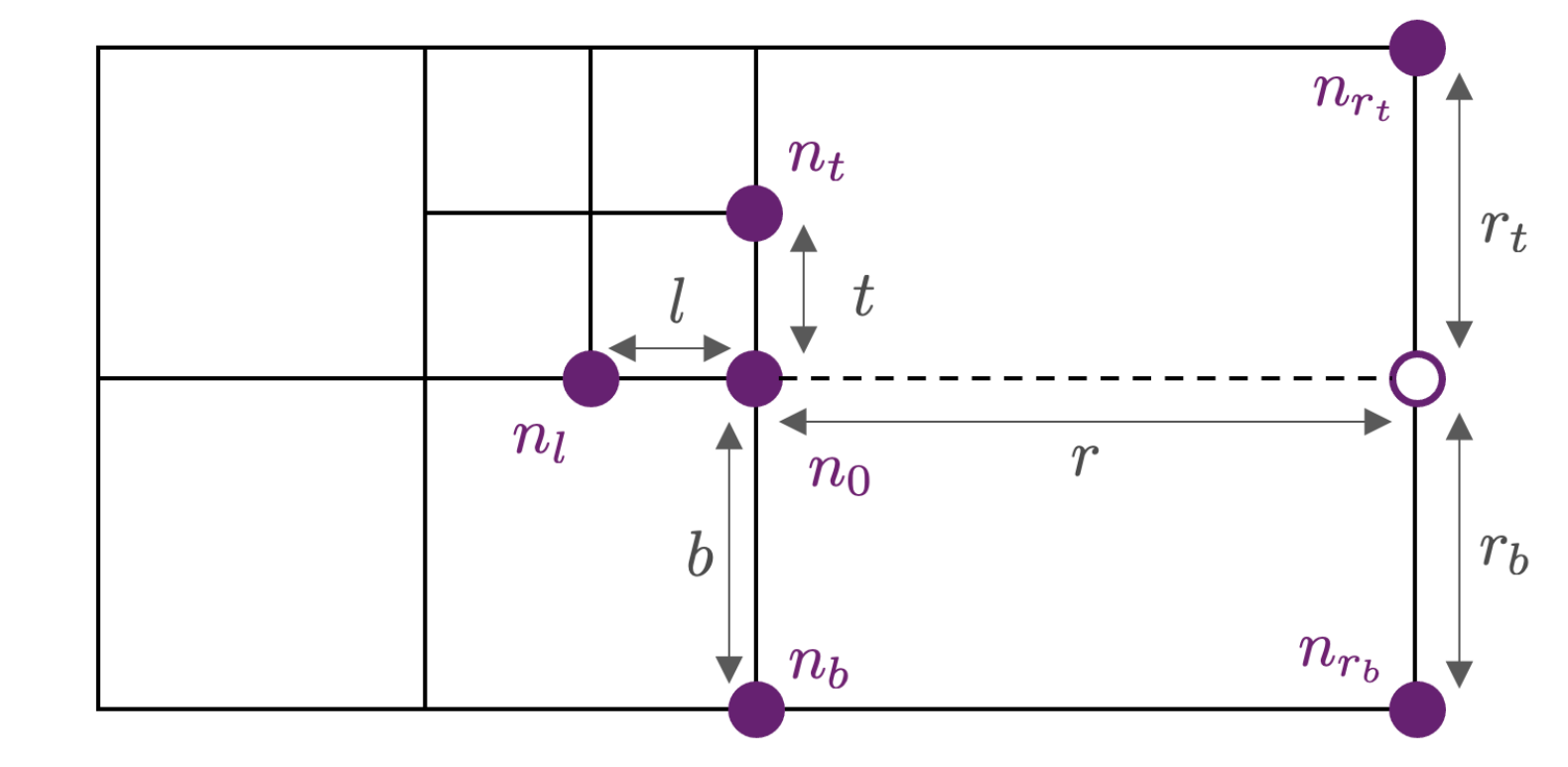}
\caption{Finite-difference discretization on a quadtree grid. Node $n_0$ has no direct right neighbor, so a ghost node $n_r$ (\includegraphics[height=0.015\textwidth]{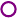}) is introduced along the edge defined by $n_{r_t}$ and $n_{r_b}$. The ghost value $\phi_r$ is computed from neighboring nodes (\includegraphics[height=0.015\textwidth]{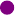}), enabling construction of standard central-difference stencils \cite{blomquist2024stable}.}
\label{fig:laplaciannodes}
\end{figure}

To resolve this, we follow the approach of Min and Gibou \cite{min2006supra} and introduce ghost nodes at T-junctions. Ghost values are defined for all nodal quantities using a third-order accurate interpolation scheme, thereby enabling consistent stencils. An example is shown in Figure \ref{fig:laplaciannodes}, where node $n_0$ has no direct right neighbor. We construct a ghost node $n_r$ along the edge connecting $n_{r_t}$ and $n_{r_b}$. For a generic nodal variable $\phi$, the ghost value $\phi_r$ is obtained from the surrounding nodes via the interpolation formula
\begin{equation}
\phi_r = \frac{r_b \phi_{r_t} + r_t \phi_{r_b}}{r_t + r_b} - \frac{r_t r_b}{t+b} \left( \frac{\phi_t - \phi_0}{t} - \frac{\phi_0 - \phi_b}{b} \right),
\label{eq:ghostnode_int}
\end{equation}
where $t$ and $b$ denote the distances from $n_0$ to its top and bottom neighbors, respectively.

\subsection{Adaptive Mesh Refinement} \label{sec:amr}
%
As was done in previous studies \cite{guittet2015stable, theillard2019sharp, bellotti2019rm, theillard2021vprm, blomquist2024stable, binswanger2025multiphase}, the quad/octree mesh is dynamically refined near interfaces (\textit{e.g.} level sets or fluid-fluid interfaces) or in areas of interesting dynamics (high velocity or vorticity gradients). Typically, at the completion of each time step, we recursively apply the chosen splitting criteria at each cell. Near interfaces, we split each cell $\mathcal{C}$ if the following criterion is met
\begin{align}
\label{eq:intf_refine}
    \min_{n \in \text{nodes} (\mathcal{C})} \left | \phi(n) \right | \leq B\cdot\text{Lip}(\phi) \cdot \text{diag}(\mathcal{C}) 
    \quad \text{and} \quad 
    \text{level}(\mathcal{C}) \leq \textrm{max}_{\textrm{level}} \text{,}
\end{align}
where $\text{Lip}(\phi)$ is an upper estimate of the minimal Lipschitz constant of the level-set function $\phi$, $\text{diag}(\mathcal{C})$ is the length of the diagonal of cell $\mathcal{C}$, $B$ is the user-specified width of the uniform band around the interface, where finer resolution is desired. We use $\max_{\textrm{level}}$ as the user-specified maximum grid level. Since the level set used to create the mesh will be reinitialized and $|\nabla \phi|=1$, we use  $\text{Lip}(\phi)=1.2$.   

The refinement based criterion for areas of high velocity gradients is the following
\begin{equation}
\label{eq:vort_refine}
    \min_{n \in \text{nodes} (\mathcal{C})} \text{diag}(\mathcal{C})\cdot \frac{\|\nabla\mathbf{u}\|}{\|\mathbf{u}\|_\infty} \geq T_V 
    \quad \text{and} \quad 
    \text{level}(\mathcal{C}) \leq \textrm{max}_{V} \text{,}     
\end{equation}
where $T_V$ is the user-specified threshold on the velocity gradient and $\textrm{max}_{V}$ is the maximum grid level allowed for velocity gradient-based refinement. We typically choose the velocity gradient-based refinement maximum level $\textrm{max}_{V}$ to be one level lower than the maximum grid level $\max_{\textrm{level}}$ due to the most significant dynamics occurring near the interface. 

Similarly, the vorticity refinement is
\begin{equation}
\label{eq:vort_refine}
    \min_{n \in \text{nodes} (\mathcal{C})} \text{diag}(\mathcal{C})\cdot \frac{\|\nabla \times \mathbf{u}\|}{\|\mathbf{u}\|_\infty} \geq T_W 
    \quad \text{and} \quad 
    \text{level}(\mathcal{C}) \leq \textrm{max}_{W} \text{,}     
\end{equation}
where $T_W$ is the user-specified threshold on the vorticity and $\textrm{max}_{W}$ is the maximum grid level allowed for vorticity-based refinement. As with the velocity refinement, we typically chose $\textrm{max}_{W}$ to be one level lower than the maximum grid level.

Finally, we ensure that a minimum resolution of $\min_{\textrm{level}}$ is maintained through the following refinement criterion,
\begin{equation}
    \text{level}(\mathcal{C}) \geq \textrm{min}_{\textrm{level}}.\label{eq:refcritminlevel}
\end{equation}
If none of these criteria are met, we merge $\mathcal{C}$ by removing all its descendants. 

\printbibliography[
heading=bibintoc,
title={References}
]

@book{boyd2001,
  title={Chebyshev and Fourier spectral methods},
  author={Boyd, John P},
  year={2001},
  publisher={Courier Corporation}
}

@article{bellotti2019rm,
  title={A coupled level-set and reference map method for interface representation with applications to two-phase flows simulation},
  author={Bellotti, Thomas and Theillard, Maxime},
  journal={Journal of Computational Physics},
  volume={392},
  pages={266--290},
  year={2019},
  publisher={Elsevier}
}

@article{cossette2014monge,
  title={The {M}onge--{A}mp{\`e}re trajectory correction for semi-{L}agrangian schemes},
  author={Cossette, Jean-Fran{\c{c}}ois and Smolarkiewicz, Piotr K and Charbonneau, Paul},
  journal={Journal of Computational Physics},
  volume={274},
  pages={208--229},
  year={2014},
  publisher={Elsevier}
}

@article{theillard2021vprm,
  title={A volume-preserving reference map method for the level set representation},
  author={Theillard, Maxime},
  journal={Journal of Computational Physics},
  volume={442},
  pages={110478},
  year={2021},
  publisher={Elsevier}
}

@article{leveque1996singlevortex,
  title={High-resolution conservative algorithms for advection in incompressible flow},
  author={Leveque, Randall J},
  journal={SIAM Journal on Numerical Analysis},
  volume={33},
  number={2},
  pages={627--665},
  year={1996},
  publisher={SIAM}
}

@article{adalsteinsson1999fast,
  title={The fast construction of extension velocities in level set methods},
  author={Adalsteinsson, David and Sethian, James A},
  journal={Journal of Computational Physics},
  volume={148},
  number={1},
  pages={2--22},
  year={1999},
  publisher={Elsevier}
}

@article{zalesak1979fully,
  title={Fully multidimensional flux-corrected transport algorithms for fluids},
  author={Zalesak, Steven T},
  journal={Journal of computational physics},
  volume={31},
  number={3},
  pages={335--362},
  year={1979},
  publisher={Elsevier}
}

@article{falcone1998slconvergence,
  title={Convergence analysis for a class of high-order semi-Lagrangian advection schemes},
  author={Falcone, Maurizio and Ferretti, Roberto},
  journal={SIAM Journal on Numerical Analysis},
  volume={35},
  number={3},
  pages={909--940},
  year={1998},
  publisher={SIAM}
}

@article{sethian1999level,
  title={Level Set Methods and Fast Marching Methods},
  author={Sethian, JA},
  journal={Cambridge University},
  year={1999}
}

@book{osher2003lsbook,
author = {Osher, Stanley. and Fedkiw, Ronald P.},
address = {New York},
booktitle = {Level set methods and dynamic implicit surfaces},
isbn = {0387954821},
language = {eng},
lccn = {2002020939},
publisher = {Springer},
series = {Applied mathematical sciences ; v. 153},
title = {Level set methods and dynamic implicit surfaces },
year = {2003},
}

@article{shu1988efficient,
  title={Efficient implementation of essentially non-oscillatory shock-capturing schemes},
  author={Shu, Chi-Wang and Osher, Stanley},
  journal={Journal of computational physics},
  volume={77},
  number={2},
  pages={439--471},
  year={1988},
  publisher={Elsevier}
}

@article{sussman1994level,
  title={A level set approach for computing solutions to incompressible two-phase flow},
  author={Sussman, Mark and Smereka, Peter and Osher, Stanley},
  journal={Journal of Computational physics},
  volume={114},
  number={1},
  pages={146--159},
  year={1994},
  publisher={Elsevier}
}

@article{sussman1999efficient,
  title={An efficient, interface-preserving level set redistancing algorithm and its application to interfacial incompressible fluid flow},
  author={Sussman, Mark and Fatemi, Emad},
  journal={SIAM Journal on scientific computing},
  volume={20},
  number={4},
  pages={1165--1191},
  year={1999},
  publisher={SIAM}
}

@article{saye2014high,
  title={High-order methods for computing distances to implicitly defined surfaces},
  author={Saye, Robert},
  journal={Communications in Applied Mathematics and Computational Science},
  volume={9},
  number={1},
  pages={107--141},
  year={2014},
  publisher={Mathematical Sciences Publishers}
}

@article{bell1989second,
  title={A second-order projection method for the incompressible Navier-Stokes equations},
  author={Bell, John B and Colella, Phillip and Glaz, Harland M},
  journal={Journal of computational physics},
  volume={85},
  number={2},
  pages={257--283},
  year={1989},
  publisher={Elsevier}
}

@article{blomquist2024stable,
  title={Stable nodal projection method on octree grids},
  author={Blomquist, Matthew and West, Scott R and Binswanger, Adam L and Theillard, Maxime},
  journal={Journal of Computational Physics},
  volume={499},
  pages={112695},
  year={2024},
  publisher={Elsevier}
}

@article{binswanger2025multiphase,
  title={Collocated Projection Method for Sharp Two-Phase Flows},
  author={Binswanger, Adam and Blomquist, Matthew and West, Scott R. and Khatri, Shilpa and Theillard, Maxime},
  journal={In Preparation},
  year={2025}
}

@article{staniforth1991semi,
  title={Semi-Lagrangian integration schemes for atmospheric models—A review},
  author={Staniforth, Andrew and C{\^o}t{\'e}, Jean},
  journal={Monthly weather review},
  volume={119},
  number={9},
  pages={2206--2223},
  year={1991}
}

@article{kamrin2012reference,
  title={Reference map technique for finite-strain elasticity and fluid--solid interaction},
  author={Kamrin, Ken and Rycroft, Chris H and Nave, Jean-Christophe},
  journal={Journal of the Mechanics and Physics of Solids},
  volume={60},
  number={11},
  pages={1952--1969},
  year={2012},
  publisher={Elsevier}
}

@article{rycroft2020reference,
  title={Reference map technique for incompressible fluid--structure interaction},
  author={Rycroft, Chris H and Wu, Chen-Hung and Yu, Yue and Kamrin, Ken},
  journal={Journal of Fluid Mechanics},
  volume={898},
  pages={A9},
  year={2020},
  publisher={Cambridge University Press}
}

@book{karniadakis2005spectral,
  title={Spectral/hp element methods for computational fluid dynamics},
  author={Karniadakis, George and Sherwin, Spencer},
  year={2005},
  publisher={OUP Oxford}
}

@article{robert1981stable,
  title={A stable numerical integration scheme for the primitive meteorological equations},
  author={Robert, Andr{\'e}},
  journal={Atmosphere-Ocean},
  volume={19},
  number={1},
  pages={35--46},
  year={1981},
  publisher={Taylor \& Francis}
}

@article{sawyer1963semi,
  title={A semi-Lagrangian method of solving the vorticity advection equation},
  author={Sawyer, John Stanley},
  journal={Tellus},
  volume={15},
  number={4},
  pages={336--342},
  year={1963},
  publisher={Taylor \& Francis}
}

@article{xiu2001semi,
  title={A semi-Lagrangian high-order method for Navier--Stokes equations},
  author={Xiu, Dongbin and Karniadakis, George Em},
  journal={Journal of computational physics},
  volume={172},
  number={2},
  pages={658--684},
  year={2001},
  publisher={Elsevier}
}

@article{rosatti2005semi,
  title={Semi-implicit, semi-Lagrangian modelling for environmental problems on staggered Cartesian grids with cut cells},
  author={Rosatti, Giorgio and Cesari, D and Bonaventura, Luca},
  journal={Journal of Computational Physics},
  volume={204},
  number={1},
  pages={353--377},
  year={2005},
  publisher={Elsevier}
}

@article{yabe2001exactly,
  title={An exactly conservative semi-Lagrangian scheme (CIP--CSL) in one dimension},
  author={Yabe, Takashi and Tanaka, Ryotaro and Nakamura, Takashi and Xiao, Feng},
  journal={Monthly Weather Review},
  volume={129},
  number={2},
  pages={332--344},
  year={2001}
}

@article{priestley1993quasi,
  title={A quasi-conservative version of the semi-Lagrangian advection scheme},
  author={Priestley, A},
  journal={Monthly Weather Review},
  volume={121},
  number={2},
  pages={621--629},
  year={1993}
}

@article{xiao2001completely,
  title={Completely conservative and oscillationless semi-Lagrangian schemes for advection transportation},
  author={Xiao, Feng and Yabe, Takashi},
  journal={Journal of computational physics},
  volume={170},
  number={2},
  pages={498--522},
  year={2001},
  publisher={Elsevier}
}

@article{giraldo2000lagrange,
  title={The Lagrange--Galerkin method for the two-dimensional shallow water equations on adaptive grids},
  author={Giraldo, Francis X},
  journal={International Journal for Numerical Methods in Fluids},
  volume={33},
  number={6},
  pages={789--832},
  year={2000},
  publisher={Wiley Online Library}
}

@article{giraldo2000lagrange2,
  title={Lagrange--Galerkin methods on spherical geodesic grids: the shallow water equations},
  author={Giraldo, Francis X},
  journal={Journal of Computational Physics},
  volume={160},
  number={1},
  pages={336--368},
  year={2000},
  publisher={Elsevier}
}

@article{donea1982arbitrary,
  title={An arbitrary Lagrangian-Eulerian finite element method for transient dynamic fluid-structure interactions},
  author={Donea, Jean and Giuliani, SHJP and Halleux, Jean-Pierre},
  journal={Computer methods in applied mechanics and engineering},
  volume={33},
  number={1-3},
  pages={689--723},
  year={1982},
  publisher={Elsevier}
}

@article{hughes1981lagrangian,
  title={Lagrangian-Eulerian finite element formulation for incompressible viscous flows},
  author={Hughes, Thomas JR and Liu, Wing Kam and Zimmermann, Thomas K},
  journal={Computer methods in applied mechanics and engineering},
  volume={29},
  number={3},
  pages={329--349},
  year={1981},
  publisher={Elsevier}
}

@article{hirt1974arbitrary,
  title={An arbitrary Lagrangian-Eulerian computing method for all flow speeds},
  author={Hirt, Cyrill W and Amsden, Anthony A and Cook, JL},
  journal={Journal of computational physics},
  volume={14},
  number={3},
  pages={227--253},
  year={1974},
  publisher={Elsevier}
}

@phdthesis{kamrin2008stochastic,
  title={Stochastic and deterministic models for dense granular flow},
  author={Kamrin, Kenneth Norman},
  year={2008},
  school={Massachusetts Institute of technology}
}

@article{cottet2008eulerian,
  title={Eulerian formulation and level set models for incompressible fluid-structure interaction},
  author={Cottet, Georges-Henri and Maitre, Emmanuel and Milcent, Thomas},
  journal={ESAIM: Mathematical Modelling and Numerical Analysis},
  volume={42},
  number={3},
  pages={471--492},
  year={2008},
  publisher={EDP Sciences}
}

@article{maitre2009applications,
  title={Applications of level set methods in computational biophysics},
  author={Maitre, Emmanuel and Milcent, Thomas and Cottet, Georges-Henri and Raoult, Annie and Usson, Yves},
  journal={Mathematical and Computer Modelling},
  volume={49},
  number={11-12},
  pages={2161--2169},
  year={2009},
  publisher={Elsevier}
}

@article{mercier2020characteristic,
  title={The characteristic mapping method for the linear advection of arbitrary sets},
  author={Mercier, Olivier and Yin, Xi-Yuan and Nave, Jean-Christophe},
  journal={SIAM Journal on Scientific Computing},
  volume={42},
  number={3},
  pages={A1663--A1685},
  year={2020},
  publisher={SIAM}
}

@article{kohno2013new,
  title={A new method for the level set equation using a hierarchical-gradient truncation and remapping technique},
  author={Kohno, Haruhiko and Nave, Jean-Christophe},
  journal={Computer Physics Communications},
  volume={184},
  number={6},
  pages={1547--1554},
  year={2013},
  publisher={Elsevier}
}

@article{nave2010gradient,
  title={A gradient-augmented level set method with an optimally local, coherent advection scheme},
  author={Nave, Jean-Christophe and Rosales, Rodolfo Ruben and Seibold, Benjamin},
  journal={Journal of Computational Physics},
  volume={229},
  number={10},
  pages={3802--3827},
  year={2010},
  publisher={Elsevier}
}

@article{liu1994weighted,
  title={Weighted essentially non-oscillatory schemes},
  author={Liu, Xu-Dong and Osher, Stanley and Chan, Tony},
  journal={Journal of computational physics},
  volume={115},
  number={1},
  pages={200--212},
  year={1994},
  publisher={Elsevier}
}

@article{min2006supra,
  title={A supra-convergent finite difference scheme for the variable coefficient Poisson equation on non-graded grids},
  author={Min, Chohong and Gibou, Fr{\'e}d{\'e}ric and Ceniceros, Hector D},
  journal={Journal of Computational Physics},
  volume={218},
  number={1},
  pages={123--140},
  year={2006},
  publisher={Elsevier}
}

@article{sussman2000coupled,
  title={A coupled level set and volume-of-fluid method for computing 3D and axisymmetric incompressible two-phase flows},
  author={Sussman, Mark and Puckett, Elbridge Gerry},
  journal={Journal of computational physics},
  volume={162},
  number={2},
  pages={301--337},
  year={2000},
  publisher={Elsevier}
}

@article{sussman2003second,
  title={A second order coupled level set and volume-of-fluid method for computing growth and collapse of vapor bubbles},
  author={Sussman, Mark},
  journal={Journal of Computational Physics},
  volume={187},
  number={1},
  pages={110--136},
  year={2003},
  publisher={Elsevier}
}

@article{olsson2005conservative,
  title={A conservative level set method for two phase flow},
  author={Olsson, Elin and Kreiss, Gunilla},
  journal={Journal of computational physics},
  volume={210},
  number={1},
  pages={225--246},
  year={2005},
  publisher={Elsevier}
}

@article{olsson2007conservative,
  title={A conservative level set method for two phase flow II},
  author={Olsson, Elin and Kreiss, Gunilla and Zahedi, Sara},
  journal={Journal of Computational Physics},
  volume={225},
  number={1},
  pages={785--807},
  year={2007},
  publisher={Elsevier}
}

@article{desjardins2008accurate,
  title={An accurate conservative level set/ghost fluid method for simulating turbulent atomization},
  author={Desjardins, Olivier and Moureau, Vincent and Pitsch, Heinz},
  journal={Journal of computational physics},
  volume={227},
  number={18},
  pages={8395--8416},
  year={2008},
  publisher={Elsevier}
}

@article{owkes2013discontinuous,
  title={A discontinuous Galerkin conservative level set scheme for interface capturing in multiphase flows},
  author={Owkes, Mark and Desjardins, Olivier},
  journal={Journal of Computational Physics},
  volume={249},
  pages={275--302},
  year={2013},
  publisher={Elsevier}
}

@article{guittet2015stable,
  title={A stable projection method for the incompressible Navier--Stokes equations on arbitrary geometries and adaptive Quad/Octrees},
  author={Guittet, Arthur and Theillard, Maxime and Gibou, Fr{\'e}d{\'e}ric},
  journal={Journal of computational physics},
  volume={292},
  pages={215--238},
  year={2015},
  publisher={Elsevier}
}

@article{theillard2019sharp,
  title={Sharp numerical simulation of incompressible two-phase flows},
  author={Theillard, Maxime and Gibou, Fr{\'e}d{\'e}ric and Saintillan, David},
  journal={Journal of Computational Physics},
  volume={391},
  pages={91--118},
  year={2019},
  publisher={Elsevier}
}

@article{zerroukat2002slice,
  title={SLICE: A Semi-Lagrangian Inherently Conserving and Efficient scheme for transport problems},
  author={Zerroukat, Mohamed and Wood, Nigel and Staniforth, Andrew},
  journal={Quarterly Journal of the Royal Meteorological Society: A journal of the atmospheric sciences, applied meteorology and physical oceanography},
  volume={128},
  number={586},
  pages={2801--2820},
  year={2002},
  publisher={Wiley Online Library}
}

@article{lauritzen2010cslam,
  title={A conservative semi-Lagrangian multi-tracer transport scheme (CSLAM) on the cubed-sphere grid},
  author={Lauritzen, Peter H and Nair, Ramachandran D and Ullrich, Paul A},
  journal={Journal of Computational Physics},
  volume={229},
  number={5},
  pages={1401--1424},
  year={2010},
  publisher={Elsevier}
}

@article{giraldo1998lagrange,
  title={The Lagrange--Galerkin spectral element method on unstructured quadrilateral grids},
  author={Giraldo, Francis X},
  journal={Journal of Computational Physics},
  volume={147},
  number={1},
  pages={114--146},
  year={1998},
  publisher={Elsevier}
}

@article{almgren1996numerical,
  title={A numerical method for the incompressible Navier-Stokes equations based on an approximate projection},
  author={Almgren, Ann S and Bell, John B and Szymczak, William G},
  journal={SIAM Journal on Scientific Computing},
  volume={17},
  number={2},
  pages={358--369},
  year={1996},
  publisher={SIAM}
}

@article{almgren2000approximate,
  title={Approximate projection methods: Part I. Inviscid analysis},
  author={Almgren, Ann S and Bell, John B and Crutchfield, William Y},
  journal={SIAM Journal on Scientific Computing},
  volume={22},
  number={4},
  pages={1139--1159},
  year={2000},
  publisher={SIAM}
}

@article{almgren1997cartesian,
  title={A Cartesian grid projection method for the incompressible Euler equations in complex geometries},
  author={Almgren, Ann S and Bell, John B and Colella, Phillip and Marthaler, Tyler},
  journal={SIAM Journal on Scientific Computing},
  volume={18},
  number={5},
  pages={1289--1309},
  year={1997},
  publisher={SIAM}
}

@article{almgren1998conservative,
  title={A conservative adaptive projection method for the variable density incompressible Navier--Stokes equations},
  author={Almgren, Ann S and Bell, John B and Colella, Phillip and Howell, Louis H and Welcome, Michael L},
  journal={Journal of computational Physics},
  volume={142},
  number={1},
  pages={1--46},
  year={1998},
  publisher={Elsevier}
}

@book{boffi2013mixed,
  title={Mixed finite element methods and applications},
  author={Boffi, Daniele and Brezzi, Franco and Fortin, Michel and others},
  volume={44},
  year={2013},
  publisher={Springer}
}

@article{vargas2025multi,
  title={Multi-material ALE remap with interface sharpening using high-order matrix-free finite element methods},
  author={Vargas, Arturo and Tomov, Vladimir Z and Skinner, M Aaron and Dobrev, Veselin and Nikl, Jan and Kolev, Tzanio and Rieben, Robert N},
  journal={Journal of Computational Physics},
  pages={114367},
  year={2025},
  publisher={Elsevier}
}

@article{dobrev2012high,
  title={High-order curvilinear finite element methods for Lagrangian hydrodynamics},
  author={Dobrev, Veselin A and Kolev, Tzanio V and Rieben, Robert N},
  journal={SIAM Journal on Scientific Computing},
  volume={34},
  number={5},
  pages={B606--B641},
  year={2012},
  publisher={SIAM}
}

@article{yin2021characteristic,
  title={A characteristic mapping method for the two-dimensional incompressible Euler equations},
  author={Yin, Xi-Yuan and Mercier, Olivier and Yadav, Badal and Schneider, Kai and Nave, Jean-Christophe},
  journal={Journal of Computational Physics},
  volume={424},
  pages={109781},
  year={2021},
  publisher={Elsevier}
}

@book{holman2024volume,
  title={A volume-preserving Characteristic Mapping Method for the 2D incompressible Euler equations},
  author={Holman-Bissegger, William},
  year={2024},
  publisher={McGill University (Canada)}
}

@inproceedings{lee2019robust,
  title={A robust volume conserving method for character-water interaction},
  author={Lee, Minjae and Hyde, David and Li, Kevin and Fedkiw, Ronald},
  booktitle={Proceedings of the 18th annual ACM SIGGRAPH/Eurographics Symposium on Computer Animation},
  pages={1--12},
  year={2019}
}

@article{bhaga1981bubbles,
  title={Bubbles in viscous liquids: shapes, wakes and velocities},
  author={Bhaga, Dahya and Weber, ME},
  journal={Journal of fluid Mechanics},
  volume={105},
  pages={61--85},
  year={1981},
  publisher={Cambridge University Press}
}

@article{yin2024characteristic,
  title={A Characteristic Mapping Method with Source Terms: Applications to Ideal Magnetohydrodynamics},
  author={Yin, Xi-Yuan and Krah, Philipp and Nave, Jean-Christophe and Schneider, Kai},
  journal={arXiv preprint arXiv:2411.13772},
  year={2024}
}

@article{mandel2020retention,
  title={Retention of rising droplets in density stratification},
  author={Mandel, Tracy L and Zhou, De Zhen and Waldrop, Lindsay and Theillard, Maxime and Kleckner, Dustin and Khatri, Shilpa},
  journal={Physical Review Fluids},
  volume={5},
  number={12},
  pages={124803},
  year={2020},
  publisher={APS}
}

@article{hortal2002development,
  title={The development and testing of a new two-time-level semi-Lagrangian scheme (SETTLS) in the ECMWF forecast model},
  author={Hortal, Mariano},
  journal={Quarterly Journal of the Royal Meteorological Society: A journal of the atmospheric sciences, applied meteorology and physical oceanography},
  volume={128},
  number={583},
  pages={1671--1687},
  year={2002},
  publisher={Wiley Online Library}
}

@book{fletcher2019semi,
  title={Semi-Lagrangian advection methods and their applications in geoscience},
  author={Fletcher, Steven J},
  year={2019},
  publisher={Elsevier}
}

@article{morton1985generalised,
  title={Generalised Galerkin methods for hyperbolic problems},
  author={Morton, KW},
  journal={Computer Methods in Applied Mechanics and Engineering},
  volume={52},
  number={1-3},
  pages={847--871},
  year={1985},
  publisher={Elsevier}
}

@article{enright2005fast,
  title={A fast and accurate semi-Lagrangian particle level set method},
  author={Enright, Douglas and Losasso, Frank and Fedkiw, Ronald},
  journal={Computers \& structures},
  volume={83},
  number={6-7},
  pages={479--490},
  year={2005},
  publisher={Elsevier}
}

\end{document}